# Statistical Mechanical Model for a closed loop plectoneme with weak helix specific forces

Dominic J. (O') Lee[1,a.)]

[1]Department of Chemistry, Imperial College London, SW7 2AZ, London, UK

## Abstract

We develop a statistical mechanical framework, based on a variational approximation, to describe closed loop plectonemes. This framework incorporates weak helix structure dependent forces into the determination of the free energy and average structure of a plectoneme. Notably, due to their chiral nature, helix structure dependent forces break the symmetry between left and right handed supercoiling. The theoretical approach, presented here, also provides a systematic way of enforcing the topological constraint of closed loop supercoiling in the variational approximation. At large plectoneme lengths, by considering correlation functions in an expansion in terms of the spatial mean twist density about its thermally averaged value, it can be argued that topological constraint may be approximated with by replacing twist and writhe by their thermal averages. A Lagrange Multiplier, containing the sum of average twist and writhe, can be added to the free energy to conveniently inforce this result. The average writhe can be calculated through the thermal average of Gauss' integral in the variational approximation. Furthermore, this approach allows for a possible way to calculate finite size corrections due to the topological constraint. Using interaction energy terms from the mean-field Kornyshev-Leikin theory, for parameter values that correspond to weak helix dependent forces, we calculate the free energy, fluctuation magnitudes and mean geometric parameters for the plectoneme. We see a slight asymmetry, where interestingly enough left handed supercoils have looser structure than right handed ones, although a lower free energy, unlike what previous ground state calculations would suggest.

## 1. Introduction

Supercoiling is important in the storage of DNA as well as gene expression. In prokaryotes DNA supercoiling is found in both plasmids and the bacterial chromosome, which are usually both formed of circular DNA. In many bacteria, like E. Coli, the DNA is negatively supercoiled [1,2]. In such organisms, the level of supercoiling has an important role to play in the expression and regulation of genes. Recently, it has been shown that different genes in E. Coli require different levels of supercoiling to be optimally expressed, and their actual positioning on the chromosome is influenced by this [3]. These levels can be controlled by the actual DNA transcription process, which generates its own supercoiling due to the under-winding of DNA [4]; and too much positive supercoiling may also inhibit transcription [5,6]. In eukaryotes supercoiling is also present. As well as DNA being coiled around histones, there has been found that there are domains of DNA with their specific own level of supercoiling [7]. These levels of supercoiling may be also important in transcription.

---

[a.)] Electronic Mail: domolee@hotmail.com

Understanding the nature of the interactions between DNA segments that make a supercoil is important in understanding how their structure changes with environment [8,9,10,11]. Supercoiling can bring distant sequences together for site specific recombination [12]. Also, the dynamics of supercoil diffusion [13], important in transcription and the expression of genes [14] (as well as replication), should also be influenced by intersegment interactions. For an equilibrium description of closed loop supercoiling in bacteria, analytical models [15,16,17] and simulations [11,18,19,20, 21,22] have been developed. Most importantly, single molecule experiments have also provided a way of probing the formation of plectoneme supercoils and the types of DNA structure (for instance, left handed L-DNA) that are formed when torsional stress is applied to the molecule[23,24,25,26,27]. These experiments have also been the topic of numerous theoretical studies [28,29,30,31,32,33,34,35,36].

An interesting line of speculation is how inter-segment interactions depending on the helix structure of a molecule might influence supercoiling. The chiral nature of helix dependent interactions results in asymmetry between left and right handed supercoils [37,38]. Such asymmetry between left and right hand supercoils may have some biological significance. Important questions are why most bacteria form negative (right handed) supercoils while hyper-thermophiles generally form positive (left handed) ones [39]; but on the other hand, why it is, under certain processes and conditions, supercoiling of the opposite handedness is observed for both hyper-thermophiles and other organisms [40]. This has attracted speculation in whether physical differences in left handed and right handed supercoils could be responsible for the choices that have been made in nature [37,38,40,41] . Particularly, Refs. [37] and [38] have speculated on the role of chiral helix specific interactions in determining this choice. A second reason to study the inclusion of such interactions into models of supercoiling is that helix specific forces might play a central role in a proposed recognition stage before recombination, where two DNA molecules with the similar base pair texts associate with each other before homologous recombination [42,43]. Here, supercoiling provides one testing ground to see if such forces exist, and are significant, due to their chiral origin. This second motivation for this study, and extending on it when considering this issue, are discussed in more detail in the final section of this paper; and what experiments, in the context of supercoiling, could be performed. Also, helix structure specific interaction may not just have relevance for DNA plectonemes; for instance, plectonemes of actin molecules have also experimentally studied [44].

How helix structure might possibly influence DNA-DNA interactions and those of other helically charged molecules was initially studied for parallel molecules, using a mean field electrostatic model [45,46], later ionic correlations were considered [47,48]. Importantly, the interaction between two parallel helical molecular segments is found to depend on the relative angle between the two DNA helices in the plane perpendicular to their long axes [45]. Furthermore, it was found that if helix forces was included for DNA molecules in a braided configuration the interaction energy was minimized by forming a left handed braid for right-handed DNA [38]. In the case of the mechanical braiding of two DNA molecules, it has been hypothesised [49] that helix structure dependent interactions might account for the small, but distinct, asymmetry in the experimental data of Ref. [50].

The strength of helix structure dependent forces may depend on how localized counter-ions are, as well as where they localize in the vicinity of the DNA molecule [42], and this should depend on ion species [42]. Also, the degree of localization may also depend on how close DNA helices are and on the degree of supercoiling. Bending and twisting of the molecules in the supercoil may help localize ions in the grooves due to increased interactions with the phosphates and base pairs, or alternately might hinder it due to steric constraints and other effects.

In determining the structure of a closed loop plectoneme, Ref. [51] investigated the ground state using the KL model of interaction [45,46]. The results of this calculation suggested that left handed supercoils have a tighter structure, and these positive supercoils may have substantially lower free energy than negatively supercoiled and relaxed DNA. Now we start to examine the effects of including both twisting fluctuations and undulations, based on previous theoretical development of the statistical mechanics [51,52,49 ,53]. Here, it is important to emphasize that the statistical mechanical calculations developed here are not just applicable to the KL theory of interaction. They could be made applicable to any more appropriate interaction model that depends on the helix symmetry of the molecule. However, the KL model provides an analytical framework for an investigation of the qualitative role of helix structure dependent forces. In this current study, we examine the case where helix structure dependent forces are considered to be weak due to thermal fluctuations, and choose an appropriate parameter range of KL model parameters to study this regime.

The paper is structured in the following way. In the next section, we start by describing the supercoiling theory. Here, we start by introducing the mathematical machinery we need to consider the statistical mechanics of closed loop supercoiling, including a description of the relative braid and helix geometry of the two stands making up the plectoneme, important when considering helix structure dependent interactions. Next, we go on to consider the elastic energies and how to include helix structure dependent forces. Lastly, we describe the full free energy used in the calculations. Its derivation is left to Appendices A and B of the supplemental material; further details can also be found in Ref. [52]. We show the results of calculations of the difference in the moment required to maintain a level of supercoiling $|\sigma|$ (modulus of the supercoiling density) between left and right handed supercoils, and the free energy. We also present plots of mean geometric parameters of the plectoneme as a function of supercoiling density, $\sigma$, and their degree of fluctuation. All of these calculations utilize the KL model of interaction, where we have varied the model parameters corresponding to: i.) the overall amount of counter-ions near the molecular surface (not accounted by Debye screening); ii.) the proportion of such ions localized near the DNA grooves; iii.) the relative ratio of the amount of ions localized near the minor DNA groove to the major one. In the main text, we present only variations of the first parameter, the charge compensation, in the results and discussion section. Results from variations in the other two parameters are given in Appendix F of the supplemental material, presented there for the interested reader, as they do not affect the main conclusions and findings of the paper. Importantly, we notice that the left-right hand symmetry is not broken to so large an extent as Ref. [51], which is in line with available experimental evidence [54,55]. The free energy for positive supercoils is still lower than that of right handed ones, the interaction between right handed helices would suggest [38,51]. However, what is interesting and slightly counter intuitive is that, here, negative supercoils form a tighter structure than positive ones. We discuss our results in the context of previous experimental work on closed looped plectonemes [10,54 ,55]. Finally, in concluding remarks and outlook, we summarize our findings, speculate on their biological significance, and point to future experimental and theoretical work.

## 2. Supercoiling theory

### 2.1 Initial theoretical considerations

Our starting point will be to divide the closed loop plectoneme into end loops and a braided section. We will define the length of the two braided sections as $L_b$, and $L$ as the total length around the plectoneme. Here, it is useful to define two coordinate systems to specify a point on the plectoneme. The first is a general coordinate $s$, which runs from $0$ to $L$, used conventionally. The coordinate $s$ may start at any reference point chosen on the molecular centre line tracing out the plectoneme; we choose it to be at the end of one segment of the braided section. The second is coordinate $\tau$ that defines exclusively a position along either of the two segments forming the braided section; $\tau$ is not used to describe any position on the end loops. This second coordinate runs from $0$ to $L_b$. In Fig. 1 we show a schematic picture of how we define the coordinates.

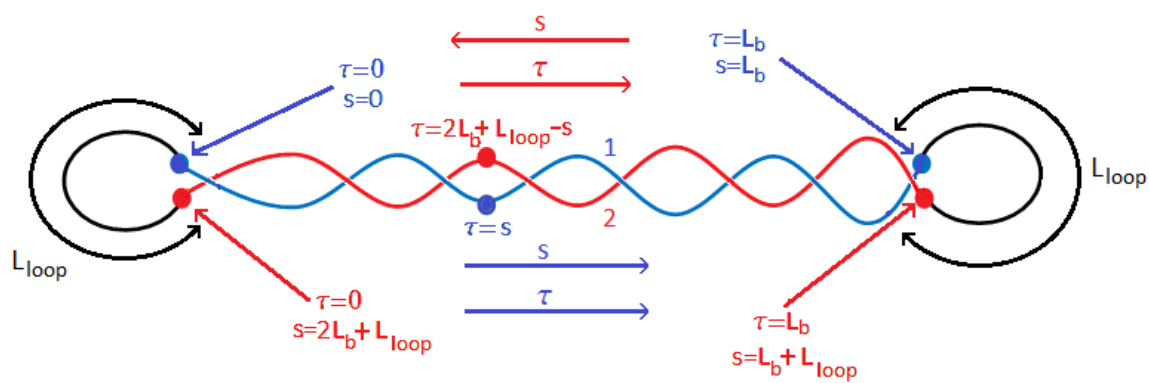

Fig.1. Schematic illustration of how the plectoneme is divided up in the model. The black parts of the plectoneme correspond to the end loop sections of length $L_{loop}$, the blue corresponds to segment 1 of braided section, and the red corresponds to segment 2. There are two sets of coordinates $s$ and $\tau$, $0 \leq s \leq 2L$ corresponds to any position around the entire closed loop, and $0 \leq \tau \leq L_b$ is defined only for the braided section. Shown as blue and red blobs are the ends of the braided section, corresponding to the sets of values $s=0, \tau=0$; $s=2L_b+L_{loop}, \tau=0$; $s=L_b, \tau=L_b$; and $s=L_b+L_{loop}, \tau=L_b$. Also shown (within the braided section) are coloured blobs corresponding to arbitrary points along the braid, where relationships between $\tau$ and $s$ are given for both segments 1 and 2. Also, the horizontal arrows show the directions of increasing $\tau$ and $s$ for both segments 1 and 2, along the braided section.

In this work we will suppose that plectoneme is sufficiently long and supercoiled that we can neglect end loops, as in previous studies [16,17]. Therefore, here, we suppose that $L \simeq 2L_b$ and neglect $L_{loop}$ (see Fig. 1). We consider an unbranched plectoneme. Investigation of effects of end loops combined with supercoil branching will be left to a later work. How to include end loops in this treatment has been discussed in Refs. [51,52], alternately for tight supercoiling a more sophisticated end loop ansatz could be used [15]. Thus, we consider the plectoneme as simply a braid formed by two sections, but still indeed subject to a topological constraint. The topological constraint is that the Linking number remains fixed. The Linking number is defined through the Fuller-White theorem

$$Lk = Wr + Tw. \qquad (2.1)$$

The first quantity is the Writhe and is computed, using Guass' integral, over the plectoneme. This integral is

$$Wr = \frac{1}{4\pi} \oint \oint ds ds' \frac{(\mathbf{r}(s) - \mathbf{r}(s')) \cdot \hat{\mathbf{t}}(s) \times \hat{\mathbf{t}}(s')}{|\mathbf{r}(s) - \mathbf{r}(s')|^{3/2}}. \qquad (2.2)$$

The position vector $\mathbf{r}(s)$ defines the position of a point along the molecular centreline in 3-D space and $\hat{\mathbf{t}}(s)$ is the tangent vector of that line (i.e. $\hat{\mathbf{t}}(s) = \mathbf{r}'(s)$, where the prime denotes differentiation with respect to the argument). The sign on the integral simply denotes a cyclic integral once round the closed loop, running from $s = 0$ to $L$, where for the full loop the position of $s = 0$ is arbitrary. Essentially, the value of the Writhe is an average of the number of times the centre line crosses its self, within a 2-D projection, over all possible 2-D projections.

The second quantity in Eq. (2.1) is the Twist, which is given by

$$Tw = \frac{1}{2\pi} \oint g(s) ds, \qquad (2.3)$$

where $g(s)$ is the local twist density. The local twist density and twist are intrinsic to the DNA double helix. The DNA double helix conformation may be most simply described by the trajectory of a curve, on the DNA surface, bisecting the centre of the minor groove. Essentially, the value of the twist is the number of times that the minor groove trajectory precesses around the molecular centre line (in the lab-frame) when we make one full cycle of the closed loop. The twist density is defined as

$$g(s) = \left( \hat{\mathbf{t}}(s) \times \hat{\mathbf{v}}(s) \right) \cdot \frac{d\hat{\mathbf{v}}(s)}{ds}. \qquad (2.4)$$

The vector $\hat{\mathbf{v}}(s)$ points away from the centreline and is chosen to bisect the minor grove at any point $s$, and is perpendicular to the tangent vector $\hat{\mathbf{t}}(s)$. Thus, $\hat{\mathbf{v}}(s)$ traces out the trajectory of the minor groove centre.

We now label the two segments making up the braid with labels $\mu = 1, 2$. For the two segments, in the braid, we define position vectors $\mathbf{r}_\mu(\tau)$ where

$$\mathbf{r}_1(\tau) = \mathbf{r}(s), \qquad\qquad 0 < s \leq L_b, \qquad (2.5)$$

$$\mathbf{r}_2(\tau) \simeq \mathbf{r}(2L_b - s), \qquad\qquad L_b \lesssim s \lesssim 2L_b. \qquad (2.6)$$

Similarly we can define tangent vectors $\hat{\mathbf{t}}_\mu(\tau) = \mathbf{r}'_\mu(\tau)$ and the vectors $\hat{\mathbf{v}}_\mu(\tau)$, where

$$\hat{\mathbf{t}}_1(\tau) = \hat{\mathbf{t}}(s), \qquad \hat{\mathbf{v}}_1(\tau) = \hat{\mathbf{v}}(s), \qquad\qquad 0 < s \leq L_b, \qquad (2.7)$$

$$\hat{\mathbf{t}}_2(\tau) \simeq -\hat{\mathbf{t}}(2L_b - s), \quad \hat{\mathbf{v}}_2(\tau) \simeq \hat{\mathbf{v}}(2L_b - s), \qquad L_b \lesssim s \lesssim 2L_b. \qquad (2.8)$$

Using these definitions (Eqs. (2.7) and (2.8)) it is possible to write twist densities for each of the segments $g_\mu(\tau)$ through Eq. (2.4). These are defined as

$$g_1(\tau) = g(s), \qquad\qquad 0 < s \leq L_b, \qquad (2.9)$$

$$g_2(\tau) = g(2L_b - s), \qquad\qquad L_b \lesssim s \lesssim 2L_b. \qquad (2.10)$$

Utilization of both Eqs. (2.9) and (2.10), allows us to rewrite Eq. (2.3) as

$$Tw \simeq \frac{1}{2\pi} \int_0^{L_b} d\tau (g_1(\tau) + g_2(\tau)), \qquad (2.11)$$

and Eqs. (2.2), (2.5), (2.6), (2.7) and (2.8) allow us to write

$$4\pi Wr \simeq \int_0^{L_b} d\tau \int_0^{L_b} d\tau' \frac{(\mathbf{r}_1(\tau) - \mathbf{r}_1(\tau')) \cdot \hat{\mathbf{t}}_1(\tau) \times \hat{\mathbf{t}}_1(\tau')}{|\mathbf{r}_1(\tau) - \mathbf{r}_1(\tau')|^{3/2}} + \int_0^{L_b} d\tau \int_0^{L_b} d\tau' \frac{(\mathbf{r}_2(\tau) - \mathbf{r}_2(\tau')) \cdot \hat{\mathbf{t}}_2(\tau) \times \hat{\mathbf{t}}_2(\tau')}{|\mathbf{r}_2(\tau) - \mathbf{r}_2(\tau')|^{3/2}}$$
$$-2 \int_0^{L_b} d\tau \int_0^{L_b} d\tau' \frac{(\mathbf{r}_1(\tau) - \mathbf{r}_2(\tau')) \cdot \hat{\mathbf{t}}_1(\tau) \times \hat{\mathbf{t}}_2(\tau')}{|\mathbf{r}_1(\tau) - \mathbf{r}_2(\tau')|^{3/2}}. \qquad (2.12)$$

In what follows is useful to define a spatial average, $g_{av}$ of the twist densities such that $g_1(\tau) = g_{av} + \delta g_1(\tau)$ and $g_2(\tau) = g_{av} + \delta g_2(\tau)$, where

$$Tw = \frac{L_b g_{av}}{\pi}, \qquad (2.13)$$

so that

$$0 \simeq \frac{1}{2\pi} \int_0^{L_b} d\tau (\delta g_1(\tau) + \delta g_2(\tau)). \qquad (2.14)$$

Note that as $g_{av}$ is the spatial average (not the thermal one), it exactly satisfies Eq. (2.13), thus from Eq. (2.11) we may write Eq. (2.14) for fluctuations in $g(s)$ that do not contribute to the twist about the loop . Thus, this separation is useful as variations $\delta g_1(\tau)$ and $\delta g_2(\tau)$ can be considered as independent from the writhe; instead they must satisfy Eq. (2.14) . Note, however, that changes in the writhe, from bending, will indeed affect the value of $g_{av}$ which is thermally fluctuating, and thus it must be considered in the thermal averaging of bending. The constraint on $\delta g_1(\tau) + \delta g_2(\tau)$, in Eq. (2.14), can be implemented by considering only non-zero Fourier harmonics that make up $g(s)$. Actually, however, $\delta g_1(\tau) + \delta g_2(\tau)$ is relatively unimportant (as we shall see later), and such a constraint on it even less important when considering large $L_b$ (as the non zero-Fourier harmonics dominate). The value of $g_{av}$ is fully constrained by the Fuller-White theorem (Eq. (2.1)) to be

$$g_{av} = \frac{\pi}{L_b}\left(Lk - Wr\right). \tag{2.15}$$

When we come to consider interactions that depend on helix structure, the variations in $\delta g_1(\tau) - \delta g_2(\tau)$, as well as the value of $g_{av}$ will be important.

There are two contributions to both $\delta g_1(\tau)$ and $\delta g_2(\tau)$. One is a thermal contribution, $\delta g_\mu^T(\tau)$ and the other describes sequence dependent variations, $\delta g_\mu^S(\tau)$. The latter contribution is to do with the imperfect stacking of base pairs [56,57]. Thus, we may write

$$\delta g_\mu(\tau) = \delta g_\mu^S(\tau) + \delta g_\mu^T(\tau). \tag{2.16}$$

Let us consider $\delta g_\mu^S(\tau)$ when $L$ is very large. We will suppose that both segments contain completely random base pair sequences, and also that the base pair sequences of the two segments are uncorrelated with each other. Thus, we may write (for a justification see Ref. [57])

$$\left\langle \delta g_\mu^{0,S}(\tau) \delta g_{\mu'}^{0,S}(\tau'))) \right\rangle_g = \frac{\delta_{\mu,\mu'}}{\lambda_c^{(0)}} \delta(s - s'), \tag{2.17}$$

$$\left\langle \delta g_\mu^{0,S}(\tau) \right\rangle_g = 0, \tag{2.18}$$

where $\delta g_\mu^{0,S}(\tau)$ is $\delta g_\mu^S(\tau)$ in the absence of any elastic strain caused by helix structure dependent DNA-DNA interactions. In Eqs. (2.17) and (2.18), the subscript $g$ on the averaging brackets denotes that they are ensemble averages over all realizations of base pair sequence.

**2.2 Specifying braid and helix geometry**

We define the braid axis as a line that cuts through the mid-points of chords connecting the two molecular centre lines. In what follows, we suppose that the axis of the braid is straight. The effects of an undulating braid axis might be considered in later work. We choose the braid axis to coincide with $z$-axis of a coordinate frame. We then can specify the braid geometry with equations for the molecular centre lines of the two segments. These read as

$$\mathbf{r}_1(\tau) = \frac{1}{2}\left(R(\tau)\cos\theta(\tau)\hat{\mathbf{i}} + R(\tau)\sin\theta(\tau)\hat{\mathbf{j}}\right) + Z(\tau)\hat{\mathbf{k}}, \tag{2.19}$$

$$\mathbf{r}_2(\tau) = -\frac{1}{2}\left(R(\tau)\cos\theta(\tau)\hat{\mathbf{i}} + R(\tau)\sin\theta(\tau)\hat{\mathbf{j}}\right) + Z(\tau)\hat{\mathbf{k}}. \tag{2.20}$$

Here, $R(\tau)$ is the distance between the centrelines of the two segments and $\theta(\tau)$ is their angle of rotation around the braid axis, tracing out the braid. These two functions can be arbitrary functions of $\tau$, as we allow the braid geometry to thermally fluctuate, though different configurations having different energy and Boltzmann weight. However, $Z(\tau)$ is constrained through $|\hat{\mathbf{t}}_\mu(\tau)| = 1$ (see below). If the braid forms an ideal double (super) helix, we have that $R(\tau) = R_0$ and $\theta(\tau) = \theta_0 - Q\tau$, where $R_0$, $\theta_0$ and $Q$ are all constants. We call such a configuration a regular braid. Although we allow for fluctuations, we will suppose that the thermally averaged braid is indeed regular, which should be the case for large $L_b$, assuming that the braid axis is indeed straight.

Along with the unitary requirement that $|\hat{\mathbf{t}}_\mu(\tau)| = 1$, we may differentiate both Eqs. (2.19) and (2.20) to write

$$\hat{\mathbf{t}}_1(\tau) = \sin\left(\frac{\eta(\tau)}{2}\right)\sin\left(\theta(\tau)+\gamma(\tau)\right)\hat{\mathbf{i}} - \sin\left(\frac{\eta(\tau)}{2}\right)\cos\left(\theta(\tau)+\gamma(\tau)\right)\hat{\mathbf{j}} + \cos\left(\frac{\eta(\tau)}{2}\right)\hat{\mathbf{k}}, \quad (2.21)$$

$$\hat{\mathbf{t}}_2(\tau) = -\sin\left(\frac{\eta(\tau)}{2}\right)\sin\left(\theta(\tau)+\gamma(\tau)\right)\hat{\mathbf{i}} + \sin\left(\frac{\eta(\tau)}{2}\right)\cos\left(\theta(\tau)+\gamma(\tau)\right)\hat{\mathbf{j}} + \cos\left(\frac{\eta(\tau)}{2}\right)\hat{\mathbf{k}}, \quad (2.22)$$

and thus require the following relations

$$Z(\tau) = \int_0^\tau d\tau' \cos\left(\frac{\eta(\tau')}{2}\right), \quad \theta(\tau) - \theta_0 = -\int_0^\tau d\tau' \frac{1}{R(\tau')}\sqrt{4\sin^2\left(\frac{\eta(\tau')}{2}\right) - \left(\frac{dR(\tau')}{d\tau'}\right)^2}, \quad (2.23)$$

as well as

$$\cos(\gamma(\tau)) = \sqrt{1 - \frac{1}{4\sin(\eta(\tau)/2)^2}\left(\frac{dR(\tau)}{d\tau}\right)^2}, \quad \sin(\gamma(\tau)) = \frac{1}{2\sin(\eta(\tau)/2)}\left(\frac{dR(\tau)}{d\tau}\right). \quad (2.24)$$

One should note that $\eta(\tau)$ is the angle between the two tangent vectors i.e. $\hat{\mathbf{t}}_1(\tau).\hat{\mathbf{t}}_2(\tau) = \cos\eta(s)$.

To consider forces that depend on helix structure we need to find a way of specifying the relative azimuthal orientation of the helices of the two segments making up the braid. We do this by constructing braid frames [58] for each segment. Both frames are described by the vector set $\{\hat{\mathbf{d}}_\mu(\tau), \hat{\mathbf{n}}_\mu(\tau), \hat{\mathbf{t}}_\mu(\tau)\}$ where

$$\hat{\mathbf{n}}_\mu(\tau) = \frac{\hat{\mathbf{t}}_\mu(\tau) \times \hat{\mathbf{d}}(\tau)}{|\hat{\mathbf{t}}_\mu(\tau) \times \hat{\mathbf{d}}(\tau)|}, \qquad \hat{\mathbf{d}}_\mu(\tau) = \hat{\mathbf{n}}_\mu(\tau) \times \hat{\mathbf{t}}_\mu(\tau), \quad (2.25)$$

and

$$\hat{\mathbf{d}}(\tau) = \frac{\mathbf{r}_1(s) - \mathbf{r}_2(s)}{|\mathbf{r}_1(s) - \mathbf{r}_2(s)|}. \quad (2.26)$$

The vectors $\hat{\mathbf{n}}_\mu(\tau)$ and $\hat{\mathbf{d}}_\mu(\tau)$ are perpendicular to $\hat{\mathbf{t}}_\mu(\tau)$. These frames allow us to specify $\hat{\mathbf{v}}_\mu(\tau)$ in terms of angles $\xi_\mu(\tau)$ in the following way

$$\hat{\mathbf{v}}_\mu(\tau) = \cos\xi_\mu(\tau)\hat{\mathbf{d}}_\mu(\tau) + \sin\xi_\mu(\tau)\hat{\mathbf{n}}_\mu(\tau). \tag{2.27}$$

Using Eqs. (2.4) and (2.27) allows us to write $g_\mu(\tau)$ in terms of $\xi_\mu(\tau)$. We may show that (see Ref. [53]), when $R'(\tau) \ll 1$,

$$g_\mu(\tau) = \frac{d\xi_\mu(\tau)}{d\tau} - \frac{\sin\eta(\tau)}{R(\tau)}. \tag{2.28}$$

The second term in Eq. (2.28) comes from the fact that as one changes $s$ the braid frames rotate.

**2.3 Elastic energies**

To describe the supercoiling, we will need to consider both the twisting and bending energies, as well as stretching fluctuations that only change $\delta g_\mu(\tau)$ (fluctuations in the twist density). It is important to note that $g_{av}$ changes only through torsional strain and is not affected by any stretching, as the supercoiled molecule is not under any mechanical stretching stress. We may first write

$$\delta g_\mu(\tau) = \frac{1}{h}\Big(\delta\Omega_\mu(\tau) - g_{av}\delta h_\mu(\tau)\Big), \tag{2.29}$$

where $\delta\Omega_\mu(\tau)$ describes patterns of deviation in the twist angles between base pairs away from their average values, which is $\bar{\Omega} = g_{av}/h$, and $\delta h_\mu(\tau)$ is the deviation in base pair rises (distance between adjacent base pairs along molecular axis) away from their average value $h$ ($h \approx 3.3\text{Å}$). Likewise, for the intrinsic base-pair dependent variations, we can write

$$\delta g_\mu^S(\tau) = \frac{1}{h}\Big(\delta\Omega_\mu^S(\tau) - g_{av}\delta h_\mu^S(\tau)\Big), \tag{2.30}$$

where $\delta\Omega_\mu^S(\tau)$ and $\delta h_\mu^S(\tau)$ are the structural contributions to $\delta\Omega_\mu(\tau)$ and $\delta h_\mu(\tau)$, respectively. If the braided section is sufficiently long, we may suppose that [59]

$$\int_0^{L_b} d\tau\Big(\delta h_1^S(\tau) + \delta h_2^S(\tau)\Big) = 0 \qquad \int_0^{L_b} d\tau\Big(\delta h_1(\tau) + \delta h_2(\tau)\Big) = 0, \tag{2.31}$$

and so it follows that (from Eq. (2.14))

$$\int_0^{L_b} d\tau\Big(\delta\Omega_1(\tau) + \delta\Omega_2(\tau)\Big) = 0, \qquad \int_0^{L_b} d\tau\Big(\delta\Omega_1^S(\tau) + \delta\Omega_2^S(\tau)\Big) = 0. \tag{2.32}$$

We will use the Elastic rod model to write down terms, for the elastic energies, in terms of $\delta\Omega_\mu(\tau)$ and $\delta h_\mu(\tau)$. First of all, the twisting energies of the two segments can be written as

$$\frac{E_{tw}}{k_B T} = \int_0^{L_b} d\tau \frac{l_{tw}}{2h^2}\left[\left(hg_{av} + \delta\Omega_1(\tau) - \frac{2\pi h}{H} - \delta\Omega_1^{0,S}(\tau)\right)^2 + \left(hg_{av} + \delta\Omega_2(\tau) - \frac{2\pi h}{H} - \delta\Omega_2^{0,S}(\tau)\right)^2\right].$$

(2.33)

Here, $\delta\Omega_\mu^{0,S}(\tau)$ are the patterns of twist angles, $\delta\Omega_\mu^S(\tau)$, in the torsionally relaxed state (no twisting strain) and $H$ is the average torsionally relaxed helical pitch ($H \approx 33.8\text{Å}$). Using Eq. (2.32), Eq. (2.33) can be rewritten as

$$\frac{E_{tw}}{k_B T} = l_{tw} L_b \left(g_{av} - \frac{2\pi}{H}\right)^2 + \int_0^{L_b} d\tau \frac{l_{tw}}{2h^2}\left[\left(\delta\Omega_1(\tau) - \delta\Omega_1^{S,0}(\tau)\right)^2 + \left(\delta\Omega_2(\tau) - \delta\Omega_2^{S,0}(\tau)\right)^2\right]. \quad (2.34)$$

Here, $l_{tw} = C/k_B T$ is the twisting persistence length, where $C$ is the twisting elastic modulus. We take $l_{tw} = 1000\text{Å}$, as determined experimentally in Ref. [60]. Next, using the elastic rod model, the elastic stretching energy can be written as

$$\frac{E_{st}}{k_B T} = \int_0^{L_b} d\tau \frac{l_{st} g_{av}^2}{2h^2}\left[\left(\delta h_1(\tau) - \delta h_1^{S,0}(\tau)\right)^2 + \left(\delta h_2(\tau) - \delta h_2^{S,0}(\tau)\right)^2\right]. \quad (2.35)$$

Here $l_{st}$ is the stretching persistence length defined by $l_{st} = Y/(g_{av}^2 k_B T)$, where $Y$ is the stretching modulus. In what follows we will approximate $l_{st} \approx HY/(2\pi k_B T)$, supposing that changes in $g_{av}$ away from $2\pi/H$ causes a second order effect (also taking into account some relative uncertainty in the determination of $Y$). Based on an experimental value of $Y \approx 1\times 10^{-4} dyn$ [61], we take $l_{st} \approx 700\text{Å}$.

There are contributions to $\delta g_\mu(\tau)$ from both twisting and stretching fluctuations, as they both depend on $\delta\Omega_\mu(\tau)$ and $\delta h_\mu(\tau)$ (see Eq. (2.29)). Using Eqs. (2.29), (2.30), (2.34) and (2.35), the sum of twisting and stretching elastic energies can be written as

$$\frac{E_{tw} + E_{st}}{k_B T} = l_{tw} L_b \left(g_{av} - \frac{2\pi}{H}\right)^2 + \frac{l_c}{2}\int_0^{L_b} d\tau \left[\left(\delta g_1(\tau) - \delta g_1^{0,S}(\tau)\right)^2 + \left(\delta g_2(\tau) - \delta g_2^{0,S}(\tau)\right)^2\right]$$
$$+ \frac{\tilde{l}_c}{2}\int_0^{L_b} d\tau \left[\left(\delta\rho_1(\tau) - \delta\rho_1^{0,S}(\tau)\right)^2 + \left(\delta\rho_2(\tau) - \delta\rho_2^{0,S}(\tau)\right)^2\right],$$

(2.36)

where

$$l_c = \frac{l_{st} l_{tw}}{(l_{st} + l_{tw})}, \qquad \tilde{l}_c = \frac{l_{st}^2}{(l_{tw} + l_{st})}, \quad (2.37)$$

$$\rho_\mu(\tau) = \frac{1}{h}\left(\frac{l_{tw}}{l_{st}}\delta\Omega_\mu(\tau) + g_{av}\delta h_\mu(\tau)\right), \qquad \rho_\mu^{0,S}(\tau) = \frac{1}{h}\left(\frac{l_{tw}}{l_{st}}\delta\Omega_\mu^{0,S}(\tau) + g_{av}\delta h_\mu^{0,S}(\tau)\right). \tag{2.38}$$

As nothing else will depend on $\delta\rho_1(\tau)$ and $\delta\rho_2(\tau)$, we may integrate these fluctuations out and neglect the last integral in Eq. (2.36). Furthermore we can rewrite the sum of the elastic energies (using Eqs. (2.28) and (2.36)) as

$$\frac{E_{tw} + E_{st}}{k_B T} = l_{tw}L_b\left(g_{av} - \frac{2\pi}{H}\right)^2 + \frac{l_c}{4}\int_0^{L_b}d\tau\left(\frac{d\Delta\xi(\tau)}{d\tau} - \Delta g_{0,S}(\tau)\right)^2 + \frac{l_c}{4}\int_0^{L_b}d\tau\left(\delta\overline{g}(\tau) - \delta\overline{g}_{0,S}(\tau)\right)^2$$

(2.39)

where $\Delta\xi(\tau) = \xi_1(\tau) - \xi_2(\tau)$, $\Delta g_{0,S}(\tau) = \delta g_1^{0,S}(\tau) - \delta g_2^{0,S}(\tau)$, $\delta\overline{g}(\tau) = \delta g_1(\tau) + \delta g_2(\tau)$, and $\delta\overline{g}_{0,S}(\tau) = \delta g_1^{0,S}(\tau) + \delta g_2^{0,S}(\tau)$. In Eq. (2.39) we use the value $l_c = 400\text{Å}$, estimated through Eq. (2.37) using the values of $l_{tw}$ and $l_{st}$ already given.

Next, we consider the bending energy which, in the elastic rod model, reads as

$$E_b = \frac{l_p}{2}\int_0^{L_b}d\tau\left[\left(\frac{d\hat{\mathbf{t}}_1(\tau)}{d\tau}\right)^2 + \left(\frac{d\hat{\mathbf{t}}_2(\tau)}{d\tau}\right)^2\right], \tag{2.40}$$

where $l_p = B/k_B T$ is the bending persistence length and $B$ is the elastic modulus. We use the commonly accepted value of $l_p = 500\text{Å}$. Using Eqs. (2.21)-(2.24), this can be approximated by (for $R'(\tau) \ll 1$ and small fluctuations in $\eta(\tau)$)

$$E_b = \int_0^{L_b}d\tau \mathcal{E}_R(R''(\tau), R'(\tau), R(\tau), \delta\eta'(\tau), \delta\eta(\tau)), \tag{2.41}$$

where

$$\begin{aligned}\mathcal{E}_R(R''(\tau), R'(\tau), R(\tau), \delta\eta'(\tau), \delta\eta(\tau)) = \\ \frac{l_p}{4}\left(\frac{d^2R(\tau)}{d\tau^2}\right)^2 + \frac{l_p}{4}\left(\frac{d\delta\eta(\tau)}{d\tau}\right)^2 - \left(\frac{dR(\tau)}{d\tau}\right)^2\frac{l_p}{R(\tau)^2}\sin^2\left(\frac{\eta_0}{2}\right) \\ + \frac{4l_p}{R(\tau)^2}\left[\sin^4\left(\frac{\eta_0}{2}\right) + \frac{\delta\eta(\tau)^2}{2}\left(3\cos^2\left(\frac{\eta_0}{2}\right)\sin^2\left(\frac{\eta_0}{2}\right) - \sin^4\left(\frac{\eta_0}{2}\right)\right)\right] \\ + \left(\frac{dR(\tau)}{d\tau}\right)\left(\frac{d\delta\eta(\tau)}{d\tau}\right)\frac{3l_p}{R(\tau)}\sin\left(\frac{\eta_0}{2}\right)\cos\left(\frac{\eta_0}{2}\right).\end{aligned} \tag{2.42}$$

Here $\delta\eta(\tau) = \eta(\tau) - \eta_0$, and $\eta_0 = \langle\eta(s)\rangle$ is the thermal average of $\eta(s)$. Details of the calculation of Eq. (2.42) can be found in Ref. [62].

## 2.3 Handling interactions that depend on helix structure

Interactions that depend on the helix structure of the molecule should depend on $\Delta\xi(\tau)$, the relative azimuthal orientation between the two helices making up the braid. Also, the effective decay range of these interactions should depend on the average twist density of the two segments. As well as that, it should also depend on both $R(\tau)$ and $\eta(\tau)$. Provided that the variations in $\Delta\xi(\tau)$, $\bar{g}(\tau)$, $R(\tau)$ and $\eta(\tau)$ are sufficiently slow (this should be the case provided that the persistence lengths $\lambda_c^{(0)}$, $l_c$ and $l_b$ are longer than decay range of the interaction) we can write [63]

$$E_{int} \simeq \int_0^{L_b} d\tau \varepsilon_{int}(\eta(\tau), R(\tau), \Delta\xi(\tau), g_{av} + \delta\bar{g}(\tau)/2). \tag{2.43}$$

A notable example of Eq. (2.43) is the mean-field electrostatic result of the KL theory [45,46]; the form for $\varepsilon_{int}$ is presented in Appendix E. Now, if it is the case that $g_{av}l_c \gg 1$ and $g_{av}\lambda_c^0 \gg 1$, we can neglect $\delta\bar{g}(\tau)$ from Eq. (2.43). Thus, we may write

$$E_{int} \simeq \int_0^{L_b} d\tau \varepsilon_{int}(\eta(\tau), R(\tau), \Delta\xi(\tau), g_{av}). \tag{2.44}$$

The form of Eq. (2.44) allows us to integrate out $\delta\bar{g}(\tau)$ and thus discard terms dependent on it from the elastic energy, Eq. (2.39).

## 2.4 Specifying the full energy functional and including steric effects

To model steric effects, we use the approach discussed before in Refs. [49,52,53,62]. In this we use a pseudo-potential of the form

$$\tilde{E}_{st} = k_BT \int_0^{L_b} d\tau \frac{\alpha_H}{2}(R(\tau) - R_0)^2 = k_BT \int_0^{L_b} d\tau \frac{\alpha_H}{2} \delta R(\tau)^2 \tag{2.45}$$

where

$$\alpha_H \approx \frac{2}{(d_{max} - d_{min})^{8/3}(l_p)^{1/3}}. \tag{2.46}$$

In Eq. (2.46), $d_{max}$ and $d_{min}$ are the maximum and minimum displacements away from $R_0$ in the braided section allowed by steric interactions. For a more detailed discussion and how $\alpha_H$ is estimated see Refs. [49,52,53,62]. In this work, we choose the conventional values of $d_{max} = -d_{min} = (R_0 - 2a)$, which yields a term similar to what has been considered in previous work (Refs. [16,17]). Considering the possibility spontaneous braiding by strong helix dependent forces [53], we argued that for a tightly braided structure the value for $d_{max}$ should be different. However in the case of strong supercoiling, where there are likely to be large elastic forces constraining $R(s)$, as well as interaction forces, this difference in choice is unlikely to matter. Here, the choice value of

$d_{\min}$ is more important, chosen to prevent unphysical and overestimation of the enhancement of the interaction terms by braid undulations, as well as inter-penetration of the two segments.

To take account of the steric interactions, we also modify the bending energy and interaction energies to be

$$\frac{\tilde{E}_B}{k_B T} = \int_0^{L_b} d\tau \big[\mathcal{E}_R(R''(\tau), R'(\tau), R(\tau), \delta\eta'(\tau), \delta\eta(\tau))\theta(\delta R(\tau) - d_{\min})\theta(d_{\max} - \delta R(\tau))$$
$$+ \mathcal{E}_R(R''(\tau), R'(\tau), R_0 + d_{\max}, \delta\eta'(\tau), \delta\eta(\tau))\theta(\delta R(\tau) - d_{\max})$$
$$+ \mathcal{E}_R(R''(\tau), R'(\tau), R_0 + d_{\min}, \delta\eta'(\tau), \delta\eta(\tau))\theta(d_{\min} - \delta R(\tau))\big], \quad (2.47)$$

$$\tilde{E}_{\text{int}} = \sum_{n=-\infty}^{\infty} \int_0^{L_b} d\tau \bar{\varepsilon}_{\text{int}}(R_0, \delta R(\tau), \eta(\tau), g_{av}, n)\exp(-in\Delta\xi(\tau)), \quad (2.48)$$

and

$$\bar{\varepsilon}_{\text{int}}(R_0, \delta R(\tau), \eta(\tau), g_{av}, n) = \frac{1}{2\pi}\int_0^{2\pi} d\Delta\xi \exp(in\Delta\xi)$$
$$\big[\varepsilon_{\text{int}}(\eta(\tau), R(\tau), \Delta\xi, g_{av})\theta(\delta R(\tau) - d_{\min})\theta(d_{\max} - \delta R(\tau))$$
$$+ \varepsilon_{\text{int}}(\eta(\tau), R_0 + d_{\max}, \Delta\xi, g_{av})\theta(\delta R(\tau) - d_{\max}) + \varepsilon_{\text{int}}(\eta(\tau), R_0 + d_{\min}, \Delta\xi, g_{av})\theta(d_{\min} - \delta R(\tau))\big],$$

(2.49)

where $\delta R(s) = R(s) - R_0$, and $R_0 = \langle R(s) \rangle$ is the thermal average of the average of the interaxial separation. In writing Eq. (2.48), we have used the fact that, through the transformation $\Delta\xi(\tau) \to \Delta\xi(\tau) + 2\pi$ (equivalent to one complete rotation of either segment about its molecular axis) the interaction should be invariant. Thus, we have expressed the interaction as a Fourier series. In modifying the bending and interaction terms we have used the prescription (discussed in previous work Refs. [49,52,53,62]) that when $\delta R(\tau) > d_{\max}$ we replace $\delta R(\tau)$ with $d_{\max}$, and when $\delta R(\tau) < d_{\max}$ we replace $\delta R(\tau)$ with $d_{\min}$. This prevents unphysical values of the bending and interaction energies contributing, as well overestimating the enhancement effect on the interaction energy due to undulations.

The full energy functional for the plectoneme is then written as

$$\frac{E_T}{k_B T} = \frac{\tilde{E}_{st} + \tilde{E}_B + \tilde{E}_{\text{int}}}{k_B T} + l_{tw} L_b \left(g_{av} - \frac{2\pi}{H}\right)^2 + \frac{l_c}{4}\int_0^{L_b} d\tau \left(\frac{d\Delta\xi(\tau)}{d\tau} - \Delta g_{0,S}(\tau)\right)^2. \quad (2.50)$$

The partition function then can be expressed as the functional integral

$$Z = \int D\Delta\xi_T(\tau) \int D\delta\eta(\tau) \int D\delta R(\tau) \exp\left(-\frac{E_T[\Delta\xi_S(\tau) + \Delta\xi_T(\tau), \delta\eta(\tau), \delta R(\tau)]}{k_B T}\right), \quad (2.51)$$

where we have divided $\Delta\xi(\tau)$ into thermal and base-pair dependent, structural contributions, $\Delta\xi_T(\tau)$ and $\Delta\xi_S(\tau)$, respectively. Note that $\Delta\xi_S(\tau)$ is assumed to satisfy the equation $\Delta\xi'_S(\tau) = \Delta g_{0,S}(\tau)$, in the case of weak helix dependent forces. Thus, we have supposed that the interaction strength is too weak to much change $\Delta\xi_S(\tau)$ through helical adaptation [64].

**2.5 The variational approximation to the free energy for weak helix dependent forces**

We want to approximate Eq. (2.51) and, therefore, the free energy for weak helix specific forces. We do this through a variational approximation, which we outline below. To do this we first write the total energy functional (Eq. (2.50)) as

$$E_T\left[\Delta\xi_S(\tau)+\Delta\xi_T(\tau),\delta\eta(\tau),\delta R(\tau)\right] = E_{T,0}\left[\delta\eta(\tau),\delta R(\tau)\right] + \frac{k_B T l_c}{4}\int_0^{L_b} d\tau \left(\frac{d\Delta\xi_T}{d\tau}\right)^2 \quad (2.52)$$
$$+ \Delta E_{int}\left[\Delta\xi_S(\tau)+\Delta\xi_T(\tau),\delta\eta(\tau),\delta R(\tau)\right],$$

where

$$\Delta E_{int}\left[\Delta\xi(\tau),\delta\eta(\tau),\delta R(\tau)\right] = \sum_{n=-\infty}^{\infty}\int_0^{L_b} d\tau(1-\delta_{n,0})\bar{\varepsilon}_{int}(R_0,\delta R(\tau),\eta(\tau),g_{av},n)\exp\left(-in\Delta\xi(\tau)\right),$$

(2.53)

and

$$E_{T,0}[\delta\eta(\tau),\delta R(\tau)] = \tilde{E}_{st} + \tilde{E}_B + \int_0^{L_b} d\tau \bar{\varepsilon}_{int}(R_0,\delta R(\tau),\eta(\tau),g_{av},0) + l_{tw} k_B T L_b \left(g_{av} - \frac{2\pi}{H}\right)^2. \quad (2.54)$$

To start with, we expand the partition function in powers of $\Delta E_{int}\left[\Delta\xi(\tau),\delta\eta(\tau),\delta R(\tau)\right]$ and then resum the partition function, after integration over $\Delta\xi_T(\tau)$, so that

$$Z \propto \int D\delta\eta(\tau)\int D\delta R(\tau)\exp\left(-\frac{E_{T,eff}\left[\Delta\xi_S(\tau),\delta\eta(\tau),\delta R(\tau)\right]}{k_B T}\right), \quad (2.55)$$

where we have an effective energy functional

$$E_{T,eff}\left[\Delta\xi_S(\tau),\delta\eta(\tau),\delta R(\tau)\right] = E_{T,0}[\delta\eta(\tau),\delta R(\tau)] - k_B T \ln\left(1-E_1+\frac{E_2}{2!}-\frac{E_3}{3!}+\frac{E_4}{4!}+...\right), \quad (2.56)$$

where

$$E_n = \frac{\left\langle \Delta E_{int}\left[\Delta\xi_T(\tau)+\Delta\xi_S(\tau),\delta\eta(\tau),\delta R(\tau)\right]^n \right\rangle_0}{(k_B T)^n}$$

$$= \frac{1}{Z_0}\int D\Delta\xi_T(\tau)\frac{\Delta E_{int}\left[\Delta\xi_T(\tau)+\Delta\xi_S(\tau),\delta\eta(\tau),\delta R(\tau)\right]^n}{(k_B T)^n}\exp\left(-\frac{l_c}{4}\int_0^{L_b}d\tau\left(\frac{d\Delta\xi_T}{d\tau}\right)^2\right),$$

(2.57)

and

$$Z_0 = \int D\Delta\xi_T(\tau)\exp\left(-\frac{l_c}{4}\int_0^{L_b}d\tau\left(\frac{d\Delta\xi_T}{d\tau}\right)^2\right).$$

(2.58)

In Appendix A we show that $E_1 = 0$. At the moment, as stated before, we have approximated $\Delta\xi_S'(\tau) \approx \Delta g_{0,S}(\tau)$. It's worth pointing out that this approximation can be relaxed by introducing the additional term into the effective energy functional

$$\frac{l_c}{4}\int_0^{L_b}d\tau\left(\frac{d\Delta\xi_S(\tau)}{d\tau}-\Delta g_{0,S}(\tau)\right)^2$$

(2.59)

into Eq. (2.56) and functionally minimizing it over $\Delta\xi_S(\tau)$, yielding a mean-field equation for $\Delta\xi_S(\tau)$. This refinement to the calculation, which takes into account some torsional adaptation, has yet to be considered, and probably should be dealt as a perturbation to $\Delta\xi_S'(\tau) = \Delta g_{0,S}(\tau)$, in the first instance, as it is a second order correction. At present, we assume that this additional contribution can be neglected.

Expanding out the logarithm in Eq. (2.56), we obtain a systematic expansion

$$\frac{E_{T,eff}\left[\delta\eta(\tau),\delta R(\tau)\right]}{k_B T} = \frac{E_{T,0}[\delta\eta(\tau),\delta R(\tau)]}{k_B T}-\frac{E_2}{2}+\frac{E_3}{3!}-\left(\frac{E_4}{4!}-\frac{E_2^2}{8}\right)+\ldots$$

(2.60)

The next step, is to perform a variational approximation to the partition function with $E_{T,eff}\left[\delta\eta(\tau),\delta R(\tau)\right]$ to deal with the functional integration over both $\delta\eta(\tau)$ and $\delta R(\tau)$. To do this we write trial energy functional of the form

$$\frac{E_T^T\left[\delta\eta(\tau),\delta R(\tau)\right]}{k_B T} = \int_0^{L_b}d\tau\left(\frac{l_p}{4}\left(\frac{d^2\delta R(\tau)}{d\tau^2}\right)^2+\frac{\beta_R}{2}\left(\frac{d\delta R(\tau)}{d\tau}\right)^2+\frac{\alpha_R}{2}\delta R(\tau)^2\right)$$
$$+\int_0^{L_b}d\tau\left(\frac{l_p}{4}\left(\frac{d\delta\eta(\tau)}{d\tau}\right)^2+\frac{\alpha_\eta}{2}\delta\eta(\tau)^2\right).$$

(2.61)

Then, in the variational approximation, we can write the approximation to the full free energy as

$$F_T = -k_B T \ln Z_T + \left\langle E_{T,\text{eff}}\left[\Delta\xi_S(\tau), \delta\eta(\tau), \delta R(\tau)\right] - E_T^T\left[\delta\eta(\tau), \delta R(\tau)\right]\right\rangle_T. \qquad (2.62)$$

Here, the subscript $T$ on the averaging bracket denotes thermal averaging using trial energy functional, Eq. (2.61), instead of Eq. (2.54). The partition function $Z_T$ is given by

$$Z_T = \int D\delta\eta(s)\int D\delta R(s) \exp\left(-\frac{E_T^T[\delta\eta(\tau), \delta R(\tau)]}{k_B T}\right). \qquad (2.63)$$

Next, we must perform an ensemble average over all base pair realizations to calculate $\langle F_T \rangle_g$, the average free energy. Details of this calculation is given in Appendices A and B (as well as also in Ref. [52]). The parameters $\beta_R$, $\alpha_R$ and $\alpha_\eta$ are chosen to minimize the free energy $\langle F_T \rangle_g$ and so obtain the best approximation to the exact free energy using Eq.(2.61).

However, it is worth commenting in the main text on one very important step in the calculation. This is how to evaluate the averages $\langle \bar{\varepsilon}_{\text{int}}(R_0, \delta R(\tau), \eta(\tau), g_{av}, n) \rangle_T$ and $\langle (g_{av} - 2\pi/H)^2 \rangle_T$. In general– as $g_{av}$ depends on the writhe (see Eq. (2.15))– these averages are very difficult to perform. However, when the braided section is much longer the correlation lengths for bending fluctuations, we make an important approximation

$$\langle \bar{\varepsilon}_{\text{int}}(R_0, \delta R(\tau), \eta(\tau), g_{av}, n) \rangle_T \approx \langle \bar{\varepsilon}_{\text{int}}(R_0, \delta R(\tau), \eta(\tau), \langle g_{av} \rangle_T, n) \rangle_T, \qquad (2.64)$$

$$\langle (g_{av} - 2\pi/H)^2 \rangle_T \approx (\langle g_{av} \rangle_T - 2\pi/H)^2. \qquad (2.65)$$

Justification of this approximation is given in Appendix A (in more detail in Ref. [52]). In essence, one expands both $\langle \bar{\varepsilon}_{\text{int}}(R_0, \delta R(\tau), \eta(\tau), g_{av}, n) \rangle_T$ and $\langle (g_{av} - 2\pi/H)^2 \rangle_T$ in a Taylor expansion about $\langle g_{av} \rangle_T$. Then, one may argue that the next to leading order terms in this expansion will scale as $1/L_b$, thus, if we consider large $L_b$, we may write Eqs. (2.64) and (2.65). Calculating the next to leading order terms in this expansion will yield finite-size corrections to both Eqs. (2.64) and (2.65) from the topological constraint (Eq. (2.1)).

## 2.6 The form for the free energy

Once the free energy has been expressed as function of $\langle g_{av} \rangle_T$ it is convenient to use a Legendre transformation to write a new (Gibbs like) free energy

$$G_T = \langle F_T \rangle_g - M\left(2\pi\langle Wr \rangle_T + 2\langle g_{av} \rangle_T L_b\right). \qquad (2.66)$$

This allows us instead of fixing $\langle g_{av} \rangle_T$ to $\pi\left(Lk - \langle Wr \rangle_T\right)/L_b$ (as required by taking the thermal average of the Fuller-White theorem (Eq. (2.1)), for a given linking number $Lk$) we may minimize

$\langle g_{av} \rangle_T$ as a free parameter, for a given moment $M$. The corresponding linking number can then be computed back through the relation $Lk = \langle Wr \rangle_T + \langle g_{av} \rangle_T L_b / \pi$.

For $\langle F_T \rangle_g$ we find that (see Appendices A and B for an outline of the calculation that follows on from the considerations of the previous subsections)

$$\langle F_T \rangle_g = F_{els} + F_{un} + F_{cor}^{(1)} + F_{cor}^{(2)} + F_{cor}^{(3)}. \tag{2.67}$$

The first term in Eq. (2.67), $F_{els}$, is the contribution from elastic energies, as well as the entropy reduction in forming the braided section of the supercoil. The term $F_{els}$ is written as

$$\frac{F_{els}}{k_B T L_b} = \frac{1}{2}\left(\frac{\alpha_\eta}{2l_p}\right)^{1/2} + \frac{1}{4l_p \theta_R^2} + \frac{l_p \theta_R^4}{d_R^2} + \frac{\alpha_H d_R^2}{2} + \frac{f_1(R_0, d_R)}{R_0^2}\left[4l_p \sin^4\left(\frac{\eta_0}{2}\right) - l_p \theta_R^2 \sin^2\left(\frac{\eta_0}{2}\right)\right.$$
$$\left. + 2\left(\frac{l_p}{2\alpha_\eta}\right)^{1/2}\left(3\cos^2\left(\frac{\eta_0}{2}\right)\sin^2\left(\frac{\eta_0}{2}\right) - \sin^4\left(\frac{\eta_0}{2}\right)\right)\right] + l_{tw} L_b\left(\langle g_{av} \rangle_T - \frac{2\pi}{H}\right)^2, \tag{2.68}$$

where

$$f_1(R_0, d_R) = \frac{R_0^2}{d_R \sqrt{2\pi}} \int_{2a-R_0}^{R_0-2a} \frac{dx}{(R_0+x)^2} \exp\left(-\frac{x^2}{2d_R^2}\right)$$
$$+ \frac{1}{2}\left(\frac{R_0^2}{4a^2}\left(1 - \text{erf}\left(\frac{R_0 - 2a}{d_R \sqrt{2}}\right)\right) + \frac{R_0^2}{4(R_0-a)^2}\left(1 - \text{erf}\left(\frac{R_0 - 2a}{d_R \sqrt{2}}\right)\right)\right), \tag{2.69}$$

with the choice $d_{max} = -d_{min} = R_0 - 2a$. Terms that depend on $\alpha_\eta$ are contributions from fluctuations in tilt of the braid $\eta(\tau)$. As the size of $\alpha_\eta$ is increased, the size of the tilt fluctuations becomes smaller. The parameters $d_R$ and $\theta_R$ quantify the size of fluctuations in $R(s)$. Here, we have $d_R^2 = \langle \delta R(\tau)^2 \rangle$ and $\theta_R^2 = \langle (dR(\tau)/d\tau)^2 \rangle$. Both of these variational parameters can be related back to the original parameters $\alpha_R$ and $\beta_R$ in Eq. (2.61), expressions that relate these two sets of parameters can be found in Appendix B of the supplemental material. The free energy is now to be minimized over $\alpha_\eta$, $d_R$ and $\theta_R$, a more convenient (but equivalent) choice than $\alpha_\eta$, $\alpha_R$ and $\beta_R$. The terms in the square bracket are contributions from to the bending energy to form the braid. The last term in Eq. (2.68) is the contribution from the twisting energy, as $\langle g_{av} \rangle_T$ is moved away from its torsionally relaxed value of $2\pi / H$.

The other terms in Eq. (2.67) are effective interaction terms between the two segments making up the braid from the pair interaction energy Eq. (2.48). This effective interaction is the interaction energy averaged over thermal fluctuations. The first of these is $F_{un}$, is the uniform rod contribution.

This is the contribution to the interaction energy from part of the interaction that does not depend on $\Delta\xi(\tau)$, namely $\bar{\varepsilon}_{int}(R_0,\delta R(\tau),\eta(\tau),g_{av},0)$. This reads as

$$F_{un} = \frac{L_b}{2\pi d_R d_\eta} \int_{-\infty}^{\infty} dr \int_{-\infty}^{\infty} d\eta \bar{\varepsilon}_{int}(R_0,r,\eta_0+\eta,\langle g_{av}\rangle_T,0)\exp\left(-\frac{r^2}{2d_R^2}\right)\exp\left(-\frac{\eta^2}{2d_\eta^2}\right). \tag{2.70}$$

Now $F_{cor}^{(1)}$ the first order correction due to the parts of the interaction that depends on $\Delta\xi(s)$ is given by

$$\frac{F_{cor}^{(1)}}{k_B T} = \frac{-\langle\langle E_2\rangle_T\rangle_g}{2} \approx F_{cor,1}^{(1)} + F_{cor,2}^{(1)} + F_{cor,3}^{(1)}, \tag{2.71}$$

$$F_{cor,1}^{(1)} = \frac{2L_b\lambda_c}{(k_B T)^2} \sum_{n=1}^{\infty} \frac{1}{n^2} \Delta_{0,0}(R_0,\eta_0,\langle g_{av}\rangle_T,d_R,d_\eta,n)^2, \tag{2.72}$$

$$F_{cor,2}^{(1)} = \frac{l_p L_b}{(k_B T)^2} \sum_{n=1}^{\infty} \Delta_{1,0}(R_0,\eta_0,\langle g_{av}\rangle_T,d_R,d_\eta,n)^2 \Omega_{1,\eta}\left(\frac{l_p n^2}{\lambda_c},l_p\alpha_\eta\right), \tag{2.73}$$

$$F_{cor,3}^{(1)} = \frac{l_p^3 L_b}{(k_B T)^2} \sum_{n=1}^{\infty} \Delta_{0,1}(R_0,\eta_0,\langle g_{av}\rangle_T,d_R,d_\eta,n)^2 \Omega_{1,R}\left(\frac{n^2 l_p}{\lambda_c},\alpha_R l_p^3,\beta_R l_p\right), \tag{2.74}$$

where

$$\Omega_{1,\eta}\left(\frac{l_p n^2}{\lambda_c},l_p\alpha_\eta\right) = \frac{1}{l_p}\int_{-\infty}^{\infty} dx \langle\delta\eta(0)\delta\eta(x)\rangle_T \exp\left(-\frac{n^2|x|}{\lambda_c}\right), \tag{2.75}$$

$$\Omega_{1,R}\left(\frac{l_p n^2}{\lambda_c},\alpha_R l_p^3,\beta_R l_p\right) = \frac{1}{l_p^3}\int_{-\infty}^{\infty} dx \langle\delta R(0)\delta R(x)\rangle_T \exp\left(-\frac{n^2|x|}{\lambda_c}\right), \tag{2.76}$$

and

$$\Delta_{j,k}(R_0,\eta_0,\langle g_{av}\rangle_T,d_R,d_\eta,n) = \frac{1}{2\pi d_R d_\eta}\int_{-\infty}^{\infty} d\eta \int_{-\infty}^{\infty} dr \frac{\eta^j}{d_\eta^{2j}} \frac{r^k}{d_R^{2k}} \exp\left(-\frac{r^2}{2d_R^2}\right)\exp\left(-\frac{\eta^2}{2d_\eta^2}\right) \tag{2.77}$$

$\bar{\varepsilon}_{int}(R_0,r,\eta_0+\eta,\langle g_{av}\rangle_T,n).$

Here, we have that $\lambda_c = l_c \lambda_c^{(0)}/(l_c + \lambda_c^{(0)})$. In writing Eqs. (2.72)-(2.77), we have supposed that $\Delta_{0,0}(R_0,\eta_0,\langle g_{av}\rangle_T,d_R,d_\eta,n) = \Delta_{0,0}(R_0,\eta_0,\langle g_{av}\rangle_T,d_R,d_\eta,-n)$, a more general expression is given in Appendix A of the supplemental material. Both $F_{cor}^{(2)}$ and $F_{cor}^{(3)}$ are higher order corrections in the expansion (Eq. (2.60)) for weak helix dependent forces. These terms are given by the expressions

$$\frac{F_{cor}^{(2)}}{k_B T} = \frac{\left\langle\left\langle E_3\right\rangle_T\right\rangle_g}{6} = \frac{2\lambda_c^2 L_b}{3(k_B T)^3} \sum_{n=-\infty}^{\infty} \sum_{m=-\infty,}^{\infty} \frac{(m^2+n^2+nm)}{n^2 m^2 (n+m)^2}(1-\delta_{n,-m})(1-\delta_{n,0})(1-\delta_{m,0})$$
$$\Delta_{0,0}\left(R_0,\eta_0,\left\langle g_{av}\right\rangle_T,d_R,d_\eta,n\right)\Delta_{0,0}\left(R_0,\eta_0,\left\langle g_{av}\right\rangle_T,d_R,d_\eta,m\right)\Delta_{0,0}\left(R_0,\eta_0,\left\langle g_{av}\right\rangle_T,d_R,d_\eta,-n-m\right),$$

(2.78)

and

$$\frac{F_{cor}^{(3)}}{k_B T} = -\frac{1}{8}\left\langle\left\langle \frac{E_4}{3} - E_2^2 \right\rangle_T\right\rangle_g = -\frac{L_b \lambda_c^3}{4(k_B T)^4} \sum_{n,m,l=-\infty}^{\infty}(1-\delta_{l,0})(1-\delta_{m,0})(1-\delta_{n,0})(1-\delta_{n+m+l,0})\left\{\bar{I}_{m,l,n}\right.$$
$$\left.-\frac{(m^2+n^2)\delta_{l,-m}}{n^2 m^2}\left[\frac{2}{n^2 m^2} + \frac{8\lambda_c^2}{\left((n^2+m^2)^2 (\lambda_c^{(0)})^2 - 4n^2 m^2 \lambda_c^2\right)}\right]\right\}\Delta_{0,0}\left(R_0,\eta_0,\left\langle g_{av}\right\rangle_T,d_R,d_\eta,n\right)$$
$$\Delta_{0,0}\left(R_0,\eta_0,\left\langle g_{av}\right\rangle_T,d_R,d_\eta,m\right)\Delta_{0,0}\left(R_0,\eta_0,\left\langle g_{av}\right\rangle_T,d_R,d_\eta,l\right)\Delta_{0,0}\left(R_0,\eta_0,\left\langle g_{av}\right\rangle_T,d_R,d_\eta,-n-m-l\right),$$

(2.79)

where

$$\bar{I}_{m,l,n} = \left(\frac{1}{(m+l+n)^2} + \frac{1}{n^2}\right)\left(\frac{1}{m^2} + \frac{1}{l^2}\right)\frac{1}{(m+l)^2}(1-\delta_{l,-m}). \tag{2.80}$$

These last two terms are needed to provide necessary repulsion that prevents an unphysical collapse of the braid to small values of $R_0$ for low values of $M$, due to $F_{cor}^{(1)}$ becoming too large. These terms were not originally considered in Ref. [52].

We minimize the free energy, defined through Eqs. (2.66)- (2.79) with respect to average geometric parameters $R_0$ and $\eta_0$, as well as $\alpha_\eta$, $d_R$, $\theta_R$ and $\left\langle g_{av}\right\rangle_T$, at fixed $M$. From the parameter values that minimize the free energy, we can compute the supercoiling density $\sigma = (Lk - Lk_0)/Lk$ as a function of $M$ by evaluating the average value of the writhe. We can invert this relationship, so that we can calculate the moment $M$, and so determine the average braid structure as a function of the supercoiling density $\sigma = (Lk - Lk_0)/Lk$, where $Lk_0 = L/H$ is the linking number in the relaxed state. How to calculate the average writhe (appearing in Eq. (2.66)) is discussed in Appendix C of the supplemental material. Also, we need specify $\bar{\varepsilon}_{int}(R_0,r,\eta_0+\eta,\left\langle g_{av}\right\rangle_T,n)$, which relies on the interaction model between DNA segments. Here we utilize the KL model; the forms for $\bar{\varepsilon}_{int}(R_0,r,\eta_0+\eta,\left\langle g_{av}\right\rangle_T,n)$ are given in Appendix E of the supplemental material.

## 3. Results and Discussion

Importantly in the KL model (see Appendix E of supplemental material), there are three important interaction parameters that we will vary to investigate their effect on the supercoiling. These are $\theta$, the fraction of phosphate charge neutralized by condensed (or bound) counter-ions;

$f_1$, the fraction of condensed (bound) ions localized in the minor groove of DNA; and $f_2$, the fraction of condensed (bound) ions localized in the major groove of DNA. There are two other important interaction parameters in the KL model, $\kappa_D$ the inverse Debye length which we fix to be $\kappa_D^{-1} = 7\text{Å}$ and the relaxed DNA helical pitch $H \simeq 33.8\text{Å}$. Both these parameters determine the decay range of the interaction terms, in the electrostatic KL model.

It is not clear what parameter values should be exactly chosen $\theta$ for $f_1$, and $f_2$; at present, their values should be fitted to experiment. However, we can motivate the range of values presented (for instance, the range $\theta = 0.4 - 0.6$). The parameter range is an appropriate choice for monovalent ions, as well as divalent ions like $Mg^{2+}$ or $Ca^{2+}$. For monovalent salt solutions, for an isolated DNA molecule, in the limit of infinite dilution (the concentration of excess salt ions being zero), the charge compensation is at its Manning value $\theta = 0.7$ [42]. For divalent ions the Manning value is $\theta = 0.8$ [42]. Based on solution of the non-linear PB equation, when salt concentration is increased the value of $\theta$ decreases [34,65]. However, one should point out, that also base pair specific ion binding effects [67], supercoiling and interactions between two segments will play a role.

In fitting the data for mechanical braiding data (in phosphate buffer solution) of Ref. [50], the values $\theta = 0.4 - 0.6$ fitted well experimental data [49], though such fitting also suggested that perhaps a lower value could be used for the higher salt concentrations. On the other hand, for divalent ions, the values would be expected to be higher; and also, for certain monovalent ions like caesium and potassium, where there is evidence for strong groove localization due to base-pair specific interactions [66,67]. Thus, we think that values of $\theta = 0.4 - 0.6$ are a realistic choice of values to investigate, to highlight trends, although lower values could also have been used. This upper range of values was chosen, as this was expected to yield the most asymmetry and still be valid for the approximation for weak helix dependent forces used.

The values $f_1$ and $f_2$ are likely to be controlled by counter-ion species [42,66], as well as base pair sequence [42,67]. For instance, for potassium ions, it is known that AT sequences localize monovalent ions in the minor groove and GC sequences localize monovalent ions in the major groove [67]. Also, the degree of localization for monovalent ions, depends on their particular species and hydration radii [66]. The smaller the hydration radii, the easier it is for ions to enter the grooves and interact with base pairs. Simulations suggest that caesium ions seem to be predominantly localized in the minor groove with a high degree of localization [67].

As the parameter space is quite large, we have investigated changing the values of the parameters in three different ways. In the first variation, we varied the value of $\theta$. In the second, we changed the overall degree of ion localization in the grooves $f_1 + f_2$ from the case of full localization $f_1 + f_2 = 1$ to that of $f_1 + f_2 = 0$, full delocalization. In the last, we looked at changing the ratio of $f_1 / f_2$. We only present the first variation, changing $\theta$, in the main text, as the other two do not seem to alter the key findings of this study, when using the KL model of interaction. The results from the other two variations are presented in Appendix F of the Supplemental Material.

When investigating these possible trends, we have chosen the default values, $\theta = 0.6$, $f_1 = 0.7$ and $f_2 = 0.3$ as a starting point. This corresponds to more ions being in the minor groove, which could correspond to the situation of monovalent ions with small hydration radii. Nevertheless, this value is choice is somewhat arbitrary, but here expansion used for weak helix forces is valid. Also, at lower values of the charge compensation, $\theta$ the interaction is found to be relatively insensitive to variations in both $f_1$ and $f_2$. In all the results, the values of $|M|$ (and so $|\sigma|$) were chosen sufficiently high for approximations used in calculating $\langle Wr \rangle$ to undoubtedly hold without considering further corrections (see Appendix C of the supplemental material for details).

We present results for $\Delta M$, the difference in magnitude of the moment $M$ between left and right handed supercoils, at a fixed value of $|\sigma|$, i.e. $\Delta M(|\sigma|) = M(\sigma) + M(-\sigma)$. Plots of the free energy are also presented. Finally, plots of the various average geometric parameters associated with the super-coil are shown, as well as their degree of fluctuation ($d_R$, $\theta_R$, and $\alpha_\eta$).

In Fig.2 we present plots of $\Delta M$, at different values of $|\sigma|$, the absolute value of supercoiling density. Due to the helix structure specific nature of the intersegment interaction, which is chiral, we expect that there is indeed a difference $\Delta M$ and left-right handed supercoil symmetry broken. As we expect, the size of $\Delta M$ should increase somewhat as the value of $|\sigma|$ is increased, as the two segments making plectoneme are brought together and helix specific interactions become stronger. However, contrary to expectations based on previous work, it does not seem to do so in a simple monotonic way, and its sign also changes. The calculations suggest that, at small values of $|\sigma|$, it is easier to build more positive supercoils, as $\Delta M$ is negative which indicates a smaller value of $|M|$ for these supercoils; but this changes at larger values $|\sigma|$, where it seems that it requires less work to increase the number of negative supercoils ($\Delta M$ positive). In fact, the left handed supercoils have lower free energy (see below in Fig. 2), however at larger values $|\sigma|$ this difference diminishes, accounting for the behaviour in $\Delta M$ (note that $M \propto dF/d\sigma$). Increasing the size of $\theta$ favours this latter behaviour. This behaviour should not be present when there sufficiently strong helix specific forces. For this latter case, not considered here, there is well-defined value of mean azimuthal orientation throughout the braid, $\langle \Delta \xi_T(\tau) \rangle$ [68]. In this case, we would expect that $\Delta M$ is always negative, and for certain values of the parameters for supercoils to spontaneously form [38,51]. The possible reason for this discrepancy with the ground state calculation is discussed later on. In addition to these results, in Appendix F of the supplemental material we present the separate moment curves for left-handed (positive) and right-handed (negative) supercoils that were used to calculate Fig. 2. It seems that most of the change in the moment curves with changing $\theta$ occurs for negative supercoiling density $\sigma$.

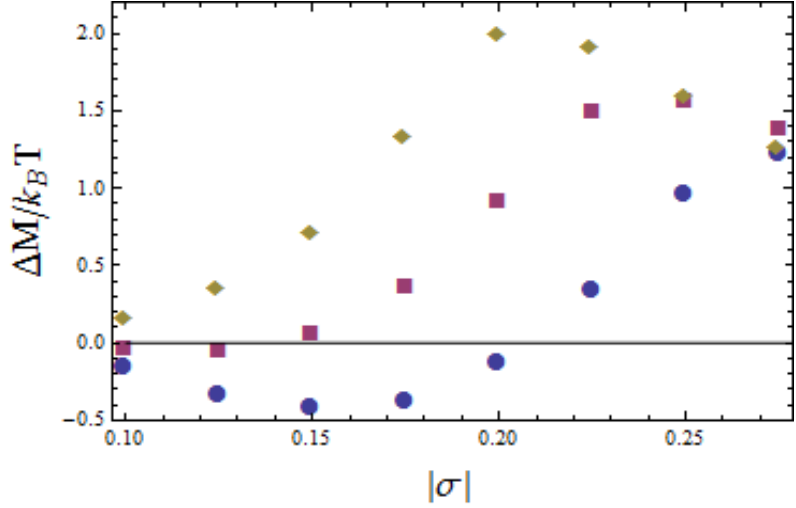

Fig. 2 Difference between moments required to produce left and right handed supercoils. The figure shows $\Delta M$, the difference between the size of the magnitude of the moment, $|M|$ for positive $\sigma$ values and that for negative ones as a function of $|\sigma|$. The solid dark yellow points, purple points and blue points correspond to the values $\theta = 0.6, 0.5$ and $0.4$, respectively. In all plots we set $f_1 = 0.7$ and $f_2 = 0.3$ in the interaction energy.

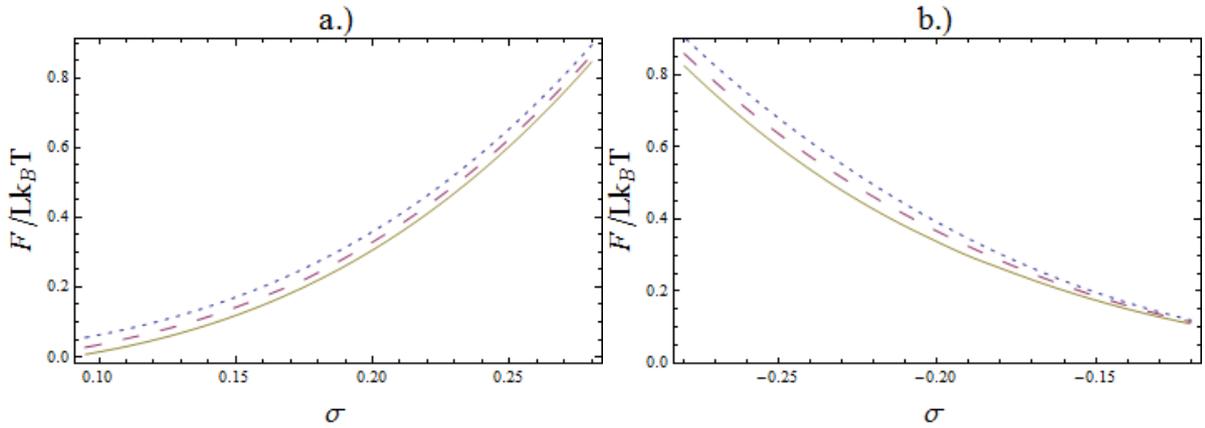

Fig. 3 Plots of $F/Lk_BT$ the supercoiling free energy per unit length and $k_BT$. On the left hand side we present curves for positive values of $\sigma$, whereas on the right we present curves for negative values. In panels a.) and b.) we investigate changing the value of $\theta$, while keeping fixed $f_1 = 0.7$ and $f_2 = 0.3$. Here, the blue short dashed line, red medium dashed line and dark yellow solid lines refer to the values $\theta = 0.4, 0.5$ and $0.6$, respectively.

We see that for the plots of the supercoiling free energy per unit length $F/Lk_BT \approx F_T/Lk_BT$, shown in Fig.3 we have significant asymmetry at small values of $|\sigma|$, however this becomes less pronounced as $|\sigma|$ is increased. As was expected, positive supercoils have a lower free energy due to positive supercoils being preferred by the chiral interaction [38]. Also, there is a slight dependence of the free energy on $\theta$ for both negative and positive $\sigma$ values. This is expected, as changing $\theta$ adjusts the amount of repulsion and so should change the overall free energy, which is indeed seen.

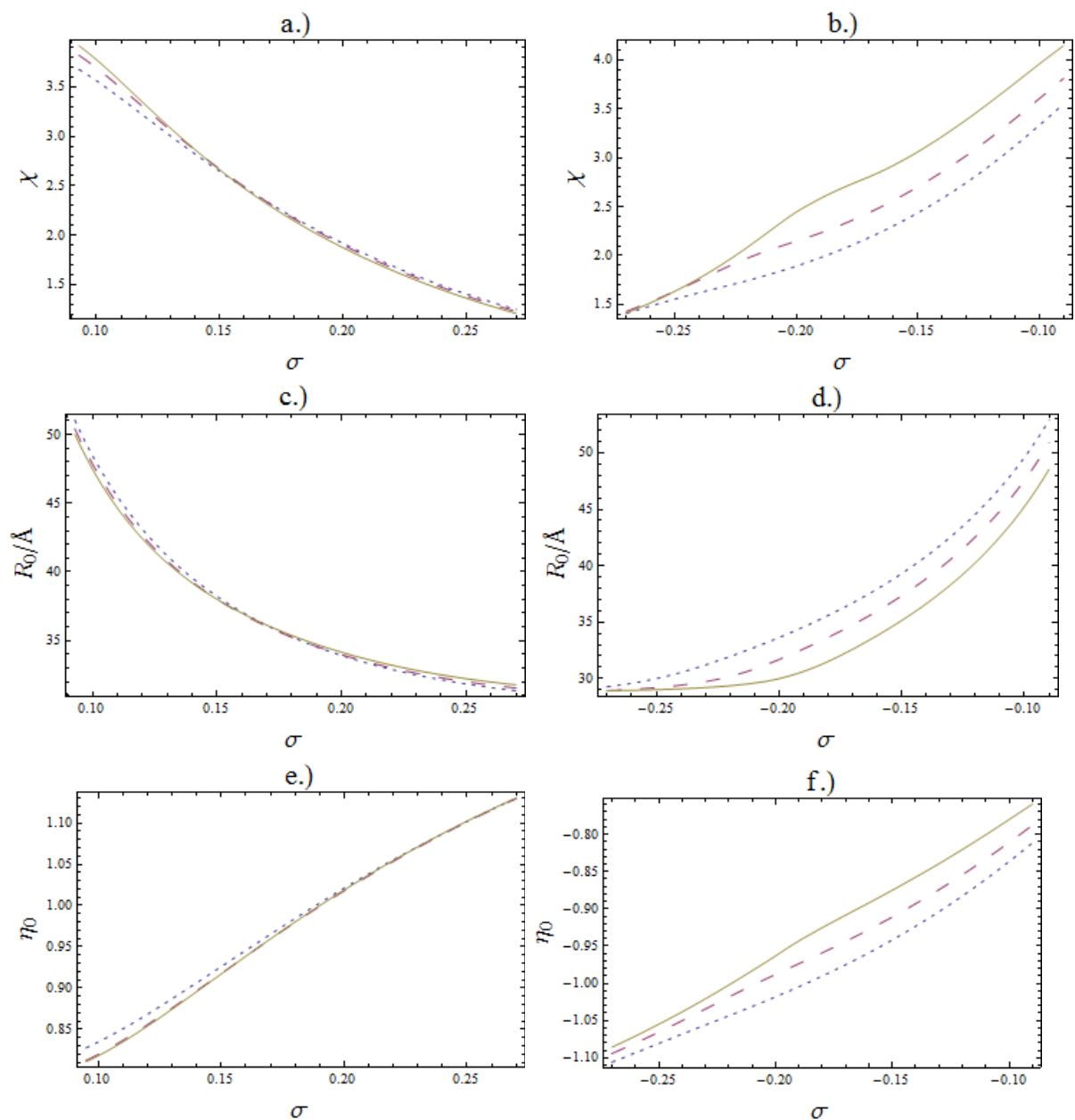

Fig. 4 Mean supercoil structural parameters as a function of supercoiling density. In all plots we set $f_1 = 0.7$ and $f_2 = 0.3$ in the interaction energy. In panels a.) (positive $\sigma$ values) and b.) (negative $\sigma$ values) we plot the ratio $\chi$ of the average writhe $\langle Wr \rangle$ to average twist difference $\langle \Delta Tw \rangle$, away from torsionally relaxed DNA. In panels c.) (positive $\sigma$) and d.) (negative $\sigma$) we plot the average inter-axial separation, $R_0$, between the two segments in the plectoneme braid. Finally in e.) (positive $\sigma$) and f.) (negative $\sigma$) we plot the average tilt angle $\eta_0$, the angle between the tangents of the molecular centre lines of the two segments in the braid. In all plots the solid dark yellow, long dashed purple and short dashed blue lines correspond to $\theta = 0.6, 0.5$ and $0.4$, respectively.

Next, we investigate the supercoil geometric parameters as functions of the supercoiling density (in Fig. 4) . Firstly, we examine the ratio $\chi = \langle Wr \rangle / \langle \Delta Tw \rangle$, where we see quite different behaviour between left and right supercoils. Though, as expected from previous work [22], the value of $\chi$ does decrease with increasing $|\sigma|$ [69]. On the other hand, if helix specific forces were strong, we might have expected a more complicated behaviour [51]. The reason for the monotonic decrease is that, though the writhe increases, the twist difference increases more on increasing $|\sigma|$. The dominant factor in reducing the rate of increase in the writhe with increasing $|\sigma|$ is the increase in repulsion between the braided segments, as their average separation decreases. This makes it less energetically favourable for linking number difference to be partitioned into writhe, as its value increases. Whereas, the optimal value of $\langle \Delta Tw \rangle$ is found to be roughly proportional to $M$, which suggests that its rate of increase get larger as $|\sigma|$ is increased, as $dM/d\sigma$ get larger (see the moment plots in Appendix F of Supplemental material).

The value of $\chi$ for right handed (negative) supercoils is affected much more by changing $\theta$ than for left handed (positive) ones. For positive values of $\sigma$, there is only a slight variation in $\chi$. Here, we see, for large values of $\sigma$, its value slightly increases as $\theta$ is increased, otherwise it stays roughly constant. For the negative values of $\sigma$, $\chi$ increases significantly with the growth of $\theta$, over a large range of $\sigma$ values, and for the value $\theta = 0.6$ it is significantly larger than that what it is for positive $\sigma$.

A major factor in the dependence of $\chi$ on $\theta$, is how $\theta$ influences the values of $R_0$ and $\eta_0$, both of which effect the writhe. In Fig.4, we see also that the values of $R_0$ and $\eta_0$ are also significantly altered by changing $\theta$ for negative $\sigma$ supercoils, whereas affected very little for positive ones. In the case of negative supercoils, smaller values of $R_0$ occur than for positive ones, at the same value of $|\sigma|$. These smaller $R_0$ values fit with the positive $\Delta M$ values, as changing $\eta_0$ produces more writhe, and thus one requires less bending energy to produce more supercoils. Negative supercoils being tighter than positive ones at fixed $|\sigma|$ is a counter-intuitive result, as the right handed structure of DNA helix specific interactions favour left handed braids. This fact was observed in Refs. [37] and [38]. Naively, one would expect, as seen in the ground state calculation, left handed supercoils would form the tighter structures. Once we have commented other features of the results, we will discuss why negative supercoils should form tighter structures. The flattening out seen for $\theta = 0.6$, as $|\sigma|$ increases, is due to a large stiffness in $R_0$ caused by short range repulsive image charge forces in the interaction model [45,46]; this is marked by a reduction in $\Delta M$. The increase in the writhe (larger $\chi$ values) with increasing $\theta$, for negative $\sigma$ values, is due to the reduction in $R_0$. Elastic forces favour a smaller magnitude of $\eta_0$ when $R_0$ is reduced, and this is likely to be responsible for the reduction seen in $\eta_0$ when $\theta$ is increased.

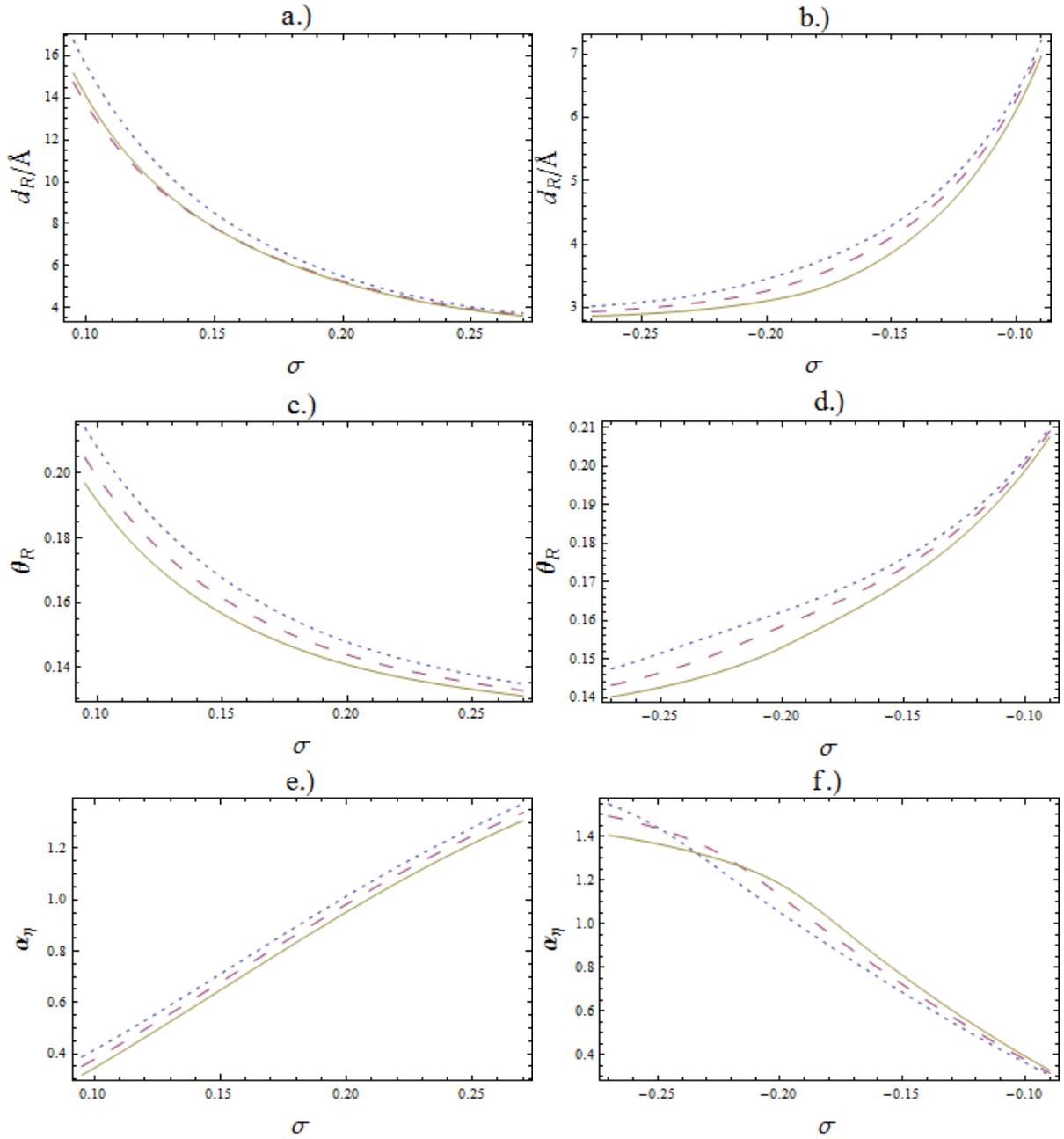

Fig. 5. Supercoiling fluctuation parameters as functions of $\sigma$, keeping fixed $f_1 = 0.7$ and $f_2 = 0.3$, but varying $\theta$. On the left hand side we present curves for positive values of $\sigma$, whereas on the right we present curves for negative values. In panels a.) and b.) we present plots of the mean squared amplitude of undulations in the supercoil $d_R = \langle \delta R(\tau)^2 \rangle^{1/2}$. In panels c.) and d.) we present the angular fluctuation amplitude $\theta_R = \langle (dR(\tau)/d\tau)^2 \rangle^{1/2}$. Finally, in panels e.) and f.) we present plots of the variational 'spring constant' $\alpha_\eta$ that determines the size of the fluctuations in the tilt angle $\eta(s)$. In all plots the blue short dashed, red medium dashed and dark yellow lines correspond to the values $\theta = 0.4, 0.5$ and $0.6$.

We have also plotted the fluctuation parameters $d_R$, $\theta_R$ and $\alpha_\eta$ in Fig 5. The most interesting of these is the plot of $d_R$, for which there is a very pronounced asymmetry between the positive and negative supercoils; the positive $\sigma$ values have a significantly larger value of $d_R$ than the negative ones. This suggests that here there is a less repulsive contribution from interaction terms, as well as a lower effective spring constant in interaxial separation, at the $R_0$ values calculated, than would be for smaller values of the inter-axial separation. For more attraction, a larger value $d_R$ would increase the effective degree of repulsion much less. Also, a larger value of $d_R$ would increase the entropy of the supercoil, and so reduce the free energy. If the two braided sections were pushed closer together to a smaller value of $R_0$, the effective spring constant of the interaction would become larger, which could result in smaller values of $d_R$ to try and reduce the amount of repulsion. Also, large values of $d_R$ are not allowed by steric repulsion, and so this would also limit $d_R$. All of this combined could create effective entropic repulsion, pushing up the free energy with decreasing $R_0$, through the entropic factors in Eq. (2.68). This reasoning could account for the larger values of $|M|$ and $R_0$ at fixed $|\sigma|$, seen for positive supercoils, and still tie in with the fact that interactions between left handed supercoils are favoured by the interaction energy [38]. Other additional effects accounting for the smaller $R_0$ values, for negative supercoils, could be the effect of changing $g_{av}$, as well as differences in the reduction in entropy due to fluctuations in $\Delta\xi(\tau)$ between left and right supercoils.

Changing both $f_1 + f_2$ and $f_1/f_2$ does not seem to alter the key findings of this paper. Firstly, for the parameter values investigated we always find that the free energy is lower for left handed supercoils than right handed ones for a given value of $|\sigma|$. Secondly, we always find that right handed supercoils form tighter structures than left handed ones.

The finding that right handed supercoils can, in fact, form tight structures through helix structure specific forces fits well with the observations of Refs. [10] and [70], where a collapsed (tightly supercoiled state) was seen for negatively supercoiled DNA . The length of the contact regions increased with increasing salt concentration and supercoiling density. The ground state analysis of Ref. [51] does not agree with this trend for tight supercoiling. Here, as well as tight negative supercoiling, the supercoiling radius decreases monotonically with increasing supercoiling density, which is in-line with these observations. However, examination of these experimental results might suggest some coexistence regime, where at a certain value of $M$ two states with different supercoiling densities would be observed, not seen for the parameter range investigated. However, this might have to do with DNA substrate interactions that restrict the thermal fluctuations perpendicular to the substrate plane, which modifies the statistical mechanics, as well as coexistence occurring at lower values of $\sigma$ than those considered here. At present, unfortunately, little seems to have been done in a detailed examination of the difference in structure between negative and positively supercoils, except for mini-circles [71] exploiting new techniques in cryo-microscopy.

Unlike the ground state study [51], the asymmetry of the supercoiling free energy is small in the results presented. It is dominated by the elastic energies required to form the supercoiling, which lead to a predominantly quadratic dependence. The asymmetry in the effective interaction between segments is a correction to this dependence. In fact, the free energy as a function of $\sigma$ can be deduced from experimental measurements. In Ref. [54] the supercoiling free energy was calculated indirectly by investigating the degree binding of ethidium bromide, which induces supercoiling, as function of concentration using a binding model. This study uses the fact binding is influenced by the supercoiling free energy. A slight asymmetry in free energy was actually observed [54], where positive supercoils indeed have a slightly lower free energy; this fits well with the findings of this paper. In another approach is to measure the gel-electrophoresis intensity profiles that can yield information about relative populations of topisomers on the closing of DNA into an un-nicked loop. In the pioneering study of Ref. [55] this was indeed done, introducing additional negative writhing under a fixed concentration ethidium bromide. The intensity profiles, in these experiments show an asymmetry in the relative populations about the most likely configuration. The most likely configuration corresponds to the closed circular DNA state before ethidium bromide is added. Indeed, however, some of this asymmetry is not due positive negative supercoil asymmetry, but that the molecule needs to be slightly twisted for closure [55]. Nevertheless, there are indications, when looking at the intensity profiles presented there, that the populations of topisomers may not be exactly Gaussian distributed in the peaks further away from the maximum, showing a further marked asymmetry. Indeed, a Gaussian distribution is what a quadratic dependence of the free energy, respecting positive-negative supercoiling symmetry, would dictate. Any observed asymmetry might be attributed to an asymmetry in the twisting elastic response of DNA. On the other hand, a slight asymmetry is again seen [53] in the mechanical braiding of two DNA molecules [50], where the molecules are nicked, and thus twisting is not coupled to braiding.

## 4. Concluding remarks and outlook

The key achievement of this work has been to build a mathematical-statistical mechanical formulism in which to deal with forces that depend on the helix structure of DNA. This analysis also contains new innovations on how to deal with Fuller-White theorem, and to fully include twisting, stretching and twisting fluctuations, thermal and intrinsic base-pair dependent. In this paper, we have focused on the situation where helix structure dependent forces may be considered weak; relative thermal twisting fluctuations of the two segments that make up the braid dominate. In the supercoil, this situation corresponds to no preferred azimuthal orientation is favoured between the two minor grooves of the braided segments. This whole approach is independent of the actual model of interaction used (provided that the helix dependent forces are weak), although in generating numerical results we have assumed the KL model of interaction; an alternate theory, or an empirical model based on simulations could also be used.

 Whether one lies in a weak or strong helix dependent regime relies on how ions are localized within the DNA grooves, at least within the KL model of interaction. Presumably, by increasing the amount that condensed and bound counter-ions that neutralize the DNA charge and increasing the amount of ions present in the major groove we can obtain results more akin to those presented in Ref. [51]. This, however, requires a different calculation– for this particular regime– than the one presented here. This would be of a similar nature to the calculation presented in Ref. [49] for mechanical braiding, but with the additional topological constraint due to closed loop supercoiling. Such a calculation supposes that there is an average angle between the two minor grooves of the segments making up the braid, i.e. $\langle \Delta \xi(\tau) \rangle$ is defined. Details of how such a calculation could be performed are presented in Ref. [52], but no numerical results have been obtained, yet. This might be presented in a later work. On the other hand, for such parameter values, the plectoneme state may compete with a condensed toroidal state of the DNA molecule [72]; or, in the case where many molecules are present, aggregation may occur. What should control the size of the charge compensation, and the groove localization, is the particular species of ions as well as their relative concentrations. Some ions are known to bind quite strongly to base pairs in the DNA grooves, while others less so. The weak helix specific force regime investigated, here, should be more appropriate for monovalent ions that have a relatively low of charge compensation; and some instances for divalent ones such as $Mg^{2+}$ and $Ca^{2+}$, where preferential interactions of ions with base pairs may weak enough to assume a value $\theta \leq 0.6$. We hypothesise that the weak helix specific regime is the correct regime in physiological conditions.

In the regime of weak helix dependent interactions, the situation is dramatically altered from the ground state considered in Ref. [51]. Here, as opposed to the large asymmetry between left and right supercoils seen there, we see a much slighter one between left and right handed supercoils. Also, we find a surprising counter intuitive result. Although, we find for positive supercoils that the free energy is slightly less than negative ones, especially at lower $|\sigma|$ values considered, it turns out that these supercoils are in fact looser structures, i.e. larger values of $R_0$ and $d_R$ are favoured. This is most likely caused by the balance of attraction and entropic effects. At a given value of $R_0$, in the left handed supercoils there is a smaller effective spring constant due to the helix specific forces; this allows for an increase in fluctuations and more entropy. If the braid is pushed to shorter average inter-axial separations, this may result in increased repulsion from the decrease of braid entropy as these enhanced fluctuations are decreased.

Any left-right handed supercoil asymmetry is important for a couple of reasons. Firstly, it may provide a signature that helix dependent interactions matter, if the degree of asymmetry is changed by using different counter-ions and salt concentrations. As discussed below, we would also expect that this asymmetry to be more pronounced when two identical DNA texts are inserted into the supercoil. In addition, this asymmetry between left and right supercoils may have biological significance. The tighter structure and smaller amount of bending fluctuations in negative supercoiling may be important for negative supercoiling as it may increase the efficiency of recombination events and the action of topoisomerases [70]. The higher conformational entropy due to significantly larger values of $d_R$ for positive supercoils may also be important, in the relation to the fact that hyper-thermophiles adopt positive supercoiling [39]. It could be because positive supercoils have more conformational entropy, as the temperature is increased, their free energy becomes significantly lower than negative ones; the positive-negative super-coil asymmetry becomes more pronounced. Therefore, it could simply be a matter that under certain conditions it becomes much thermodynamically easier to compact DNA in positive supercoils than negative ones. Alternately, there could be other reasons. The study of Ref. [41] suggested that positive and negative DNA denature equally as well at temperatures of 107°C and concentrations of either 50mM sodium phosphate or 0.5mM potassium phosphate. However, a difference might still be observed at higher salt concentrations, closer to physiological ones. Also, the handedness of supercoiling may have still some influence formation of long lasting localized DNA bubbles, which have been observed to in supercoiled DNA [73]. The formation of these bubbles is also affected by temperature [73] and their formation may important in certain biological processes, such as transcription and the binding of certain proteins [74]. In positive supercoils, the lower free energy due to helix specific interactions between double stranded DNA, and higher bending fluctuation entropy, might hinder the formation of DNA bubbles, and this might have some bearing on the choice between positive and negative supercoils. Bubbles might form more readily in negative supercoils more than positive ones. It could be that certain biological processes can run only efficiently at sufficiently high temperatures for positive supercoils, due to the temperature dependence of these bubbles. Whereas, the relative ease of bubble formation in negative supercoils could become detrimental at too high temperatures. These are intriguing ideas that need further investigation.

Recent biotechnical advances allow for positive and negative supercoils to be manufactured without the use of Ethidium Bromide or other intercalating agents [75]. This may allow better understanding of differences between left and right supercoils, as supercoiling can be induced without any perturbing factors and on a large scale [75]. Through electrophoresis experiments similar to Ref. [55], the free energy profile should be re-examined under various conditions and any asymmetry fitted to available theory. Indeed, the experiments should be performed in various types of monovalent salt ranging from Caesium to Sodium salts, in various concentrations, to determine how the degree of asymmetry changes. Changing the salt type is indeed interesting, as the degree of groove localization, and which groove the ions localize at, is expected to depend on salt type [66], as well as sequence [67]. ATM and new techniques in cryro-ET [71] would also be useful. In the case of cryro-ET, through rapid adiabatic freezing, the degree of fluctuation in the (as well as the average) supercoiling radius could be measured and compared between left and right hand supercoils. Also, how temperature influences any asymmetry should also be investigated. To test what contribution to asymmetry is from elastic constants twisting experiments of the form of [76] should be done in various ionic conditions corresponding to the actual supercoiling experiments (as was advocated in Ref. [51]), as well as mechanical braiding experiments [50].

Theoretically, one obvious extension to this work is to consider end loops and supercoil branching. Naturally, entropy favours the branching of supercoils; however, what limits their number is elastic energy of the formation of end loops and junctions. The cost to form branches competes with the energy of braiding. Certain conditions, like low salt concentration, favour increased branching [13,36], as the relative energy cost to form a branch is low. Indeed, it would be interesting to see what effect helix specific forces might have on the average number of supercoil branches.

It has been suggested that supercoiling might useful in probing the nature of possible forces responsible for homology recognition [51]. There is evidence to suggest that two DNA segments with two identical sequences may associate more, due to reduced repulsion between them, than for two un-alike ones [77,78,79,80,81], or even cause some attraction: through the existence of recognition forces. A mechanism for this phenomenon has been suggested using the KL model of interaction [82], as well as other mechanisms [81,83]. It was suggested in Ref. [51], that special supercoil constructs could be used to further probe this difference. These constructs would contain two sequences that are identical to each other of the same length. We would expect that, if helix structure specific forces are indeed the origin of recognition forces, the supercoiling asymmetry should be more pronounced in these constructs than in natural plectonemes. By changing the supercoiling density, it should be possible to probe how the difference in interaction energy between alike and non-alike segments changes as a function of average inter-axial separation $R_0$ and average tilt angle $\eta_0$. Additionally, if such preferential interactions were present, they must affect the degree of supercoiling branching; less branches would be favoured in the constructs, as the preferential interaction between like sequences for such constructs would be lost in the formation of any super-coil branches. A theoretical investigation of these effects, based on this study, might be interesting and may prompt experimental work, using such constructs, to probe the nature of recognition forces. A grasp of the nature of recognition forces could be important in understanding biological processes involving DNA in the cell.

## Acknowledgements

D.J. Lee would like to acknowledge useful discussions with R. Cortini , A. Korte, A. A. Korynshev, E. L. Starostin and G.H.M. van der Heijden, which stimulated this work. He is also thankful of the continuing support of the Department of Chemistry at Imperial College.

## Outline of the supplemental Material

Appendix A contains explicit details of calculations to do with the expansion of the free energy for weak helix forces, as outlined in Subsection 2.5. Appendix B contains additional details in deriving the expressions for the free energy contained in Subsection 2.6. Appendix C gives an outline of how the average writhe is calculated, along with various integrals contained in Appendix D. In Appendix E it is shown how, using the KL model of interaction, the various expressions for the free energy simplify. Additional numerical results are presented in Appendix F.

[59] The assumption in writing Eq. (2.31) comes from the fact that it is expected that $\langle \delta h_1(\tau) \rangle = \langle \delta h_2(\tau) \rangle = 0$ and $\langle \delta h_1^S(\tau) \rangle = \langle \delta h_2^S(\tau) \rangle = 0$; stretching fluctuations thermal or base pair dependent do not affect the average length of the molecule. Also, the first order and higher Fourier harmonics of $\delta h(s)$ (both $\delta h_1(\tau)$ and $\delta h_2(\tau)$ analytically continued together into one function with the coordinate $s$ around the loop) around the loop satisfy Eq.(2.31), whereas only the zeroth order harmonic does not. The first order harmonic has periodicity $\approx 2L_b$, and the second order mode $\approx L_b$ and so on (note that we always require $\delta h(L \approx 2L_b) = \delta h(0)$). Also, note that the smallest periodicity for the modes is $h$, the average base pair

spacing. For $L_b \gg h$, the contribution to the free energy from the zeroth mode (which doesn't satisfy Eq. (2.31)) is overwhelmed from the contribution from the other modes, as their entropy contribution is far larger. The assumption can be relaxed by considering the Fourier modes in $\delta h(s)$ and $\delta\Omega(s)$ around the loop and including their zeroth order modes.

# Supplemental Material

## A. The weak helix forces expansion of the free energy

From Eqs. (2.61) and (2.63) of the main text we may write

$$\langle F_T \rangle_g \approx -k_B T \ln Z_T + \langle E_{T,0}[\delta\eta(\tau), \delta R(\tau)] - E_T^T[\delta\eta(\tau), \delta R(\tau)] \rangle_T$$
$$+ k_B T \left\langle \langle E_1 \rangle_g - \frac{\langle E_2 \rangle_g}{2} + \frac{\langle E_3 \rangle_g}{3!} - \left( \frac{\langle E_4 \rangle_g}{4!} - \frac{\langle E_2^2 \rangle_g}{8} \right) \right\rangle_T .$$

(A.1)

We start by using Eqs. (2.58) and (2.59) of the main text, and dividing $\Delta\xi(\tau)$ into structural and thermal contributions, namely $\Delta\xi(\tau) = \Delta\xi_S(\tau) + \Delta\xi_T(\tau)$. This allows us to write for each term in the expansion (Eq. (A.1)) the following expressions (using Eq. (2.58) of the main text)

$$\langle E_1 \rangle_g = \frac{1}{\tilde{Z}_0(k_B T)} \sum_{n=-\infty,}^{\infty} \int D\Delta\xi_T(\tau) \int_0^{L_b} d\tau \bar{\varepsilon}_{\text{int}}(R_0, \delta R(\tau), \eta(\tau), g_{av}, n)(1-\delta_{n,0})$$
$$\exp(-in\Delta\xi_T(\tau)) \langle \exp(-in\Delta\xi_S(\tau)) \rangle_g \exp\left(-\frac{E_{tw,0}[\Delta\xi_T(\tau)]}{k_B T}\right),$$

(A.2)

$$\langle E_2 \rangle_g = \frac{1}{\tilde{Z}_0(k_B T)^2} \sum_{n=-\infty,}^{\infty} \sum_{m=-\infty,}^{\infty} \int D\Delta\xi_T(\tau) \int_0^{L_b} d\tau \int_0^{L_b} d\tau' (1-\delta_{n,0})(1-\delta_{m,0}) \bar{\varepsilon}_{\text{int}}(R_0, \delta R(\tau), \eta(\tau), g_{av}, n)$$
$$\bar{\varepsilon}_{\text{int}}(R_0, \delta R(\tau'), \eta(\tau'), g_{av}, m) \exp(-i(n\Delta\xi_T(\tau) + m\Delta\xi_T(\tau'))) \langle \exp(-i(n\Delta\xi_S(\tau) + m\Delta\xi_S(\tau'))) \rangle_g$$
$$\exp\left(-\frac{E_{tw,0}[\Delta\xi_T(\tau)]}{k_B T}\right),$$

(A.3)

$$\langle E_3 \rangle_g = \frac{1}{\tilde{Z}_0(k_B T)^3} \sum_{n=-\infty,}^{\infty} \sum_{m=-\infty,}^{\infty} \sum_{l=-\infty,}^{\infty} \int D\Delta\xi_T(\tau) \int_0^{L_b} d\tau \int_0^{L_b} d\tau' \int_0^{L_b} d\tau'' (1-\delta_{n,0})(1-\delta_{m,0})(1-\delta_{l,0})$$
$$\bar{\varepsilon}_{\text{int}}(R_0, \delta R(\tau), \eta(\tau), g_{av}, n) \bar{\varepsilon}_{\text{int}}(R_0, \delta R(\tau'), \eta(\tau'), g_{av}, m) \bar{\varepsilon}_{\text{int}}(R_0, \delta R(\tau''), \eta(\tau''), g_{av}, l)$$
$$\exp(-i(n\Delta\xi_T(\tau) + m\Delta\xi_T(\tau') + l\Delta\xi_T(\tau''))) \langle \exp(-i(n\Delta\xi_S(\tau) + m\Delta\xi_S(\tau') + l\Delta\xi_S(\tau''))) \rangle_g$$
$$\exp\left(-\frac{E_{tw,0}[\Delta\xi_T(\tau)]}{k_B T}\right),$$

(A.4)

$$\langle E_4 \rangle_g = \frac{1}{\tilde{Z}_0 (k_B T)^4} \sum_{n=-\infty}^{\infty} \sum_{m=-\infty}^{\infty} \sum_{l=-\infty}^{\infty} \sum_{p=-\infty}^{\infty} \int D\Delta\xi_T(\tau) \int_0^{L_b} d\tau \int_0^{L_b} d\tau' \int_0^{L_b} d\tau'' \int_0^{L_b} d\tau''' (1-\delta_{n,0})(1-\delta_{m,0})$$
$$(1-\delta_{l,0})(1-\delta_{p,0}) \bar{\varepsilon}_{int}(R_0, \delta R(\tau), \eta(\tau), g_{av}, n) \bar{\varepsilon}_{int}(R_0, \delta R(\tau'), \eta(\tau'), g_{av}, m) \bar{\varepsilon}_{int}(R_0, \delta R(\tau''), \eta(\tau''), g_{av}, l)$$
$$\bar{\varepsilon}_{int}(R_0, \delta R(\tau'''), \eta(\tau'''), g_{av}, p) \exp\left(-i(n\Delta\xi_T(\tau) + m\Delta\xi_T(\tau') + l\Delta\xi_T(\tau'') + p\Delta\xi_T(\tau'''))\right)$$
$$\left\langle \exp\left(-i(n\Delta\xi_S(\tau) + m\Delta\xi_S(\tau') + l\Delta\xi_S(\tau'') + p\Delta\xi_S(\tau'''))\right) \right\rangle_g \exp\left(-\frac{E_{tw,0}[\Delta\xi_T(\tau)]}{k_B T}\right),$$

(A.5)

where

$$\tilde{Z}_0 = \int D\Delta\xi_T(\tau) \exp\left(-\frac{E_{tw,0}[\Delta\xi_T(\tau)]}{k_B T}\right), \tag{A.6}$$

$$\frac{E_{tw,0}[\Delta\xi_T(\tau)]}{k_B T} = \frac{l_c}{4} \int_{-\infty}^{\infty} d\tau \left(\frac{d\Delta\xi_T(\tau)}{d\tau}\right)^2. \tag{A.7}$$

We start our evaluation of Eqs. (A.3)-(A.5) by considering the thermal averages

$$\left\langle \exp(-in\Delta\xi_T(\tau)) \right\rangle_\xi = \frac{1}{\tilde{Z}_0} \int D\Delta\xi_T(\tau) \exp(-in\Delta\xi_T(\tau)) \exp\left(-\frac{E_{tw,0}[\Delta\xi_T(\tau)]}{k_B T}\right), \tag{A.8}$$

$$\left\langle \exp(-i(n\Delta\xi_T(\tau) + m\Delta\xi_T(\tau'))) \right\rangle_\xi = \frac{1}{\tilde{Z}_0} \int D\Delta\xi_T(\tau) \exp(-i(n\Delta\xi_T(\tau) + m\Delta\xi_T(\tau'))) \exp\left(-\frac{E_{tw,0}[\Delta\xi_T(\tau)]}{k_B T}\right),$$

(A.9)

$$\left\langle \exp(-i(n\Delta\xi_T(\tau) + m\Delta\xi_T(\tau') + l\Delta\xi_T(\tau''))) \right\rangle_\xi =$$
$$\frac{1}{\tilde{Z}_0} \int D\Delta\xi_T(\tau) \exp(-i(n\Delta\xi_T(\tau) + m\Delta\xi_T(\tau') + l\Delta\xi_T(\tau''))) \exp\left(-\frac{E_{tw,0}[\Delta\xi_T(\tau)]}{k_B T}\right), \tag{A.10}$$

$$\left\langle \exp(-i(n\Delta\xi_T(\tau) + m\Delta\xi_T(\tau') + l\Delta\xi_T(\tau'') + p\Delta\xi_T(\tau'''))) \right\rangle_\xi =$$
$$\frac{1}{\tilde{Z}_0} \int D\Delta\xi_T(\tau) \exp(-i(n\Delta\xi_T(\tau) + m\Delta\xi_T(\tau') + l\Delta\xi_T(\tau'') + p\Delta\xi_T(\tau'''))) \exp\left(-\frac{E_{tw,0}[\Delta\xi_T(\tau)]}{k_B T}\right).$$

(A.11)

Let us first consider the evaluation of Eq. (A.8). This can be written as (using Fourier transforms)

$$\langle \exp(-in\Delta\xi_T(\tau))\rangle_\xi = \frac{1}{\tilde{Z}_0}\int D\Delta\tilde{\xi}_T(k)\exp\left(-\frac{l_c}{4}\frac{1}{2\pi}\int_{-\infty}^{\infty}dk k^2 \Delta\tilde{\xi}_T(k)\Delta\tilde{\xi}_T(-k)\right)$$
$$\exp\left(-\frac{in}{2}\left(\Delta\tilde{\xi}_T(k)\exp(-ik\tau') + \Delta\tilde{\xi}_T(-k)\exp(ik\tau')\right)\right). \quad (A.12)$$

On performing the integration, we find that

$$\langle \exp(-in\Delta\xi_T(\tau))\rangle_\xi = \exp\left(-\frac{1}{l_c}\frac{1}{2\pi}\int_{-\infty}^{\infty}\frac{dk}{k^2}\right) = 0, \quad (A.13)$$

as the integral over $k$ in Eq. (A.13) is singular. Thus, we find that $\langle E_1\rangle_g = 0$.

Next, let us consider the evaluation of Eq. (A.9). Using Fourier transforms (here we have assumed that $L_b$ is sufficiently long that we can neglect finite size effects), we may write this equation as

$$\langle \exp(-i(n\Delta\xi_T(\tau) + m\Delta\xi_T(\tau')))\rangle_\xi = \frac{1}{\tilde{Z}_0}\int D\Delta\tilde{\xi}_T(k)\exp\left(-\frac{in}{2}\left(\Delta\tilde{\xi}_T(k)\exp(-ik\tau) + \Delta\tilde{\xi}_T(-k)\exp(ik\tau)\right)\right)$$
$$\exp\left(-\frac{im}{2}\left(\Delta\tilde{\xi}_T(k)\exp(-ik\tau') + \Delta\tilde{\xi}_T(-k)\exp(ik\tau')\right)\right)\exp\left(-\frac{l_c}{4}\frac{1}{2\pi}\int_{-\infty}^{\infty}dk k^2\Delta\tilde{\xi}_T(k)\Delta\tilde{\xi}_T(-k)\right).$$
$$(A.14)$$

On performing the functional integration over $\Delta\tilde{\xi}_T(k)$, we find for the average

$$\langle \exp(-i(n\Delta\xi_T(\tau) + m\Delta\xi_T(\tau')))\rangle_\xi$$
$$= \exp\left(-\frac{1}{l_c}\frac{1}{2\pi}\int_{-\infty}^{\infty}\frac{dk}{k^2}(n\exp(-ik\tau) + m\exp(-ik\tau'))(n\exp(ik\tau) + m\exp(ik\tau'))\right) \quad (A.15)$$
$$= \exp\left(-\frac{1}{l_c}\frac{1}{2\pi}\int_{-\infty}^{\infty}\frac{dk}{k^2}(n^2 + m^2 + 2nm\cos(k(\tau-\tau')))\right).$$

The average expressed by Eq. (A.15) is only not zero when $n = -m$. This is because when $n \neq -m$ the integral inside the exponential in Eq. (A.15) again diverges. For $n = -m$, the integral evaluates to

$$\frac{1}{2\pi}\int_{-\infty}^{\infty}\frac{dk}{k^2}(2n^2 - 2n^2\cos(k(\tau-\tau'))) = n^2|\tau-\tau'|. \quad (A.16)$$

Thus, we may write

$$\langle \exp(-i(n\Delta\xi_T(\tau) + m\Delta\xi_T(\tau')))\rangle_\xi = \delta_{n,-m}\exp\left(-\frac{n^2|\tau-\tau'|}{l_p}\right). \quad (A.17)$$

Using similar steps one may show

$$\left\langle \exp\left(-i(n\Delta\xi_T(\tau)+m\Delta\xi_T(\tau')+l\Delta\xi_T(\tau''))\right)\right\rangle_T$$

$$=\exp\left(-\frac{1}{l_c}\frac{1}{2\pi}\int_{-\infty}^{\infty}\frac{dk}{k^2}\left(n\exp(-ik\tau)+m\exp(-ik\tau')+l\exp(-ik\tau'')\right)\left(n\exp(ik\tau)+m\exp(ik\tau')+l\exp(ik\tau'')\right)\right)$$

$$=\exp\left(-\frac{1}{l_c}\frac{1}{2\pi}\int_{-\infty}^{\infty}\frac{dk}{k^2}\left(n^2+m^2+l^2+2nm\cos(k(\tau-\tau'))+2nl\cos(k(\tau-\tau''))+2ml\cos(k(\tau'-\tau''))\right)\right).$$

(A.18)

Only when $l=-n-m$ is the average not equal to zero (again due to divergent integrals). In this case we have the integral to evaluate

$$\frac{1}{2\pi}\int_{-\infty}^{\infty}\frac{dk}{k^2}\left(n^2+m^2+(n+m)^2+2nm\cos(k(\tau-\tau'))-2n(n+m)\cos(k(\tau-\tau''))-2m(n+m)\cos(k(\tau'-\tau''))\right)$$

$$=n(n+m)|\tau-\tau''|+m(n+m)|\tau'-\tau''|-nm|\tau-\tau'|.$$

(A.19)

On substitution of Eq. (A.19) into Eq. (A.18) we obtain

$$\left\langle \exp\left(-i(n\Delta\xi_T(\tau)+m\Delta\xi_T(\tau')+l\Delta\xi_T(\tau''))\right)\right\rangle_\xi = \delta_{l,-n,-m}$$

$$\exp\left(-\frac{1}{l_p}\left(n(n+m)|\tau-\tau''|+m(n+m)|\tau'-\tau''|-nm|\tau-\tau'|\right)\right).$$

(A.20)

Again, using similar analysis as before, we may also write from Eq. (A.11)

$$\left\langle \exp\left(-i(n\Delta\xi_T(\tau)+m\Delta\xi_T(\tau')+l\Delta\xi_T(\tau'')+p\Delta\xi_T(\tau'''))\right)\right\rangle_\xi$$

$$=\exp\left(-\frac{1}{l_c}\frac{1}{2\pi}\int_{-\infty}^{\infty}\frac{dk}{k^2}\left(n\exp(-ik\tau)+m\exp(-ik\tau')+l\exp(-ik\tau'')+p\exp(-ik\tau''')\right)\right.$$

$$\left(n\exp(ik\tau)+m\exp(ik\tau')+l\exp(ik\tau'')+p\exp(ik\tau''')\right)\Big)$$

$$=\exp\left(-\frac{1}{l_c}\frac{1}{2\pi}\int_{-\infty}^{\infty}\frac{dk}{k^2}\left(n^2+m^2+l^2+p^2+2nm\cos(k(\tau-\tau'))+2nl\cos(k(\tau-\tau''))\right.\right.$$

$$+2np\cos(k(\tau-\tau'''))+2ml\cos(k(\tau'-\tau''))+2mp\cos(k(\tau'-\tau'''))+2lp\cos(k(\tau''-\tau'''))\Big)\Big).$$

(A.21)

Only when $p=-n-m-l$ is the average not equal to zero. When this condition is satisfied, we have that

$$\frac{1}{2\pi}\int_{-\infty}^{\infty}\frac{dk}{k^2}\left(n^2+m^2+l^2+(n+m+l)^2+2nm\cos(k(\tau-\tau'))+2nl\cos(k(\tau-\tau''))+2ml\cos(k(\tau'-\tau''))\right.$$

$$-2n(n+m+l)\cos(k(\tau-\tau'''))-2m(n+m+l)\cos(k(\tau'-\tau'''))-2l(n+m+l)\cos(k(\tau''-\tau'''))\Big)$$

$$=n(n+m+l)|\tau-\tau'''|+m(n+m+l)|\tau'-\tau'''|+l(n+m+l)|\tau''-\tau'''|-nm|\tau-\tau'|-nl|\tau-\tau''|-ml|\tau'-\tau''|,$$

(A.22)

and thus, substituting Eq. (A.22) into (A.21), we have for the average

$$\left\langle \exp\left(-i(n\Delta\xi_T(\tau)+m\Delta\xi_T(\tau')+l\Delta\xi_T(\tau'')+p\Delta\xi_T(\tau'''))\right)\right\rangle_\xi = \exp\left(\frac{nm|\tau-\tau'|+nl|\tau-\tau''|+ml|\tau'-\tau''|}{l_p}\right)$$

$$\exp\left(-\frac{\left(n(n+m+l)|\tau-\tau'''|+m(n+m+l)|\tau'-\tau'''|+l(n+m+l)|\tau''-\tau'''|\right)}{l_p}\right)\delta_{p,-n,-m-l}.$$

(A.23)

Using these results (Eqs.(A.17), (A.20) and (A.23)) we may re-express Eqs. (A.3)-(A.5) as

$$\left\langle E_2 \right\rangle_g = \frac{1}{(k_B T)^2}\sum_{n=-\infty}^{\infty}\int_0^{L_p}d\tau\int_0^{L_p}d\tau'\left(1-\delta_{n,0}\right)\bar{\varepsilon}_{\text{int}}(R_0,\delta R(\tau),\eta(\tau),g_{av},n)$$

$$\bar{\varepsilon}_{\text{int}}(R_0,\delta R(\tau'),\eta(\tau'),g_{av},-n)\exp\left(-\frac{n^2|\tau-\tau'|}{l_p}\right)\left\langle \exp\left(-in(\Delta\xi_S(\tau)-\Delta\xi_S(\tau'))\right)\right\rangle_g,$$

(A.24)

$$\left\langle E_3 \right\rangle_g = \frac{1}{(k_B T)^3}\sum_{n=-\infty}^{\infty}\sum_{m=-\infty}^{\infty}\int_0^{L_p}d\tau\int_0^{L_p}d\tau'\int_0^{L_p}d\tau''\left(1-\delta_{n,0}\right)\left(1-\delta_{m,0}\right)\left(1-\delta_{n,-m}\right)$$

$$\bar{\varepsilon}_{\text{int}}(R_0,\delta R(\tau),\eta(\tau),g_{av},n)\bar{\varepsilon}_{\text{int}}(R_0,\delta R(\tau'),\eta(\tau'),g_{av},m)\bar{\varepsilon}_{\text{int}}(R_0,\delta R(\tau''),\eta(\tau''),g_{av},-n-m)$$

$$\exp\left(-\frac{1}{l_p}\left(n(n+m)|\tau-\tau''|+m(n+m)|\tau'-\tau''|-nm|\tau-\tau'|\right)\right)$$

$$\left\langle \exp\left(-i(n\Delta\xi_S(\tau)+m\Delta\xi_S(\tau')-(n+m)\Delta\xi_S(\tau''))\right)\right\rangle_g,$$

(A.25)

$$\left\langle E_4 \right\rangle_g = \frac{1}{(k_B T)^4}\sum_{n=-\infty}^{\infty}\sum_{m=-\infty}^{\infty}\sum_{l=-\infty}^{\infty}\int_0^{L_p}d\tau\int_0^{L_p}d\tau'\int_0^{L_p}d\tau''\int_0^{L_p}d\tau'''\left(1-\delta_{n,0}\right)\left(1-\delta_{m,0}\right)\left(1-\delta_{l,0}\right)\left(1-\delta_{l,-n-m}\right)$$

$$\bar{\varepsilon}_{\text{int}}(R_0,\delta R(\tau),\eta(\tau),g_{av},n)\bar{\varepsilon}_{\text{int}}(R_0,\delta R(\tau'),\eta(\tau'),g_{av},m)\bar{\varepsilon}_{\text{int}}(R_0,\delta R(\tau''),\eta(\tau''),g_{av},l)$$

$$\bar{\varepsilon}_{\text{int}}(R_0,\delta R(\tau'''),\eta(\tau'''),g_{av},-n-m-l)\exp\left(\frac{nm|\tau-\tau'|+nl|\tau-\tau''|+ml|\tau'-\tau''|}{l_p}\right)$$

(A.26)

$$\exp\left(-\frac{1}{l_p}\left(n(n+m+l)|\tau-\tau'''|+m(n+m+l)|\tau'-\tau'''|+l(n+m+l)|\tau''-\tau'''|\right)\right)$$

$$\left\langle \exp\left(-i(n\Delta\xi_S(\tau)+m\Delta\xi_S(\tau')+l\Delta\xi_S(\tau'')-(n+m+l)\Delta\xi_S(\tau'''))\right)\right\rangle_g.$$

Also, at this stage, it useful to write down an expression for $\left\langle E_2^2 \right\rangle_g$ (the last term in Eq. (A.1))

$$\left\langle E_2^2 \right\rangle_g = \frac{1}{(k_B T)^4}\sum_{n=-\infty}^{\infty}\sum_{m=-\infty}^{\infty}\int_0^{L_p}d\tau\int_0^{L_p}d\tau'\int_0^{L_p}d\tau''\int_0^{L_p}d\tau'''\left(1-\delta_{n,0}\right)\left(1-\delta_{m,0}\right)\bar{\varepsilon}_{\text{int}}(R_0,\delta R(\tau),\eta(\tau),g_{av},n)$$

$$\bar{\varepsilon}_{\text{int}}(R_0,\delta R(\tau'),\eta(\tau'),g_{av},-n)\bar{\varepsilon}_{\text{int}}(R_0,\delta R(\tau''),\eta(\tau''),g_{av},m)\bar{\varepsilon}_{\text{int}}(R_0,\delta R(\tau'''),\eta(\tau'''),g_{av},-m)$$

$$\exp\left(-\frac{n^2|\tau-\tau'|+m^2|\tau''-\tau'''|}{l_p}\right)\left\langle \exp\left(-i(n\Delta\xi_S(\tau)-n\Delta\xi_S(\tau')+m\Delta\xi_S(\tau'')-m\Delta\xi_S(\tau'''))\right)\right\rangle_g.$$

Next we evaluate the ensemble averages over base pair realisations. Using Gaussian statistics and Eqs. (2.17) and (2.18) of the main text, we may express

$$\langle \exp(-in(\Delta\xi_S(\tau) - \Delta\xi_S(\tau')))\rangle_g$$
$$= \frac{1}{Z_g} \int D\delta g_1^{0,S}(\tau) \int D\delta g_2^{0,S}(\tau) P[g_1^{0,S}(\tau)] P[g_2^{0,S}(\tau)] \exp(-in(\Delta\xi_S(\tau) - \Delta\xi_S(\tau'))), \quad (A.28)$$

$$Z_g = \int D\delta g_1^{0,S}(\tau) \int D\delta g_2^{0,S}(\tau) P[g_1^{0,S}(\tau)] P[g_2^{0,S}(\tau)]. \quad (A.29)$$

Here, the Gaussian probability distributions are given by

$$P[g_\mu^{0,S}(\tau)] = \exp\left(-\frac{\lambda_c^{(0)}}{2} \int_0^{L_b} d\tau\, g_\mu^{0,S}(\tau)^2\right), \quad (A.30)$$

and the functional integrals in Eqs. (A.28) and (A.29) sum over all possible realization of $\delta g_1^{0,S}(\tau)$ and $\delta g_2^{0,S}(\tau)$. All the averages in Eqs. (A.24)-(A.27) depend only on the difference between $g_1^{0,S}(\tau)$ and $g_2^{0,S}(\tau)$ (as well as relative distance between two points along the molecules, $\tau - \tau'$). Therefore, we can write the average in Eq. (A.29) as

$$\langle \exp(-in(\Delta\xi_S(\tau) - \Delta\xi_S(\tau')))\rangle_g$$
$$= \frac{1}{Z_\Delta} \int D\Delta\xi_S(\tau) \exp(-in(\Delta\xi_S(\tau) - \Delta\xi_S(\tau'))) P_\Delta[\Delta\xi_S(\tau)], \quad (A.31)$$

where

$$Z_\Delta = \int D\Delta\xi_S(\tau) P_\Delta[\Delta\xi_S(\tau)], \quad (A.32)$$

$$P_\Delta[\Delta\xi_S(\tau)] = \exp\left[-\frac{\lambda_c^{(0)}}{4} \int_{-\infty}^{\infty} d\tau \left(\frac{d\Delta\xi_S(\tau)}{d\tau}\right)^2\right]. \quad (A.33)$$

Similarly, we may express for the ensemble averages

$$\langle \exp(-i(n\Delta\xi_S(\tau) + m\Delta\xi_S(\tau') - (n+m)\Delta\xi_S(\tau'')))\rangle_g$$
$$= \frac{1}{Z_\Delta} \int D\Delta\xi_S(\tau) \exp(-i(n\Delta\xi_S(\tau) + m\Delta\xi_S(\tau') - (n+m)\Delta\xi_S(\tau''))) P_\Delta[\Delta\xi_S(\tau)], \quad (A.34)$$

$$\left\langle \exp\left(-i(n\Delta\xi_S(\tau)+m\Delta\xi_S(\tau')+l\Delta\xi_S(\tau'')-(n+m+l)\Delta\xi_S(\tau'''))\right)\right\rangle_g$$

$$=\frac{1}{Z_\Delta}\int D\Delta\xi_S(\tau)\exp\left(-i(n\Delta\xi_S(\tau)+m\Delta\xi_S(\tau')+l\Delta\xi_S(\tau'')-(n+m+l)\Delta\xi_S(\tau'''))\right)P_\Delta[\Delta\xi_S(\tau)],$$

(A.35)

and

$$\left\langle \exp\left(-i(n\Delta\xi_S(\tau)-n\Delta\xi_S(\tau')+m\Delta\xi_S(\tau'')-m\Delta\xi_S(\tau'''))\right)\right\rangle_g$$

$$=\frac{1}{Z_\Delta}\int D\Delta\xi_S(\tau)\exp\left(-i(n\Delta\xi_S(\tau)+n\Delta\xi_S(\tau')+m\Delta\xi_S(\tau'')-m\Delta\xi_S(\tau'''))\right)P_\Delta[\Delta\xi_S(\tau)].$$

(A.36)

The first three averages (Eqs.(A.31), (A.34) and (A.35)) simply give us back Eqs. (A.17), (A.20) and (A.23), but with $l_p$ replaced by $\lambda_c^{(0)}$. The last average is a special case of Eq. (A.35), and can be written as

$$\left\langle \exp\left(-i(n\Delta\xi_T(\tau)-n\Delta\xi_T(\tau')+m\Delta\xi_T(\tau'')-m\Delta\xi_T(\tau''')) \right)\right\rangle_\xi =$$
$$\exp\left(-\frac{1}{\lambda_c^{(0)}}\left(n^2|\tau-\tau'|+m^2|\tau''-\tau'''|+nm\left(|\tau-\tau'''|+|\tau'-\tau''|-|\tau'-\tau'''|-|\tau-\tau''|\right)\right)\right).$$

(A.37)

Eqs. (A.24)-(A.27) can now be written as

$$\langle E_2\rangle_g = \frac{1}{(k_BT)^2}\sum_{n=-\infty}^{\infty}\int_0^{L_p}d\tau\int_0^{L_p}d\tau'(1-\delta_{n,0})\bar{\varepsilon}_{\text{int}}(R_0,\delta R(\tau),\eta(\tau),g_{av},n)$$
$$\bar{\varepsilon}_{\text{int}}(R_0,\delta R(\tau'),\eta(\tau'),g_{av},-n)\exp\left(-\frac{n^2|\tau-\tau'|}{\lambda_c}\right),$$

(A.38)

$$\langle E_3\rangle_g = \frac{1}{(k_BT)^3}\sum_{n=-\infty}^{\infty}\sum_{m=-\infty}^{\infty}\int_0^{L_p}d\tau\int_0^{L_p}d\tau'\int_0^{L_p}d\tau''(1-\delta_{n,0})(1-\delta_{m,0})(1-\delta_{m,-n})$$
$$\bar{\varepsilon}_{\text{int}}(R_0,\delta R(\tau),\eta(\tau),g_{av},n)\bar{\varepsilon}_{\text{int}}(R_0,\delta R(\tau'),\eta(\tau'),g_{av},m)\bar{\varepsilon}_{\text{int}}(R_0,\delta R(\tau''),\eta(\tau''),g_{av},-n-m)$$ (A.39)
$$\exp\left(-\frac{1}{\lambda_c}\left(n(n+m)|\tau-\tau''|+m(n+m)|\tau'-\tau''|-nm|\tau-\tau'|\right)\right),$$

$$\langle E_4\rangle_g = \frac{1}{(k_BT)^4}\sum_{n=-\infty}^{\infty}\sum_{m=-\infty}^{\infty}\sum_{l=-\infty}^{\infty}\int_0^{L_p}d\tau\int_0^{L_p}d\tau'\int_0^{L_p}d\tau''\int_0^{L_p}d\tau'''(1-\delta_{n,0})(1-\delta_{m,0})(1-\delta_{l,0})(1-\delta_{l,-n-m})$$
$$\bar{\varepsilon}_{\text{int}}(R_0,\delta R(\tau),\eta(\tau),g_{av},n)\bar{\varepsilon}_{\text{int}}(R_0,\delta R(\tau'),\eta(\tau'),g_{av},m)\bar{\varepsilon}_{\text{int}}(R_0,\delta R(\tau''),\eta(\tau''),g_{av},l)$$
$$\bar{\varepsilon}_{\text{int}}(R_0,\delta R(\tau'''),\eta(\tau'''),g_{av},-n-m-l)\exp\left(\frac{nm|\tau-\tau'|+nl|\tau-\tau''|+ml|\tau'-\tau''|}{\lambda_c}\right)$$ (A.40)
$$\exp\left(-\frac{(n(n+m+l)|\tau-\tau'''|+m(n+m+l)|\tau'-\tau'''|+l(n+m+l)|\tau''-\tau'''|)}{\lambda_c}\right),$$

$$\langle E_2^2 \rangle_g = \frac{1}{(k_B T)^4} \sum_{n=-\infty}^{\infty} \sum_{m=-\infty}^{\infty} \int_0^{L_b} d\tau \int_0^{L_b} d\tau' \int_0^{L_b} d\tau'' \int_0^{L_b} d\tau''' (1-\delta_{n,0})(1-\delta_{m,0}) \bar{\varepsilon}_{\text{int}}(R_0, \delta R(\tau), \eta(\tau), g_{av}, n)$$

$$\bar{\varepsilon}_{\text{int}}(R_0, \delta R(\tau'), \eta(\tau'), g_{av}, -n) \bar{\varepsilon}_{\text{int}}(R_0, \delta R(\tau''), \eta(\tau''), g_{av}, m) \bar{\varepsilon}_{\text{int}}(R_0, \delta R(\tau'''), \eta(\tau'''), g_{av}, -m)$$

$$\exp\left(-\frac{n^2|\tau-\tau'|+m^2|\tau''-\tau'''|}{\lambda_c}\right) \exp\left(-\frac{nm(|\tau-\tau'''|+|\tau'-\tau''|-|\tau'-\tau'''|-|\tau-\tau''|)}{\lambda_c^{(0)}}\right).$$

(A.41)

The next part of the evaluation is to evaluate the average over $\delta R(\tau)$ and $\delta\eta(\tau)$ bending fluctuations using the trial functional given by Eq. (2.62) of the main text. To do this we need to consider the averages

$$\langle \bar{\varepsilon}_{\text{int}}(R_0, \delta R(\tau), \eta(\tau), g_{av}, n) \bar{\varepsilon}_{\text{int}}(R_0, \delta R(\tau'), \eta(\tau'), g_{av}, -n) \rangle_T$$

$$= \frac{1}{Z_T} \int D\delta R(\tau) \int D\eta(\tau) \bar{\varepsilon}_{\text{int}}(R_0, \delta R(\tau), \eta(\tau), g_{av}, n) \bar{\varepsilon}_{\text{int}}(R_0, \delta R(\tau'), \eta(\tau'), g_{av}, -n) \qquad \text{(A.42)}$$

$$\exp\left(-\frac{E_T^T[\delta\eta(\tau), \delta R(\tau)]}{k_B T}\right),$$

$$\langle \bar{\varepsilon}_{\text{int}}(R_0, \delta R(\tau), \eta(\tau), g_{av}, n) \bar{\varepsilon}_{\text{int}}(R_0, \delta R(\tau'), \eta(\tau'), g_{av}, m) \bar{\varepsilon}_{\text{int}}(R_0, \delta R(\tau''), \eta(\tau''), g_{av}, -n-m) \rangle_T$$

$$= \frac{1}{Z_T} \int D\delta R(\tau) \int D\eta(\tau) \bar{\varepsilon}_{\text{int}}(R_0, \delta R(\tau), \eta(\tau), g_{av}, n) \bar{\varepsilon}_{\text{int}}(R_0, \delta R(\tau'), \eta(\tau'), g_{av}, m)$$

$$\bar{\varepsilon}_{\text{int}}(R_0, \delta R(\tau''), \eta(\tau''), g_{av}, -n-m) \exp\left(-\frac{E_T^T[\delta\eta(\tau), \delta R(\tau)]}{k_B T}\right),$$

(A.43)

$$\langle \bar{\varepsilon}_{\text{int}}(R_0, \delta R(\tau), \eta(\tau), g_{av}, n) \bar{\varepsilon}_{\text{int}}(R_0, \delta R(\tau'), \eta(\tau'), g_{av}, m) \bar{\varepsilon}_{\text{int}}(R_0, \delta R(\tau''), \eta(\tau''), g_{av}, l)$$

$$\bar{\varepsilon}_{\text{int}}(R_0, \delta R(\tau'''), \eta(\tau'''), g_{av}, -n-m-l) \rangle_T = \frac{1}{Z_T} \int D\delta R(\tau) \int D\eta(\tau) \bar{\varepsilon}_{\text{int}}(R_0, \delta R(\tau), \eta(\tau), g_{av}, n)$$

$$\bar{\varepsilon}_{\text{int}}(R_0, \delta R(\tau'), \eta(\tau'), g_{av}, m) \bar{\varepsilon}_{\text{int}}(R_0, \delta R(\tau''), \eta(\tau''), g_{av}, l) \bar{\varepsilon}_{\text{int}}(R_0, \delta R(\tau'''), \eta(\tau'''), g_{av}, -n-m-l)$$

$$\exp\left(-\frac{E_T^T[\delta\eta(\tau), \delta R(\tau)]}{k_B T}\right).$$

(A.44)

The averages given by Eqs. (A.42)-(A.44) are very difficult to perform in general, due to the fact that they depend on $g_{av}$, which in turn depends on the writhe that depends on braid geometry. However, if the braided section of the plectoneme is sufficiently long, we may expand out in powers of $g_{av} - \langle g_{av} \rangle_T = \pi(\langle Wr \rangle_T - Wr)/L_b$. All averages with a functional dependence on $g_{av}$ can be expanded out in the form

$$\langle A(g_{av})\rangle_T = \langle A(\langle g_{av}\rangle_T)\rangle_T + \langle A'(\langle g_{av}\rangle_T)g_{av}\rangle_T - \langle A'(\langle g_{av}\rangle_T)\rangle_T \langle g_{av}\rangle_T + \ldots$$
$$= \langle A(\langle g_{av}\rangle_T)\rangle_T + \frac{\pi}{L_b}\left[\langle A'(\langle g_{av}\rangle_T)Wr\rangle_T - \langle A'(\langle g_{av}\rangle_T)\rangle_T \langle Wr\rangle_T\right] + \ldots$$
(A.45)

Here $A(g_{av})$ is an arbitrary function of $g_{av}$, that is also independently of that a functional of $\delta R(\tau)$ and $\delta\eta(\tau)$; and $A'(g_{av})$ is its derivative with respect to the displayed argument. One can argue that (for more details see Ref [1]) that cumulent $\langle A'(\langle g_{av}\rangle_T)Wr\rangle_T - \langle A'(\langle g_{av}\rangle_T)\rangle_T \langle Wr\rangle_T$ for large $L_b$ scales with respect $L_b$ as a constant, in the same way as $\langle A(\langle g_{av}\rangle_T)\rangle_T$. This is provided that $L_b \gg \lambda_R, \lambda_\eta$, where $\lambda_R$ and $\lambda_\eta$ are the correlation ranges for fluctuations in both $\delta R$ and $\eta$. Thus, each order of term in the expansion, given by Eq. (A.45), corresponds effectively to corrections to a particular order in $1/L_b$. For example, in the expansion, we may write (up to order $1/L_b$)

$$\langle \bar{\varepsilon}_{int}(R_0,\delta R(\tau),\eta(\tau),g_{av},n)\bar{\varepsilon}_{int}(R_0,\delta R(\tau'),\eta(\tau'),g_{av},-n)\rangle_T \approx \frac{1}{Z_T}\int D\delta R(\tau)\int D\eta(\tau)$$

$$\bar{\varepsilon}_{int}(R_0,\delta R(\tau),\eta(\tau),\langle g_{av}\rangle_T,n)\bar{\varepsilon}_{int}(R_0,\delta R(\tau'),\eta(\tau'),\langle g_{av}\rangle_T,-n)\exp\left(-\frac{E_T^T[\delta\eta(\tau),\delta R(\tau)]}{k_B T}\right)$$

$$+\frac{\pi}{L_b Z_T}\int D\delta R(\tau)\int D\eta(\tau)\left[\frac{\partial\bar{\varepsilon}_{int}(R_0,\delta R(\tau),\eta(\tau),\langle g_{av}\rangle_T,n)}{\partial\langle g_{av}\rangle_T}\bar{\varepsilon}_{int}(R_0,\delta R(\tau'),\eta(\tau'),\langle g_{av}\rangle_T,-n)\right.$$

$$\left.+\bar{\varepsilon}_{int}(R_0,\delta R(\tau),\eta(\tau),\langle g_{av}\rangle_T,n)\frac{\partial\bar{\varepsilon}_{int}(R_0,\delta R(\tau'),\eta(\tau'),\langle g_{av}\rangle_T,-n)}{\partial\langle g_{av}\rangle_T}\right]$$

$$\left(Wr[\delta\eta(\tau),\delta R(\tau)]-\langle Wr\rangle_T\right)\exp\left(-\frac{E_T^T[\delta\eta(\tau),\delta R(\tau)]}{k_B T}\right)+O(L_b^{-2}).$$

(A.46)

This expansion is a systematic way to take account of the finite-size effects of the twist-writhe coupling that is required by the Fuller-White theorem. However, in all the averages, we will assume $L_b$ to be very large and neglect the $1/L_b$ corrections (and higher orders). Thus, we are able approximate Eqs. (A.42)-(A.44) with

$$\langle \bar{\varepsilon}_{int}(R_0, \delta R(\tau), \eta(\tau), g_{av}, n) \bar{\varepsilon}_{int}(R_0, \delta R(\tau'), \eta(\tau'), g_{av}, -n) \rangle_T$$

$$\approx \frac{1}{Z_T} \int_{-\infty}^{\infty} dr_1 \int_{-\infty}^{\infty} d\eta_1 \int_{-\infty}^{\infty} dr_2 \int_{-\infty}^{\infty} d\eta_2 \bar{\varepsilon}_{int}(R_0, r_1, \eta_1, \langle g_{av} \rangle_T, n) \bar{\varepsilon}_{int}(R_0, r_2, \eta_2, \langle g_{av} \rangle_T, -n) \int D\delta R(\tau) \int D\eta(\tau)$$

$$\delta(r_1 - \delta R(s)) \delta(r_2 - \delta R(s')) \delta(\eta_1 - \eta(s)) \delta(\eta_2 - \eta(s')) \exp\left(-\frac{E_T^T[\delta\eta(\tau), \delta R(\tau)]}{k_B T}\right)$$

$$\equiv \int_{-\infty}^{\infty} dr_1 \int_{-\infty}^{\infty} d\eta_1 \int_{-\infty}^{\infty} dr_2 \int_{-\infty}^{\infty} d\eta_2 \bar{\varepsilon}_{int}(R_0, r_1, \eta_1, \langle g_{av} \rangle_T, n) \bar{\varepsilon}_{int}(R_0, r_2, \eta_2, \langle g_{av} \rangle_T, -n) \Xi_{\eta,2}(\eta_1, \eta_2; \tau - \tau') \Xi_{R,2}(r_1, r_2; \tau - \tau'),$$

(A.47)

$$\langle \bar{\varepsilon}_{int}(R_0, \delta R(\tau), \eta(\tau), g_{av}, n) \bar{\varepsilon}_{int}(R_0, \delta R(\tau'), \eta(\tau'), g_{av}, m) \bar{\varepsilon}_{int}(R_0, \delta R(\tau''), \eta(\tau''), g_{av}, -n-m) \rangle_T$$

$$\approx \frac{1}{Z_T} \int_{-\infty}^{\infty} dr_1 \int_{-\infty}^{\infty} d\eta_1 \int_{-\infty}^{\infty} dr_2 \int_{-\infty}^{\infty} d\eta_2 \int_{-\infty}^{\infty} dr_3 \int_{-\infty}^{\infty} d\eta_3 \bar{\varepsilon}_{int}(R_0, r_1, \eta_1, \langle g_{av} \rangle_T, n) \bar{\varepsilon}_{int}(R_0, r_2, \eta_2, \langle g_{av} \rangle_T, m)$$

$$\bar{\varepsilon}_{int}(R_0, r_3, \eta_3, \langle g_{av} \rangle_T, -n-m) \int D\delta R(\tau) \int D\eta(\tau) \delta(r_1 - \delta R(s)) \delta(r_2 - \delta R(s')) \delta(r_3 - \delta R(s''))$$

$$\delta(\eta_1 - \eta(s)) \delta(\eta_2 - \eta(s')) \delta(\eta_3 - \eta(s'')) \exp\left(-\frac{E_T^T[\delta\eta(\tau), \delta R(\tau)]}{k_B T}\right)$$

$$\equiv \int_{-\infty}^{\infty} dr_1 \int_{-\infty}^{\infty} d\eta_1 \int_{-\infty}^{\infty} dr_2 \int_{-\infty}^{\infty} d\eta_2 \int_{-\infty}^{\infty} dr_3 \int_{-\infty}^{\infty} d\eta_3 \bar{\varepsilon}_{int}(R_0, r_1, \eta_1, \langle g_{av} \rangle_T, n) \bar{\varepsilon}_{int}(R_0, r_2, \eta_2, \langle g_{av} \rangle_T, m)$$

$$\bar{\varepsilon}_{int}(R_0, r_3, \eta_3, \langle g_{av} \rangle_T, -n-m) \Xi_{\eta,3}(\eta_1, \eta_2, \eta_3; \tau - \tau', \tau - \tau'') \Xi_{R,3}(r_1, r_2, r_3; \tau - \tau', \tau - \tau''),$$

(A.48)

$$\langle \bar{\varepsilon}_{int}(R_0, \delta R(\tau), \eta(\tau), g_{av}, n) \bar{\varepsilon}_{int}(R_0, \delta R(\tau'), \eta(\tau'), g_{av}, m) \bar{\varepsilon}_{int}(R_0, \delta R(\tau''), \eta(\tau''), g_{av}, l)$$

$$\bar{\varepsilon}_{int}(R_0, \delta R(\tau'''), \eta(\tau'''), g_{av}, -n-m-l) \rangle_T = \frac{1}{Z_T} \int_{-\infty}^{\infty} dr_1 \int_{-\infty}^{\infty} d\eta_1 \int_{-\infty}^{\infty} dr_2 \int_{-\infty}^{\infty} d\eta_2 \int_{-\infty}^{\infty} dr_3 \int_{-\infty}^{\infty} d\eta_3 \int_{-\infty}^{\infty} dr_4 \int_{-\infty}^{\infty} d\eta_4$$

$$\bar{\varepsilon}_{int}(R_0, r_1, \eta_1, g_{av}, n) \bar{\varepsilon}_{int}(R_0, r_2, \eta_2, g_{av}, m) \bar{\varepsilon}_{int}(R_0, r_3, \eta_3, g_{av}, l) \bar{\varepsilon}_{int}(R_0, r_4, \eta_4, g_{av}, -n-m-l)$$

$$\int D\delta R(\tau) \int D\eta(\tau) \delta(r_1 - \delta R(s)) \delta(r_2 - \delta R(s')) \delta(r_3 - \delta R(s'')) \delta(r_4 - \delta R(s''')) \delta(\eta_1 - \eta(s))$$

$$\delta(\eta_2 - \eta(s')) \delta(\eta_3 - \eta(s'')) \delta(\eta_4 - \eta(s''')) \exp\left(-\frac{E_T^T[\delta\eta(\tau), \delta R(\tau)]}{k_B T}\right)$$

$$= \int_{-\infty}^{\infty} dr_1 \int_{-\infty}^{\infty} d\eta_1 \int_{-\infty}^{\infty} dr_2 \int_{-\infty}^{\infty} d\eta_2 \int_{-\infty}^{\infty} dr_3 \int_{-\infty}^{\infty} d\eta_3 \int_{-\infty}^{\infty} dr_4 \int_{-\infty}^{\infty} d\eta_4 \bar{\varepsilon}_{int}(R_0, r_1, \eta_1, g_{av}, n) \bar{\varepsilon}_{int}(R_0, r_2, \eta_2, g_{av}, m)$$

$$\bar{\varepsilon}_{int}(R_0, r_3, \eta_3, g_{av}, l) \bar{\varepsilon}_{int}(R_0, r_4, \eta_4, g_{av}, -n-m-l) \Xi_{\eta,4}(\eta_1, \eta_2, \eta_3, \eta_4; \tau - \tau', \tau - \tau'', \tau - \tau''')$$

$$\Xi_{R,4}(r_1, r_2, r_3, r_4; \tau - \tau', \tau - \tau'', \tau - \tau''').$$

(A.49)

Now we have that assuming that $L_b \gg \lambda_R, \lambda_\eta$ (so that we can make the limits of integration infinite, with respect to $\tau$, in the energy functional, Eq. (2.62) of the main text) we have the following

$$\Xi_{\eta,2}(\eta_1,\eta_2;\tau-\tau') = \frac{1}{Z_\eta}\frac{1}{(2\pi)^2}\int_{-\infty}^{\infty}dp_1\int_{-\infty}^{\infty}dp_2\int D\eta(\tau)\exp(ip_1(\eta_1-\eta(\tau)))$$
$$\exp(ip_2(\eta_2-\eta(\tau')))\exp\left(-\int_{-\infty}^{\infty}d\tau\left(\frac{l_p}{4}\left(\frac{d\delta\eta(\tau)}{d\tau}\right)^2+\frac{\alpha_\eta}{2}\delta\eta(\tau)^2\right)\right),$$
(A.50)

$$\Xi_{\eta,3}(\eta_1,\eta_2,\eta_3;\tau-\tau',\tau-\tau'') = \frac{1}{Z_\eta}\frac{1}{(2\pi)^3}\int_{-\infty}^{\infty}dp_1\int_{-\infty}^{\infty}dp_2\int_{-\infty}^{\infty}dp_3\int D\eta(\tau)\exp(ip_1(\eta_1-\eta(\tau)))$$
$$\exp(ip_2(\eta_2-\eta(\tau')))\exp(ip_3(\eta_3-\eta(\tau'')))\exp\left(-\int_{-\infty}^{\infty}d\tau\left(\frac{l_p}{4}\left(\frac{d\delta\eta(\tau)}{d\tau}\right)^2+\frac{\alpha_\eta}{2}\delta\eta(\tau)^2\right)\right),$$
(A.51)

$$\Xi_{\eta,4}(\eta_1,\eta_2,\eta_3,\eta_4;\tau-\tau',\tau-\tau'',\tau-\tau''') = \frac{1}{Z_\eta}\frac{1}{(2\pi)^4}\int_{-\infty}^{\infty}dp_1\int_{-\infty}^{\infty}dp_2\int_{-\infty}^{\infty}dp_3\int_{-\infty}^{\infty}dp_4\int D\eta(\tau)$$
$$\exp(ip_1(\eta_1-\eta(\tau)))\exp(ip_2(\eta_2-\eta(\tau')))\exp(ip_3(\eta_3-\eta(\tau'')))\exp(ip_4(\eta_4-\eta(\tau''')))$$
$$\exp\left(-\int_{-\infty}^{\infty}d\tau\left(\frac{l_p}{4}\left(\frac{d\delta\eta(\tau)}{d\tau}\right)^2+\frac{\alpha_\eta}{2}\delta\eta(\tau)^2\right)\right),$$
(A.52)

$$\Xi_{R,2}(r_1,r_2;\tau-\tau') = \frac{1}{Z_R}\frac{1}{(2\pi)^2}\int_{-\infty}^{\infty}dk_1\int_{-\infty}^{\infty}dk_2\int D\delta R(\tau)\exp(ik_1(r_1-\delta R(\tau)))$$
$$\exp(ik_2(r_2-\delta R(\tau')))\exp\left(-\int_0^{L_b}d\tau\left(\frac{l_p}{4}\left(\frac{d^2\delta R(\tau)}{d\tau^2}\right)^2+\frac{\beta_R}{2}\left(\frac{d\delta R(\tau)}{d\tau}\right)^2+\frac{\alpha_R}{2}\delta R(\tau)^2\right)\right),$$
(A.53)

$$\Xi_{R,3}(r_1,r_2,r_3;\tau-\tau',\tau-\tau'') = \frac{1}{Z_R}\frac{1}{(2\pi)^3}\int_{-\infty}^{\infty}dk_1\int_{-\infty}^{\infty}dk_2\int_{-\infty}^{\infty}dk_3\int D\delta R(\tau)$$
$$\exp(ik_1(r_1-\delta R(\tau)))\exp(ik_2(r_2-\delta R(\tau')))\exp(ik_3(r_3-\delta R(\tau'')))$$
$$\exp\left(-\int_0^{L_b}d\tau\left(\frac{l_p}{4}\left(\frac{d^2\delta R(\tau)}{d\tau^2}\right)^2+\frac{\beta_R}{2}\left(\frac{d\delta R(\tau)}{d\tau}\right)^2+\frac{\alpha_R}{2}\delta R(\tau)^2\right)\right),$$
(A.54)

$$\Xi_{R,4}(r_1,r_2,r_3,r_4;\tau-\tau',\tau-\tau'',\tau-\tau''') = \frac{1}{Z_R}\frac{1}{(2\pi)^4}\int_{-\infty}^{\infty}dk_1\int_{-\infty}^{\infty}dk_2\int_{-\infty}^{\infty}dk_3\int_{-\infty}^{\infty}dk_4\int D\delta R(\tau)$$
$$\exp(ik_1(r_1-\delta R(\tau)))\exp(ik_2(r_2-\delta R(\tau')))\exp(ik_3(r_3-\delta R(\tau'')))\exp(ik_4(r_4-\delta R(\tau''')))$$
$$\exp\left(-\int_0^{L_b}d\tau\left(\frac{l_p}{4}\left(\frac{d^2\delta R(\tau)}{d\tau^2}\right)^2+\frac{\beta_R}{2}\left(\frac{d\delta R(\tau)}{d\tau}\right)^2+\frac{\alpha_R}{2}\delta R(\tau)^2\right)\right),$$
(A.55)

where

$$Z_\eta = \int D\eta(\tau)\exp\left(-\int_{-\infty}^{\infty}d\tau\left(\frac{l_p}{4}\left(\frac{d\delta\eta(\tau)}{d\tau}\right)^2+\frac{\alpha_\eta}{2}\delta\eta(\tau)^2\right)\right),$$
(A.56)

and

$$Z_R = \int D\delta R(\tau) \exp\left(-\int_0^{L_b} d\tau \left(\frac{l_p}{4}\left(\frac{d^2\delta R(\tau)}{d\tau^2}\right)^2 + \frac{\beta_R}{2}\left(\frac{d\delta R(\tau)}{d\tau}\right)^2 + \frac{\alpha_R}{2}\delta R(\tau)^2\right)\right). \quad (A.57)$$

Here, note that $\delta\eta(\tau) = \eta(\tau) - \eta_0$, where $\eta_0 = \langle\eta(\tau)\rangle_T$. The functional integrals in Eqs. (A.50)-(A.55) are straightforward to evaluate. By way of illustration we will consider $\Xi_{\eta,2}(\eta_1,\eta_2;\tau-\tau')$ and $\Xi_{R,2}(r_1,r_2;\tau-\tau')$. We may first write Eqs. (A.50) and (A.53) as

$$\Xi_{\eta,2}(\eta_1,\eta_2;\tau-\tau') = \frac{1}{Z_\eta}\frac{1}{(2\pi)^2}\int_{-\infty}^{\infty} dp_1 \int_{-\infty}^{\infty} dp_2 \exp\left(i\left(p_1(\eta_1-\eta_0)+p_2(\eta_2-\eta_0)\right)\right)$$

$$\int D\delta\tilde\eta(k)\exp\left(-\frac{ip_1}{4\pi}\int_{-\infty}^{\infty} dk\left(\delta\tilde\eta(k)e^{-ik\tau}+\delta\tilde\eta(-k)e^{ik\tau}\right)\right)\exp\left(-\frac{ip_2}{4\pi}\int_{-\infty}^{\infty} dk\left(\delta\tilde\eta(k)e^{-ik\tau'}+\delta\tilde\eta(-k)e^{ik\tau'}\right)\right)$$

$$\exp\left(-\frac{1}{4\pi}\int_{-\infty}^{\infty} d\tau\delta\tilde\eta(k)G_\eta(k)^{-1}\delta\tilde\eta(-k)\right),$$

(A.58)

$$\Xi_{R,2}(r_1,r_2;\tau-\tau') = \frac{1}{Z_R}\frac{1}{(2\pi)^2}\int_{-\infty}^{\infty} dk_1 \int_{-\infty}^{\infty} dk_2 \exp\left(i(k_1 r_1 + k_2 r_2)\right)$$

$$\int D\delta\tilde R(k)\exp\left(-\frac{ik_1}{4\pi}\int_{-\infty}^{\infty} dk\left(\delta\tilde R(k)e^{-ik\tau}+\delta\tilde R(-k)e^{ik\tau}\right)\right)\exp\left(-\frac{ik_2}{4\pi}\int_{-\infty}^{\infty} dk\left(\delta\tilde R(k)e^{-ik\tau'}+\delta\tilde R(-k)e^{ik\tau'}\right)\right)$$

$$\exp\left(-\frac{1}{4\pi}\int_{-\infty}^{\infty} d\tau\delta\tilde R(k)G_R(k)^{-1}\delta\tilde R(-k)\right),$$

(A.59)

where

$$G_\eta(k) = \frac{2}{l_p k^2 + 2\alpha_\eta} \quad \text{and} \quad G_R(k) = \frac{2}{l_p k^4 + 2\beta_R k^2 + 2\alpha_\eta}. \quad (A.60)$$

The functional integrals may be evaluated leading to

$$\Xi_{\eta,2}(\eta_1,\eta_2;\tau-\tau') = \frac{1}{(2\pi)^2}\int_{-\infty}^{\infty} dp_1 \int_{-\infty}^{\infty} dp_2 \exp\left(i\left(p_1(\eta_1-\eta_0)+p_2(\eta_2-\eta_0)\right)\right)$$

$$\exp\left(-\frac{d_\eta^2\left(p_1^2+p_2^2\right)}{2}\right)\exp\left(p_1 p_2 G_\eta(\tau-\tau')\right),$$

(A.61)

$$\Xi_{R,2}(\eta_1,\eta_2;\tau-\tau') = \frac{1}{(2\pi)^2}\int_{-\infty}^{\infty} dk_1 \int_{-\infty}^{\infty} dk_2 \exp\left(i(k_1 r_1 + k_2 r_2)\right)\exp\left(-\frac{d_R^2(k_1^2+k_2^2)}{2}\right) \quad \text{(A.62)}$$

$$\exp\left(k_1 k_2 G_R(\tau-\tau')\right),$$

where $d_R^2$ and $d_\eta^2$ are defined by the integrals

$$d_\eta^2 = \frac{1}{\pi}\int_{-\infty}^{\infty}\frac{dk}{l_p k^2 + 2\alpha_\eta} \quad \text{and} \quad d_R^2 = \frac{1}{\pi}\int_{-\infty}^{\infty}\frac{dk}{l_p k^4 + 2\beta_R k^2 + 2\alpha_R}, \quad \text{(A.63)}$$

and also we may define

$$\theta_R^2 = \frac{1}{\pi}\int_{-\infty}^{\infty}\frac{k^2 dk}{l_p k^4 + 2\beta_R k^2 + 2\alpha_R}. \quad \text{(A.64)}$$

Also, note that $G_\eta(\tau-\tau')$ and $G_R(\tau-\tau')$ may be expressed as

$$G_\eta(\tau-\tau') = \frac{1}{\pi}\int_{-\infty}^{\infty}\frac{\exp(ik(\tau-\tau'))dk}{l_p k^2 + 2\alpha_\eta}, \quad G_R(\tau-\tau') = \frac{1}{\pi}\int_{-\infty}^{\infty}\frac{\exp(ik(\tau-\tau'))dk}{l_p k^4 + 2\beta_R k^2 + 2\alpha_R}. \quad \text{(A.65)}$$

Similarly, we find that

$$\Xi_{\eta,3}(\eta_1,\eta_2,\eta_3;\tau-\tau',\tau-\tau'') = \frac{1}{(2\pi)^3}\int_{-\infty}^{\infty}dp_1\int_{-\infty}^{\infty}dp_2\int_{-\infty}^{\infty}dp_3 \exp\left(i(p_1(\eta_1-\eta_0)+p_2(\eta_2-\eta_0)+p_3(\eta_3-\eta_0))\right)$$

$$\exp\left(-\frac{d_\eta^2(p_1^2+p_2^2+p_3^2)}{2}\right)\exp\left(p_1 p_2 G_\eta(\tau-\tau')\right)\exp\left(p_1 p_3 G_\eta(\tau-\tau'')\right)\exp\left(p_2 p_3 G_\eta(\tau'-\tau'')\right),$$

$$\text{(A.66)}$$

$$\Xi_{\eta,4}(\eta_1,\eta_2,\eta_3,\eta_4;\tau-\tau',\tau-\tau'',\tau-\tau''') = \frac{1}{(2\pi)^4}\int_{-\infty}^{\infty}dp_1\int_{-\infty}^{\infty}dp_2\int_{-\infty}^{\infty}dp_3\int_{-\infty}^{\infty}dp_4 \exp\left(-\frac{d_\eta^2(p_1^2+p_2^2+p_3^2+p_4^2)}{2}\right)$$

$$\exp\left(i(p_1(\eta_1-\eta_0)+p_2(\eta_2-\eta_0)+p_3(\eta_3-\eta_0)+p_4(\eta_4-\eta_0))\right)\exp\left(p_1 p_2 G_\eta(\tau-\tau')\right)\exp\left(p_1 p_3 G_\eta(\tau-\tau'')\right)$$

$$\exp\left(p_1 p_4 G_\eta(\tau-\tau''')\right)\exp\left(p_2 p_3 G_\eta(\tau'-\tau'')\right)\exp\left(p_2 p_4 G_\eta(\tau'-\tau''')\right)\exp\left(p_3 p_4 G_\eta(\tau''-\tau''')\right),$$

$$\text{(A.67)}$$

$$\Xi_{R,3}(r_1,r_2,r_3;\tau-\tau',\tau-\tau'') = \frac{1}{(2\pi)^3}\int_{-\infty}^{\infty}dk_1\int_{-\infty}^{\infty}dk_2\int_{-\infty}^{\infty}dk_3 \exp\left(i(k_1 r_1+k_2 r_2+k_3 r_3)\right)$$

$$\exp\left(-\frac{d_R^2(k_1^2+k_2^2+k_3^2)}{2}\right)\exp\left(k_1 k_2 G_R(\tau-\tau')\right)\exp\left(k_1 k_3 G_R(\tau-\tau'')\right)\exp\left(k_2 k_3 G_R(\tau'-\tau'')\right),$$

$$\text{(A.68)}$$

$$\Xi_{R,4}(r_1,r_2,r_3,r_4;\tau-\tau',\tau-\tau'',\tau-\tau''') = \frac{1}{(2\pi)^4}\int_{-\infty}^{\infty}dk_1\int_{-\infty}^{\infty}dk_2\int_{-\infty}^{\infty}dk_3\int_{-\infty}^{\infty}dk_4\exp\left(i(k_1r_1+k_2r_2+k_3r_3+k_4r_4)\right)$$

$$\exp\left(-\frac{d_R^2(k_1^2+k_2^2+k_3^2+k_4^2)}{2}\right)\exp(k_1k_2G_R(\tau-\tau'))\exp(k_1k_3G_R(\tau-\tau''))\exp(k_1k_4G_R(\tau-\tau'''))$$

$$\exp(k_2k_3G_R(\tau'-\tau''))\exp(k_2k_4G_R(\tau'-\tau'''))\exp(k_3k_4G_R(\tau''-\tau''')).$$

(A.69)

The remaining integrals in Eqs. (A.61), (A.62), (A.66)-(A.69) may be evaluated using matrices and vector representations of the variables $\{p_j,\delta\eta_j,k_j,r_j\}$. In general, we may write for such integrals

$$\Xi_{\eta,j}(\delta\boldsymbol{\eta}_j;\tau_1-\tau_2,\tau_1-\tau_3\ldots\tau_1-\tau_j)=\frac{1}{(2\pi)^j}\int d\mathbf{p}_j\exp\left(i\frac{(\mathbf{p}_j^T.\delta\boldsymbol{\eta}_j+\delta\boldsymbol{\eta}_j^T.\mathbf{p}_j)}{2}\right)\exp\left(-\frac{\mathbf{p}_j^T.\mathbf{M}_j.\mathbf{p}_j}{2}\right),$$

(A.70)

$$\Xi_{R,j}(\mathbf{r}_j;\tau_1-\tau_2,\tau_1-\tau_3\ldots\tau_1-\tau_j)=\frac{1}{(2\pi)^j}\int d\mathbf{k}_j\exp\left(i\frac{(\mathbf{k}_j^T.\mathbf{r}_j+\mathbf{r}_j^T.\mathbf{k}_j)}{2}\right)\exp\left(-\frac{\mathbf{k}_j^T.\mathbf{N}_j.\mathbf{k}_j}{2}\right),$$

(A.71)

where

$$\delta\boldsymbol{\eta}_j=\begin{pmatrix}\eta_1-\eta_0\\ \ldots\\ \ldots\\ \eta_j-\eta_0\end{pmatrix},\quad \mathbf{p}_j=\begin{pmatrix}p_1\\ \ldots\\ \ldots\\ p_j\end{pmatrix},\quad \mathbf{M}_j=\begin{pmatrix}d_\eta^2 & G_\eta(\tau_1-\tau_2) & \ldots & G_\eta(\tau_1-\tau_j)\\ G_\eta(\tau_1-\tau_2) & d_\eta^2 & \ldots & \ldots\\ \ldots & \ldots & \ldots & \ldots\\ G_\eta(\tau_1-\tau_j) & \ldots & \ldots & d_\eta^2\end{pmatrix},\quad (A.72)$$

$$\mathbf{r}_j=\begin{pmatrix}r_1\\ \ldots\\ \ldots\\ r_j\end{pmatrix},\quad \mathbf{k}_j=\begin{pmatrix}k_1\\ \ldots\\ \ldots\\ k_j\end{pmatrix},\quad \mathbf{N}_j=\begin{pmatrix}d_R^2 & G_R(\tau_1-\tau_2) & \ldots & G_R(\tau_1-\tau_j)\\ G_R(\tau_1-\tau_2) & d_R^2 & \ldots & \ldots\\ \ldots & \ldots & \ldots & \ldots\\ G_R(\tau_1-\tau_j) & \ldots & \ldots & d_R^2\end{pmatrix},\quad (A.73)$$

and $\delta\boldsymbol{\eta}_j^T$, $\mathbf{p}_j^T$, $\mathbf{r}_j^T$ and $\mathbf{k}_j^T$ are the transposes, as well as the integration measures (in Eqs.(A.70) and (A.71) ) are defined as $d\mathbf{p}_j=dp_1\ldots dp_j$ and $d\mathbf{k}_j=dk_1\ldots dk_j$. The integrations are then easily performed yielding

$$\Xi_{\eta,j}(\delta\boldsymbol{\eta}_j;\tau_1-\tau_2,\tau_1-\tau_3\ldots\tau_1-\tau_j)\approx\frac{1}{(2\pi)^{j/2}}\frac{1}{\sqrt{\det[\mathbf{M}_j]}}\exp\left(-\delta\boldsymbol{\eta}_j^T.\mathbf{M}_j^{-1}.\delta\boldsymbol{\eta}_j\right),\quad (A.74)$$

$$\Xi_{R,j}(\mathbf{r}_j;\tau_1-\tau_2,\tau_1-\tau_3\ldots\tau_1-\tau_j) \approx \frac{1}{(2\pi)^{j/2}} \frac{1}{\sqrt{\det[\mathbf{N}_j]}} \exp\left(-\mathbf{r}_j^T.\mathbf{N}_j^{-1}.\mathbf{r}_j\right), \quad (A.75)$$

where $\mathbf{M}_j^{-1}$ and $\mathbf{N}_j^{-1}$ are the inverse matrices of $\mathbf{M}_j$ and $\mathbf{N}_j$, respectively, as well as $\det[\mathbf{M}_j]$ and $\det[\mathbf{N}_j]$ being their determinants. As way of illustration, one can write Eqs. (A.74) and (A.75) explicitly for the particular case of both $\Xi_{\eta,2}$ and $\Xi_{R,2}$. Here, we may write

$$\Xi_{\eta,2}(\eta_1,\eta_2;\tau-\tau') = \frac{1}{(2\pi)} \frac{1}{\sqrt{d_\eta^4 - G_\eta(\tau-\tau')^2}} \exp\left(-\frac{\left((\eta_1-\eta_0)^2 + (\eta_2-\eta_0)^2\right)d_\eta^2}{2\left(d_\eta^4 - G_\eta(\tau-\tau')^2\right)}\right)$$
$$\exp\left(-\frac{(\eta_1-\eta_0)(\eta_2-\eta_0)G_\eta(\tau-\tau')}{\left(d_\eta^4 - G_\eta(\tau-\tau')^2\right)}\right), \quad (A.76)$$

$$\Xi_{R,2}(r_1,r_2;\tau-\tau') = \frac{1}{(2\pi)} \frac{1}{\sqrt{d_R^4 - G_R(\tau-\tau')^2}} \exp\left(-\frac{\left(r_1^2 + r_2^2\right)d_R^2}{2\left(d_R^4 - G_R(\tau-\tau')^2\right)}\right)$$
$$\exp\left(-\frac{r_1 r_2 G_R(\tau-\tau')}{\left(d_R^4 - G_R(\tau-\tau')^2\right)}\right). \quad (A.77)$$

However, in what follows, we will assume that the correlation ranges due to bending fluctuations are small, so that we may approximate Eqs. (A.61), (A.62), (A.66)-(A.69) with

$$\Xi_{\eta,2}(\eta_1,\eta_2;\tau-\tau') \approx \frac{1}{(2\pi)} \frac{1}{d_\eta^2} \left(1 + \frac{G_\eta(\tau-\tau')}{d_\eta^4}(\eta_1-\eta_0)(\eta_2-\eta_0)\right) \exp\left(-\frac{(\eta_1-\eta_0)^2 + (\eta_2-\eta_0)^2}{2d_\eta^2}\right),$$
$$(A.78)$$

$$\Xi_{R,2}(r_1,r_2;\tau-\tau') \approx \frac{1}{(2\pi)} \frac{1}{d_R^2} \left(1 + \frac{G_R(\tau-\tau')}{d_R^4} r_1 r_2\right) \exp\left(-\frac{r_1^2 + r_1^2}{2d_R^2}\right), \quad (A.79)$$

$$\Xi_{\eta,3}(\eta_1,\eta_2,\eta_3;\tau-\tau',\tau-\tau'') \approx \frac{1}{(2\pi)^{3/2}} \frac{1}{d_\eta^3} \exp\left(-\frac{(\eta_1-\eta_0)^2 + (\eta_2-\eta_0)^2 + (\eta_3-\eta_0)^2}{2d_\eta^2}\right), \quad (A.80)$$

$$\Xi_{\eta,4}(\eta_1,\eta_2,\eta_3,\eta_4;\tau-\tau',\tau-\tau'',\tau-\tau''') \approx$$
$$\frac{1}{(2\pi)^2} \frac{1}{d_\eta^4} \exp\left(-\frac{(\eta_1-\eta_0)^2 + (\eta_2-\eta_0)^2 + (\eta_3-\eta_0)^2 + (\eta_4-\eta_0)^2}{2d_\eta^2}\right), \quad (A.81)$$

$$\Xi_{R,3}(r_1,r_2,r_3;\tau-\tau',\tau-\tau'') \approx \frac{1}{(2\pi)^{3/2}} \frac{1}{d_R^3} \exp\left(-\frac{r_1^2 + r_2^2 + r_3^2}{2d_R^2}\right), \quad (A.82)$$

$$\Xi_{R,4}(r_1,r_2,r_3,r_4;\tau-\tau',\tau-\tau'',\tau-\tau''') \approx \frac{1}{(2\pi)^2}\frac{1}{d_R^4}\exp\left(-\frac{r_1^2+r_2^2+r_3^2+r_4^2}{2d_R^2}\right). \tag{A.83}$$

Thus from Eqs. (A.78)-(A.83), combined with Eqs. (A.47)- (A.49), we may write

$$\begin{aligned}
&\langle \bar{\varepsilon}_{\text{int}}(R_0,\delta R(\tau),\eta(\tau),g_{av},n)\bar{\varepsilon}_{\text{int}}(R_0,\delta R(\tau'),\eta(\tau'),g_{av},-n)\rangle_T \\
&\approx \Delta_{0,0}\left(R_0,\eta_0,\langle g_{av}\rangle_T,d_R,d_\eta,n\right)\Delta_{0,0}\left(R_0,\eta_0,\langle g_{av}\rangle_T,d_R,d_\eta,-n\right) \\
&+G_\eta(\tau-\tau')\Delta_{1,0}\left(R_0,\eta_0,\langle g_{av}\rangle_T,d_R,d_\eta,n\right)\Delta_{1,0}\left(R_0,\eta_0,\langle g_{av}\rangle_T,d_R,d_\eta,-n\right) \\
&+G_R(\tau-\tau')\Delta_{0,1}\left(R_0,\eta_0,\langle g_{av}\rangle_T,d_R,d_\eta,n\right)\Delta_{0,1}\left(R_0,\eta_0,\langle g_{av}\rangle_T,d_R,d_\eta,-n\right),
\end{aligned} \tag{A.84}$$

$$\begin{aligned}
&\langle \bar{\varepsilon}_{\text{int}}(R_0,\delta R(\tau),\eta(\tau),g_{av},n)\bar{\varepsilon}_{\text{int}}(R_0,\delta R(\tau'),\eta(\tau'),g_{av},m)\bar{\varepsilon}_{\text{int}}(R_0,\delta R(\tau''),\eta(\tau''),g_{av},-n-m)\rangle_T \\
&\approx \Delta_{0,0}\left(R_0,\eta_0,\langle g_{av}\rangle_T,d_R,d_\eta,n\right)\Delta_{0,0}\left(R_0,\eta_0,\langle g_{av}\rangle_T,d_R,d_\eta,m\right)\Delta_{0,0}\left(R_0,\eta_0,\langle g_{av}\rangle_T,d_R,d_\eta,-n-m\right),
\end{aligned} \tag{A.85}$$

$$\begin{aligned}
&\langle \bar{\varepsilon}_{\text{int}}(R_0,\delta R(\tau),\eta(\tau),g_{av},n)\bar{\varepsilon}_{\text{int}}(R_0,\delta R(\tau'),\eta(\tau'),g_{av},m)\bar{\varepsilon}_{\text{int}}(R_0,\delta R(\tau''),\eta(\tau''),g_{av},l) \\
&\bar{\varepsilon}_{\text{int}}(R_0,\delta R(\tau'''),\eta(\tau'''),g_{av},-n-m-l)\rangle_T \approx \Delta_{0,0}\left(R_0,\eta_0,\langle g_{av}\rangle_T,d_R,d_\eta,n\right) \\
&\Delta_{0,0}\left(R_0,\eta_0,\langle g_{av}\rangle_T,d_R,d_\eta,m\right)\Delta_{0,0}\left(R_0,\eta_0,\langle g_{av}\rangle_T,d_R,d_\eta,l\right)\Delta_{0,0}\left(R_0,\eta_0,\langle g_{av}\rangle_T,d_R,d_\eta,-n-m-l\right),
\end{aligned} \tag{A.86}$$

where

$$\Delta_{j,k}\left(R_0,\eta_0,\langle g_{av}\rangle_T,d_R,d_\eta,n\right) = \frac{1}{2\pi d_R d_\eta}\int_{-\infty}^{\infty}d\eta\int_{-\infty}^{\infty}dr\,\frac{\eta^j}{d_\eta^{2j}}\frac{r^k}{d_R^{2k}}\exp\left(-\frac{r^2}{2d_R^2}\right)\exp\left(-\frac{\eta^2}{2d_\eta^2}\right) \tag{A.87}$$

$\bar{\varepsilon}_{\text{int}}(R_0,r,\eta_0+\eta,\langle g_{av}\rangle_T,n).$

Next, we need to consider the remaining $\tau$ integrations in Eqs. (A.38)-(A.41), were we have to evaluate the integrals

$$I_1 = (1-\delta_{n,0})\int_0^{L_b}d\tau\int_0^{L_b}d\tau'\exp\left(-\frac{n^2|\tau-\tau'|}{\lambda_c}\right), \tag{A.88}$$

$$\begin{aligned}
I_2 &= (1-\delta_{n,0})(1-\delta_{m,0})(1-\delta_{n,-m})\int_0^{L_b}d\tau\int_0^{L_b}d\tau'\int_0^{L_b}d\tau'' \\
&\exp\left(-\frac{1}{\lambda_c}\left(n(n+m)|\tau-\tau''|+m(n+m)|\tau'-\tau''|-nm|\tau-\tau'|\right)\right),
\end{aligned} \tag{A.89}$$

$$I_3 = (1-\delta_{n,0})(1-\delta_{m,0})(1-\delta_{l,0})(1-\delta_{l,-n-m}) \int_0^{L_b} d\tau \int_0^{L_b} d\tau' \int_0^{L_b} d\tau'' \int_0^{L_b} d\tau'''$$
$$\exp\left(-\frac{1}{\lambda_c}\left(n(n+m+l)|\tau-\tau'''|+m(n+m+l)|\tau'-\tau'''|+l(n+m+l)|\tau''-\tau'''|\right)\right) \quad (A.90)$$
$$\exp\left(\frac{1}{\lambda_c}\left(nm|\tau-\tau'|+nl|\tau-\tau''|+ml|\tau'-\tau''|\right)\right),$$

$$\tilde{I}_3 = (1-\delta_{n,0})(1-\delta_{m,0}) \int_0^{L_b} d\tau \int_0^{L_b} d\tau' \int_0^{L_b} d\tau'' \int_0^{L_b} d\tau''' \exp\left(-\frac{n^2|\tau-\tau'|+m^2|\tau''-\tau'''|}{\lambda_c}\right)$$
$$\exp\left(-\frac{nm(|\tau-\tau'''|+|\tau'-\tau''|-|\tau'-\tau'''|-|\tau-\tau''|)}{\lambda_c^{(0)}}\right), \quad (A.91)$$

as well as

$$\Omega_{1,\eta}\left(\frac{l_p n^2}{\lambda_c}, l_p \alpha_\eta\right) = \frac{(1-\delta_{n,0})}{l_p} \int_{-\infty}^{\infty} dx \, G_\eta(x) \exp\left(-\frac{n^2|x|}{\lambda_c}\right), \quad (A.92)$$

$$\Omega_{1,R}\left(\frac{l_p n^2}{\lambda_c}, \alpha_R l_p^3, \beta_R l_p\right) = \frac{(1-\delta_{n,0})}{l_p^3} \int_{-\infty}^{\infty} dx \, G_R(x) \exp\left(-\frac{n^2|x|}{\lambda_c}\right). \quad (A.93)$$

In Eqs. (A.92) and (A.93) we have already set the limits of integration to infinity, valid when $L_b \gg \lambda_b$. The evaluation of Eqs. (A.88)-(A.91) are straightforward, but the last two sets of integrals are rather complicated due to the number of terms and integrations, but simplify considerably when $L_b \gg \lambda_b$. For $L_b \gg \lambda_b$, Eqs. (A.88)-(A.91) evaluate to

$$I_1 = \frac{2 L_b \lambda_c (1-\delta_{n,0})}{n^2}, \quad (A.94)$$

$$I_2 = \frac{4 \lambda_c^2 L_b (1-\delta_{n,0})(1-\delta_{m,0})(1-\delta_{m,-n})(m^2+n^2+nm)}{n^2 m^2 (n+m)^2}, \quad (A.95)$$

$$I_3 = (1-\delta_{n,0})(1-\delta_{m,0})(1-\delta_{l,0})(1-\delta_{l,-n-m})\left[2 L_b \lambda_c^3 \{\bar{I}_{n,m,l} + \bar{I}_{l,m,n} + \bar{I}_{n,l,m}\}\right.$$
$$+4 L_b^2 \lambda_c^2 \left[\frac{1}{n^2 m^2}\{\delta_{l,-n}+\delta_{l,-m}\}+\frac{1}{l^2 n^2}\delta_{m,-n}\right]-8 L_b \lambda_c^3 \left[\frac{(n^2+m^2)\{\delta_{l,-m}+\delta_{l,-n}\}}{m^4 n^4}+\frac{(l^2+n^2)\delta_{m,-n}}{n^4 l^4}\right]\right],$$
$$(A.96)$$

where
$$\bar{I}_{n,m,l} = \left(\frac{1}{(n+m+l)^2}+\frac{1}{l^2}\right)\left(\frac{1}{n^2}+\frac{1}{m^2}\right)\frac{1}{(n+m)^2}(1-\delta_{m,-n}), \quad (A.97)$$

and

$$\tilde{I}^3 = (1-\delta_{n,0})(1-\delta_{m,0})\left\{L_b\lambda_c^3\left[\frac{16\lambda_c^2(m^2+n^2)}{n^2m^2\left(\left(n^2+m^2\right)^2(\lambda_c^{(0)})^2 - 4n^2m^2\lambda_c^2\right)} - \frac{4(m^2+n^2)}{n^4m^4}\right] + \frac{4\lambda_c^2 L_b^2}{n^2m^2}\right\}.$$

(A.98)

Eqs. (A.92) and (A.93) evaluate to (see Ref. [1] for details of the calculation)

$$\Omega_{1,\eta}\left(\frac{l_p n^2}{\lambda_c}, \alpha_\eta l_p\right) = 2\left(\frac{1}{2\alpha_\eta l_p}\right)^{1/2} \frac{1-\delta_{n,0}}{\left(\frac{l_p n^2}{\lambda_c}\right) + \left(2\alpha_\eta l_p\right)^{1/2}},$$

(A.99)

$$\Omega_{1,R}\left(\frac{n^2 l_p}{\lambda_c}, \alpha_R l_p^3, \beta_R l_p\right) \equiv (1-\delta_{n,0})\frac{4n^2 l_p}{\lambda_c}\left(2\alpha_R l_p^3\right)^{-5/4} \tilde{I}_1\left(2\beta_R l_p\left(\frac{1}{2\alpha_R l_p^3}\right)^{1/2}, \left(\frac{n^2 l_p}{\lambda_c}\right)^2\left(\frac{1}{2\alpha_R l_p^3}\right)^{1/2}\right),$$

(A.100)

Where

$$\tilde{I}_1(\gamma,\delta) = \frac{1}{\delta^2 - \gamma\delta + 1}\left(\frac{1}{2\sqrt{\delta}} - \frac{1}{4}\sqrt{2+\gamma} + \frac{(\delta-\gamma/2)}{2\sqrt{2+\gamma}}\right).$$

(A.101)

This allows us to write (from Eqs. (A.38), (A.84), (A.92)-(A.94))

$$-\frac{\langle\langle E_2\rangle_T\rangle_g}{2} = \frac{F_{cor}^{(1)}}{k_B T} \approx (F_{cor,1}^{(1)} + F_{cor,2}^{(1)} + F_{cor,3}^{(1)}),$$

(A.102)

where

$$F_{cor,1}^{(1)} = \frac{L_b \lambda_c}{(k_B T)^2}\sum_{n=-\infty}^{\infty}\frac{(1-\delta_{n,0})}{n^2}\Delta_{0,0}\left(R_0,\eta_0,\langle g_{av}\rangle_T,d_R,d_\eta,n\right)\Delta_{0,0}\left(R_0,\eta_0,\langle g_{av}\rangle_T,d_R,d_\eta,-n\right),$$

(A.103)

$$F_{cor,2}^{(1)} = \frac{l_p L_b}{2(k_B T)^2}\sum_{n=-\infty}^{\infty}\Delta_{1,0}\left(R_0,\eta_0,\langle g_{av}\rangle_0,d_R,d_\eta,n\right)^2 \Omega_{1,\eta}\left(\frac{l_p n^2}{\lambda_c}, l_p \alpha_\eta\right),$$

(A.104)

$$F_{cor,3}^{(1)} = \frac{l_p^3 L_b}{2(k_B T)^2}\sum_{n=-\infty}^{\infty}\Delta_{0,1}\left(R_0,\eta_0,\langle g_{av}\rangle_T,d_R,d_\eta,n\right)\Delta_{0,1}\left(R_0,\eta_0,\langle g_{av}\rangle_T,d_R,d_\eta,-n\right)\Omega_{1,R}\left(\frac{n^2 l_p}{\lambda_c}, \alpha_R l_p^3, \beta_R l_p\right).$$

(A.105)

Eqs. (A.102)-(A.105) are presented in the main text (Eqs. (2.72)-(2.75)). For the other terms in the expansion (Eq. (A.1)) we may write (from Eqs. (A.39)-(A.41), (A.85), (A.86) and (A.95)-(A.98))

$$\frac{F_{cor}^{(2)}}{k_B T} = \frac{\langle\langle E_3 \rangle_T \rangle_g}{6} = \frac{2\lambda_c^2 L_b}{3(k_B T)^3} \sum_{n=-\infty, m=-\infty,}^{\infty} \frac{(m^2 + n^2 + nm)}{n^2 m^2 (n+m)^2} (1-\delta_{n,-m})(1-\delta_{n,0})(1-\delta_{m,0})$$
$$\Delta_{0,0}(R_0, \eta_0, \langle g_{av} \rangle_T, d_R, d_\eta, n) \Delta_{0,0}(R_0, \eta_0, \langle g_{av} \rangle_T, d_R, d_\eta, m) \Delta_{0,0}(R_0, \eta_0, \langle g_{av} \rangle_T, d_R, d_\eta, -n-m),$$

(A.106)

$$\frac{F_{cor}^{(3)}}{k_B T} = -\frac{1}{8}\left\langle\left\langle \frac{E_4}{3} - E_2^2 \right\rangle_T \right\rangle_g = -\frac{L_b \lambda_c^3}{4(k_B T)^4} \sum_{n,m,l=-\infty}^{\infty} (1-\delta_{l,0})(1-\delta_{m,0})(1-\delta_{n,0})(1-\delta_{n+m+l,0})\{\overline{I}_{m,l,n}(1-\delta_{l,-m})$$
$$-\frac{(m^2+n^2)\delta_{l,-m}}{n^2 m^2}\left[\frac{2}{n^2 m^2} + \frac{8\lambda_c^2}{\left((n^2+m^2)^2 (\lambda_c^{(0)})^2 - 4n^2 m^2 \lambda_c^2\right)}\right]\Delta_{0,0}(R_0, \eta_0, \langle g_{av} \rangle_T, d_R, d_\eta, n)$$
$$\Delta_{0,0}(R_0, \eta_0, \langle g_{av} \rangle_T, d_R, d_\eta, m) \Delta_{0,0}(R_0, \eta_0, \langle g_{av} \rangle_T, d_R, d_\eta, l) \Delta_{0,0}(R_0, \eta_0, \langle g_{av} \rangle_T, d_R, d_\eta, -n-m-l).$$

(A.107)

Also, Eqs.(A.106) and (A.107) are also presented as Eqs. (2.79) and (2.80) of the main text.

## B. Evaluation of other terms in the Free energy

Next in the evaluation of Eq. (A.1), we need to consider
$\langle E_{T,0}[\delta\eta(\tau), \delta R(\tau)] - E_T^T[\delta\eta(\tau), \delta R(\tau)]\rangle_T$. This can be written as (using Eq. (2.54) of the main text)

$$\langle E_{T,0}[\delta\eta(\tau), \delta R(\tau)] - E_T^T[\delta\eta(\tau), \delta R(\tau)]\rangle_T \simeq \langle \tilde{E}_{st}[\delta\eta(\tau), \delta R(\tau)] + \tilde{E}_B[\delta\eta(\tau), \delta R(\tau)] - E_T^T[\delta\eta(\tau), \delta R(\tau)]\rangle_T$$
$$+ \int_0^{L_b} d\tau \langle \overline{\varepsilon}_{int}(R_0, \delta R(\tau), \eta_0 + \delta\eta(\tau), \langle g_{av} \rangle_T, 0)\rangle_T + l_{tw} L_b \left(\langle g_{av} \rangle_T - \frac{2\pi}{H}\right)^2.$$

(B.1)

Here, $\tilde{E}_{st}[\delta\eta(\tau), \delta R(\tau)]$ and $\tilde{E}_B[\delta\eta(\tau), \delta R(\tau)]$ are the steric and bending energy contributions discussed in the main text. The second to last term is part of the interaction energy that does not depend on $\Delta\xi(\tau)$, and the last term is the increase in twisting elastic energy from changing $g_{av}$ away from $2\pi/H$ averaged over bending fluctuations. In writing these two last terms in Eq. (B.1) we have already approximated

$$\left\langle \left(g_{av} - \frac{2\pi}{H}\right)\right\rangle_T \approx \left(\langle g_{av} \rangle_T - \frac{2\pi}{H}\right)^2,$$

(B.2)

$$\langle \bar{\varepsilon}_{\text{int}}(R_0, \delta R(\tau), \eta_0 + \delta\eta(\tau), g_{av}, 0)\rangle_T \approx \langle \bar{\varepsilon}_{\text{int}}(R_0, \delta R(\tau), \eta_0 + \delta\eta(\tau), \langle g_{av}\rangle_T, 0)\rangle_T, \quad (B.3)$$

which should be valid for large $L_b$ (see the previous Appendix and Ref. [1] for arguments). Now, let us examine the terms contained in Eq. (B.1). First, we may express

$$\begin{aligned}
&\langle \bar{\varepsilon}_{\text{int}}(R_0, \bar{\varepsilon}_{\text{int}}(R_0, \delta R(\tau), \eta_0 + \delta\eta(\tau), \langle g_{av}\rangle_T, 0)\rangle_T \\
&= \int_{-\infty}^{\infty} dr \int_{-\infty}^{\infty} d\eta \, \bar{\varepsilon}_{\text{int}}(R_0, r, \eta_0 + \eta, \langle g_{av}\rangle_T, 0) \langle \delta(\delta R(\tau) - r)\delta(\delta\eta(\tau) - \eta)\rangle_T \\
&= \int_{-\infty}^{\infty} dr \int_{-\infty}^{\infty} d\eta \, \bar{\varepsilon}_{\text{int}}(R_0, r, \eta_0 + \eta, \langle g_{av}\rangle_T, 0) \Xi_{\eta,1}(\eta) \Xi_{R,1}(r),
\end{aligned} \quad (B.4)$$

where

$$\begin{aligned}
\Xi_{\eta,1}(\eta) &= \frac{1}{Z_\eta} \frac{1}{(2\pi)} \int_{-\infty}^{\infty} dp \int D\delta\eta(\tau) \exp(ip(\eta - \delta\eta(\tau))) \\
&\quad \exp\left(-\int_{-\infty}^{\infty} d\tau \left(\frac{l_p}{4}\left(\frac{d\delta\eta(\tau)}{d\tau}\right)^2 + \frac{\alpha_\eta}{2}\delta\eta(\tau)^2\right)\right),
\end{aligned} \quad (B.5)$$

and

$$\begin{aligned}
\Xi_{R,1}(r) &= \frac{1}{Z_R} \frac{1}{(2\pi)} \int_{-\infty}^{\infty} dk \int D\delta R(\tau) \exp(ik(r - \delta R(\tau))) \\
&\quad \exp\left(-\int_{0}^{L_b} d\tau \left(\frac{l_p}{4}\left(\frac{d^2\delta R(\tau)}{d\tau^2}\right)^2 + \frac{\beta_R}{2}\left(\frac{d\delta R(\tau)}{d\tau}\right)^2 + \frac{\alpha_R}{2}\delta R(\tau)^2\right)\right).
\end{aligned} \quad (B.6)$$

Here note that $Z_\eta$ and $Z_R$ are given by Eqs. (A.56) and (A.57). The functional integrations are easily performed yielding

$$\Xi_{\eta,1}(\eta) = \frac{1}{(2\pi)} \int_{-\infty}^{\infty} dp \exp(ip\eta) \exp\left(-\frac{d_\eta^2 p^2}{2}\right) = \frac{1}{d_\eta \sqrt{2\pi}} \exp\left(-\frac{\eta^2}{2d_\eta^2}\right), \quad (B.7)$$

$$\Xi_{R,1}(r) = \frac{1}{(2\pi)} \int_{-\infty}^{\infty} dk \exp(ikr) \exp\left(-\frac{d_R^2 k^2}{2}\right) = \frac{1}{d_R \sqrt{2\pi}} \exp\left(-\frac{r^2}{2d_R^2}\right). \quad (B.8)$$

Thus, we can write the $\Delta\xi(\tau)$ independent contribution from the interaction energy, to the free energy, as

$$F_{un} = \int_0^{L_b} d\tau \left\langle \bar{\varepsilon}_{int}(R_0, \delta R(\tau), \eta_0 + \delta\eta(\tau), \langle g_{av}\rangle_T, 0)\right\rangle_T$$

$$= \frac{L_b}{2\pi d_R d_\eta} \int_{-\infty}^{\infty} dr \int_{-\infty}^{\infty} d\eta \bar{\varepsilon}_{int}\left(R_0, r, \eta_0 + \eta, \langle g_{av}\rangle_0, 0\right) \exp\left(-\frac{r^2}{2d_R^2}\right) \exp\left(-\frac{\eta^2}{2d_\eta^2}\right).$$ (B.9)

This is Eq. (2.71) of the main text. Now let us deal with $\left\langle \tilde{E}_{st}[\delta\eta(\tau), \delta R(\tau)] + \tilde{E}_B[\delta\eta(\tau), \delta R(\tau)] - E_T^T[\delta\eta(\tau), \delta R(\tau)]\right\rangle_T$. This can be written using Eqs. (2.45), (2.47) and (2.62) of the main text as

$$\frac{\left\langle \tilde{E}_{st}[\delta\eta(\tau), \delta R(\tau)] + \tilde{E}_B[\delta\eta(\tau), \delta R(\tau)] - E_T^T[\delta\eta(\tau), \delta R(\tau)]\right\rangle_T}{k_B T} =$$

$$\int_0^{L_b} d\tau \frac{\alpha_H - \alpha_R}{2}\left\langle \delta R(\tau)^2\right\rangle_T - \int_0^{L_b} d\tau \frac{\beta_R}{2}\left\langle \left(\frac{d\delta R(\tau)}{d\tau}\right)^2\right\rangle_T - \int_0^{L_b} d\tau \frac{\alpha_\eta}{2}\left\langle \delta\eta(\tau)^2\right\rangle_T$$

$$+ \int_0^{L_b} d\tau \left[\left\langle \mathcal{E}_R(0, R'(\tau), R(\tau), 0, \delta\eta(\tau))\theta(\delta R(\tau) - d_{min})\theta(d_{max} - \delta R(\tau))\right\rangle_T\right.$$ (B.10)

$$+ \left\langle \mathcal{E}_R(0, R'(\tau), R_0 + d_{max}, 0, \delta\eta(\tau))\theta(\delta R(\tau) - d_{max})\right\rangle_T$$

$$+ \left.\left\langle \mathcal{E}_R(0, R'(\tau), R_0 + d_{min}, 0, \delta\eta(\tau))\theta(d_{min} - \delta R(\tau))\right\rangle_T\right].$$

In writing Eq. (B.10), we have anticipated that the term linear in $\delta\eta'(\tau)R'(\tau)$ (contained within Eq. (2.42)) averages to zero.

Let us consider the first two terms of Eq. (B.10) using the definitions for $d_\eta^2$, $d_R^2$ and $\theta_R^2$ (Eqs. (A.63) and (A.64)) we can write

$$\frac{1}{L_b}\int_0^{L_b} d\tau \frac{\alpha_H - \alpha_R}{2}\left\langle \delta R(\tau)^2\right\rangle_T - \frac{1}{L_b}\int_0^{L_b} d\tau \frac{\beta_R}{2}\left\langle \left(\frac{d\delta R(\tau)}{d\tau}\right)^2\right\rangle_T - \frac{1}{L_b}\int_0^{L_b} d\tau \frac{\alpha_\eta}{2}\left\langle \delta\eta(\tau)^2\right\rangle_T$$

$$= \frac{\alpha_H - \alpha_R}{2} d_R^2 - \frac{\beta_R}{2}\theta_R^2 - \frac{\alpha_\eta}{2} d_\eta^2.$$ (B.11)

Also, evaluating the integrals in Eqs. (A.63) and (A.64) yields the following interrelations between the parameters

$$d_\eta^2 = \frac{1}{(2l_p\alpha_\eta)^{1/2}}, \qquad \alpha_R = \frac{l_p}{2}\left(\frac{\theta_R}{d_R}\right)^4 \quad \text{and} \quad \beta_R = \frac{l_p}{2}\left(\frac{1}{l_p^2\theta_R^4} - 2\left(\frac{\theta_R}{d_R}\right)^2\right). \qquad (B.12)$$

The expressions in Eq. (B.12) can be used to eliminate both $\alpha_R$ and $\beta_R$ in Eq. (B.11) in favour of $d_R$ and $\theta_R$. Next, in Eq. (B.10) let us consider the bending energy (last) term. First of all, we can express

$$\int_0^{L_b} d\tau \Big[ \langle \mathcal{E}_R(0, R'(\tau), R(\tau), 0, \delta\eta(\tau))\theta(\delta R(\tau) - d_{\min})\theta(d_{\max} - \delta R(\tau)) \rangle_T$$
$$+ \langle \mathcal{E}_R(0, R'(\tau), R_0 + d_{\max}, 0, \delta\eta(\tau))\theta(\delta R(\tau) - d_{\max}) \rangle_T$$
$$+ \langle \mathcal{E}_R(0, R'(\tau), R_0 + d_{\min}, 0, \delta\eta(\tau))\theta(d_{\min} - \delta R(\tau)) \rangle_T \Big]$$
$$= \int_0^{L_b} d\tau \int_{-\infty}^{\infty} dr \int_{-\infty}^{\infty} dr' \int_{-\infty}^{\infty} d\eta \Big[ \mathcal{E}_R(0, r', R_0 + r, 0, \eta)\theta(r - d_{\min})\theta(d_{\max} - r) + \mathcal{E}_R(0, r', R_0 + d_{\max}, 0, \eta)\theta(r - d_{\max})$$
$$+ \mathcal{E}_R(0, r', R_0 + d_{\min}, 0, \eta)\theta(d_{\min} - r) \Big] \langle \delta(\delta R(\tau) - r)\delta(\delta R'(\tau) - r')\delta(\delta\eta(\tau) - \eta) \rangle_T$$
$$= \int_0^{L_b} d\tau \int_{-\infty}^{\infty} dr \int_{-\infty}^{\infty} dr' \int_{-\infty}^{\infty} d\eta \Big[ \mathcal{E}_R(0, r', R_0 + r, 0, \eta)\theta(r - d_{\min})\theta(d_{\max} - r) + \mathcal{E}_R(0, r', R_0 + d_{\max}, 0, \eta)\theta(r - d_{\max})$$
$$+ \mathcal{E}_R(0, r', R_0 + d_{\min}, 0, \eta)\theta(d_{\min} - r) \Big] \tilde{\Xi}_{R,1}(r, r')\Xi_{\eta,1}(\eta),$$

(B.13)

where

$$\tilde{\Xi}_{R,1}(r, r') = \frac{1}{Z_R} \frac{1}{(2\pi)^2} \int_{-\infty}^{\infty} dp \int_{-\infty}^{\infty} dp' \int D\delta R(\tau) \exp(ip(r - \delta R(\tau))) \exp(ip'(r' - \delta R'(\tau)))$$
$$\exp\left( -\int_0^{L_b} d\tau \left( \frac{l_p}{4}\left(\frac{d^2\delta R(\tau)}{d\tau^2}\right)^2 + \frac{\beta_R}{2}\left(\frac{d\delta R(\tau)}{d\tau}\right)^2 + \frac{\alpha_R}{2}\delta R(\tau)^2 \right) \right)$$
$$= \frac{1}{Z_R} \frac{1}{(2\pi)^2} \int_{-\infty}^{\infty} dp \int_{-\infty}^{\infty} dp' \exp(i(pr + p'r'))$$
$$\int D\delta\tilde{R}(k)\exp\left(-\frac{ip}{4\pi}\int_{-\infty}^{\infty} dk\left(\delta\tilde{R}(k)e^{-ik\tau} + \delta\tilde{R}(-k)e^{ik\tau}\right)\right)\exp\left(-\frac{ip'}{4\pi}\int_{-\infty}^{\infty} dk\left(-ik\delta\tilde{R}(k)e^{-ik\tau} + ik\delta\tilde{R}(-k)e^{ik\tau}\right)\right)$$
$$\exp\left(-\frac{1}{4\pi}\int_{-\infty}^{\infty} d\tau \delta\tilde{R}(k)G_R(k)^{-1}\delta\tilde{R}(-k)\right).$$

(B.14)

The functional integral in Eq. (B.14) can be performed leaving

$$\tilde{\Xi}_{R,1}(r,r') = \frac{1}{(2\pi)^2} \int_{-\infty}^{\infty} dp \int_{-\infty}^{\infty} dp' \exp(irp + ir'p') \exp\left(-\frac{d_R^2 p^2}{2}\right) \exp\left(-\frac{\theta_R^2 p'^2}{2}\right)$$
$$= \frac{1}{(2\pi)d_R \theta_R} \exp\left(-\frac{r^2}{2d_R^2}\right) \exp\left(-\frac{r'^2}{2\theta_R^2}\right).$$
(B.15)

Indeed, any term odd in $r'$ in Eq. (B.13) will integrate to zero. Using Eqs.(B.7), (B.13) and (B.15) (as well as the form for $\mathcal{E}_R$ given in the main text) allows us to write

$$\frac{1}{L_b} \int_0^{L_b} d\tau \Big[ \langle \mathcal{E}_R(0, R'(\tau), R(\tau), 0, \delta\eta(\tau)) \theta(\delta R(\tau) - d_{\min}) \theta(d_{\max} - \delta R(\tau)) \rangle_T$$
$$+ \langle \mathcal{E}_R(0, R'(\tau), R_0 + d_{\max}, 0, \delta\eta(\tau)) \theta(\delta R(\tau) - d_{\max}) \rangle_T$$
$$+ \langle \mathcal{E}_R(0, R'(\tau), R_0 + d_{\min}, 0, \delta\eta(\tau)) \theta(d_{\min} - \delta R(\tau)) \rangle_T \Big] =$$
$$\frac{\tilde{f}_1(R_0, d_R, d_{\max}, d_{\min})}{R_0^2} \left[ 4l_p \sin^4\left(\frac{\eta_0}{2}\right) - l_p \theta_R^2 \sin^2\left(\frac{\eta_0}{2}\right) \right.$$
$$\left. + 2 \left(\frac{l_p}{2\alpha_\eta}\right)^{1/2} \left( 3\cos^2\left(\frac{\eta_0}{2}\right) \sin^2\left(\frac{\eta_0}{2}\right) - \sin^4\left(\frac{\eta_0}{2}\right) \right) \right],$$
(B.16)

where

$$\tilde{f}_1(R_0, d_R, d_{\max}, d_{\min}) = \frac{R_0^2}{d_R \sqrt{2\pi}} \int_{d_{\min}}^{d_{\max}} \frac{dx}{(R_0 + x)^2} \exp\left(-\frac{x^2}{2d_R^2}\right)$$
$$+ \frac{1}{2} \left( \frac{R_0^2}{(R_0 + d_{\min})^2} \left(1 - \text{erf}\left(-\frac{d_{\min}}{d_R \sqrt{2}}\right)\right) + \frac{R_0^2}{(R_0 + d_{\max})^2} \left(1 - \text{erf}\left(\frac{d_{\max}}{d_R \sqrt{2}}\right)\right) \right).$$
(B.17)

This is Eq. (2.70) of the main text where we have also made the choice $d_{\max} = -d_{\min} = R_0 - 2a$.

Now, let's consider $\ln Z_T$ (in Eq. (A.1)). First of all, we can write

$$-\left(\frac{\partial \ln Z_T}{\partial \alpha_R}\right)_{\beta_R} = \frac{L_b}{2} \langle \delta R(\tau)^2 \rangle_0 = \frac{d_R^2 L_b}{2}, \quad -\left(\frac{\partial \ln Z_T^T}{\partial \beta_R}\right)_{\alpha_R} = \frac{L_b}{2} \left\langle \left(\frac{d\delta R(\tau)}{d\tau}\right)^2 \right\rangle_0 = \frac{\theta_R^2 L_b}{2},$$

$$-\frac{\partial \ln Z_T^T}{\partial \alpha_\eta} = \frac{L_b}{2} \langle \delta\eta(\tau)^2 \rangle_0 = \frac{L_b}{2^{3/2} \alpha_\eta^{1/2} l_b^{1/2}}.$$
(B.18)

Utilizing the relationships between the variables $\alpha_R$, $\beta_R$ and $\theta_R$, $d_R$ (Eq. (B.12)) and solving the differential equations in Eq. (B.18) (see Ref. [1] for further details) yields (up to a constant of integration that can be chosen to be zero)

$$-\ln Z_T = \frac{\alpha_\eta^{1/2} L_b}{2^{1/2} l_p^{1/2}} + \frac{L_b}{2 l_p \theta_R^2} + \frac{L_b}{2\lambda_h}.$$
(B.19)

Eq. (A.1), (B.1),(B.9), (B.11), (B.12), (B.16) and (B.19) combined allow us to write the total free energy

$$\langle F_T \rangle_g = F_{els} + F_{un} + F_{cor}^{(1)} + F_{cor}^{(2)} + F_{cor}^{(3)}, \tag{B.20}$$

where

$$\frac{F_{els}}{k_B T L_b} = \frac{1}{2}\left(\frac{\alpha_\eta}{2l_p}\right)^{1/2} + \frac{1}{4l_p \theta_R^2} + \frac{l_p \theta_R^4}{d_R^2} + \frac{\alpha_H d_R^2}{2} + \frac{\tilde{f}_1(R_0, d_R, d_{\max}, d_{\min})}{R_0^2}\left[4l_p \sin^4\left(\frac{\eta_0}{2}\right) - l_p \theta_R^2 \sin^2\left(\frac{\eta_0}{2}\right)\right.$$

$$\left. +2\left(\frac{l_p}{2\alpha_\eta}\right)^{1/2}\left(3\cos^2\left(\frac{\eta_0}{2}\right)\sin^2\left(\frac{\eta_0}{2}\right) - \sin^4\left(\frac{\eta_0}{2}\right)\right)\right] + l_{tw} L_b \left(\langle g_{av} \rangle_T - \frac{2\pi}{H}\right)^2$$

(B.21)

In Eq. (B.21), $F_{un}$, $F_{cor}^{(1)}$, $F_{cor}^{(2)}$, and $F_{cor}^{(3)}$ are given by Eqs.(B.9) and (A.102)-(A.107). All of these results are presented in the main text as Eqs. (2.68)-(2.81).

## C. Computing the average writhe

One of the last analytical tasks to perform is to approximate the average writhe. This has already been discussed in Ref. [1]. Here, we will repeat the analysis somewhat, but will make a further simplifying approximation. Following Ref. [1], we may start by writing

$$4\pi Wr \approx Wr_{1,1} + Wr_{2,2} - 2Wr_{1,2}, \tag{C.1}$$

where

$$Wr_{1,1} = \int_0^{L_b} d\tau \int_0^{L_b} d\tau' \frac{(\mathbf{r}_1(\tau) - \mathbf{r}_1(\tau')) \cdot \hat{\mathbf{t}}_1(\tau) \times \hat{\mathbf{t}}_1(\tau')}{|\mathbf{r}_1(\tau) - \mathbf{r}_1(\tau')|^{3/2}}, \tag{C.2}$$

$$Wr_{2,2} = \int_0^{L_b} d\tau \int_0^{L_b} d\tau' \frac{(\mathbf{r}_2(\tau) - \mathbf{r}_2(\tau')) \cdot \hat{\mathbf{t}}_2(\tau) \times \hat{\mathbf{t}}_2(\tau')}{|\mathbf{r}_2(\tau) - \mathbf{r}_2(\tau')|^{3/2}}, \tag{C.3}$$

$$Wr_{1,2} = \int_0^{L_b} d\tau \int_0^{L_b} d\tau' \frac{(\mathbf{r}_1(\tau) - \mathbf{r}_2(\tau')) \cdot \hat{\mathbf{t}}_1(\tau) \times \hat{\mathbf{t}}_2(\tau')}{|\mathbf{r}_1(\tau) - \mathbf{r}_2(\tau')|^{3/2}}. \tag{C.4}$$

Through substitution of Eqs. (2.19), (2.20), (2.21) and (2.22) of the main text into Eqs. (C.1)-(C.4) we can write

$$Wr_1 \equiv Wr_{1,1} = Wr_{2,2} = \int_0^{L_b} d\tau \int_0^{L_b} d\tau' \frac{\omega_{1,1}(\tau,\tau')}{\left(\frac{R(\tau)^2}{4} + \frac{R(\tau')^2}{4} - \frac{R(\tau)R(\tau')}{2}\cos(\theta(\tau)-\theta(\tau')) + (Z(\tau)-Z(\tau'))^2\right)^{3/2}},$$

(C.5)

$$Wr_2 \equiv Wr_{1,2} = \int_0^{L_b} d\tau \int_0^{L_b} d\tau' \frac{\omega_{1,2}(\tau,\tau')}{\left(\frac{R(\tau)^2}{4} + \frac{R(\tau')^2}{4} + \frac{R(\tau)R(\tau')}{2}\cos(\theta(\tau)-\theta(\tau')) + (Z(\tau)-Z(\tau'))^2\right)^{3/2}},$$

(C.6)

and

$$\begin{aligned}\omega_{1,j}(\tau,\tau') = &-\frac{R(\tau)}{2}\cos\left(\frac{\eta(\tau')}{2}\right)\sqrt{\sin^2\left(\frac{\eta(\tau)}{2}\right) - \frac{1}{4}\left(\frac{dR(\tau)}{d\tau}\right)^2} \\ &-\frac{R(\tau')}{2}\cos\left(\frac{\eta(\tau)}{2}\right)\sqrt{\sin^2\left(\frac{\eta(\tau')}{2}\right) - \frac{1}{4}\left(\frac{dR(\tau')}{d\tau'}\right)^2} \\ &-(-1)^j \sin\left(\frac{\eta(\tau)}{2}\right)\cos\left(\frac{\eta(\tau')}{2}\right)\frac{R(\tau')}{2}\cos(\gamma(\tau)+\theta(\tau)-\theta(\tau')) \\ &-(-1)^j \sin\left(\frac{\eta(\tau')}{2}\right)\cos\left(\frac{\eta(\tau)}{2}\right)\frac{R(\tau)}{2}\cos(\gamma(\tau')+\theta(\tau')-\theta(\tau)) \\ &+(-1)^j (Z(\tau)-Z(\tau'))\sin\left(\frac{\eta(\tau)}{2}\right)\sin\left(\frac{\eta(\tau')}{2}\right)\sin(\gamma(\tau)-\gamma(\tau')+\theta(\tau)-\theta(\tau')).\end{aligned}$$

(C.7)

To compute the averages $\langle Wr_j \rangle_T$, which we require in our evaluation of the free energy, we make the approximations that

$$\theta(\tau) \approx \theta_0 - \langle Q \rangle_T \tau, \qquad Z(\tau) = \langle \cos(\eta(\tau)/2) \rangle_T \tau. \tag{C.8}$$

As discussed in Ref. [1], these should be valid when the pitch of the braided section is larger than the correlation lengths of $\delta R(\tau)$ and $\delta\eta(\tau)$ fluctuations, which should be valid for a sufficiently tightly supercoiled braid. These key approximations (Eq. (C.8)) allow us to write

$$\langle Wr_j\rangle_0 \approx \int_{-\infty}^{\infty} d\eta_1 \int_{-\infty}^{\infty} d\eta_2 \int_{-\infty}^{\infty} dr'_1 \int_{-\infty}^{\infty} dr'_2 \int_{-\infty}^{\infty} dr_1 \int_{-\infty}^{\infty} dr_2 \int_0^{L_b} d\tau \int_0^{L_b} d\tau' \Xi_{\eta,2}(\eta_1,\eta_2;\tau-\tau')\tilde{\Xi}_{R,2}(r_1,r_2,r'_1,r'_2;\tau-\tau')$$

$$\tilde{\omega}_j(\tau-\tau'; R_0+r_1, R_0+r_2, r'_1, r'_2, \eta_0+\eta_1, \eta_0+\eta_2) \left( \frac{R_0^2}{2}\left(1+(-1)^j \cos(\langle Q\rangle_T (\tau-\tau'))\right) + \frac{r_1^2}{4} + \frac{r_2^2}{4} \right.$$

$$\left. + \frac{(r_1+r_2)R_0}{2}\left(1+(-1)^j \cos(\langle Q\rangle_T (\tau-\tau'))\right) + (-1)^j \frac{r_1 r_2}{2}\cos(\langle Q\rangle_T (\tau-\tau')) + \left\langle \cos\left(\frac{\eta(\tau)}{2}\right)\right\rangle_T^2 (\tau-\tau')^2 \right)^{-3/2},$$

(C.9)

where

$$\tilde{\omega}_j(\tau-\tau'; r_0+r_1, r_0+r_2, r'_1, r'_2, \eta_0+\eta_1, \eta_0+\eta_2) =$$

$$-\frac{r_0+r_1}{2}\cos\left(\frac{\eta_0+\eta_2}{2}\right)\sqrt{\sin^2\left(\frac{\eta_0+\eta_1}{2}\right) - \frac{r'^2_1}{4}} - \frac{r_0+r_2}{2}\cos\left(\frac{\eta_0+\eta_1}{2}\right)\sqrt{\sin^2\left(\frac{\eta_0+\eta_2}{2}\right) - \frac{r'^2_2}{4}}$$

$$-(-1)^j \sin\left(\frac{\eta_0+\eta_1}{2}\right)\cos\left(\frac{\eta_0+\eta_2}{2}\right)\frac{r_0+r_2}{2}\cos(\gamma(\tau) - \langle Q\rangle_T(\tau-\tau'))$$ (C.10)

$$-(-1)^j \sin\left(\frac{\eta_0+\eta_2}{2}\right)\cos\left(\frac{\eta_0+\eta_1}{2}\right)\frac{r_0+r_1}{2}\cos(\gamma(\tau') - \langle Q\rangle_T(\tau'-\tau))$$

$$+(-1)^j (\tau-\tau')\left\langle\cos\left(\frac{\eta(\tau)}{2}\right)\right\rangle_T \sin\left(\frac{\eta_0+\eta_1}{2}\right)\sin\left(\frac{\eta(\tau')}{2}\right)\sin(\gamma(\tau)-\gamma(\tau')-\langle Q\rangle_T(\tau-\tau')),$$

$$\tilde{\Xi}_{R,2}(r_1,r_2,r'_1,r'_2;\tau-\tau') = \frac{1}{Z_R}\frac{1}{(2\pi)^4}\int_{-\infty}^{\infty} dp_1 \int_{-\infty}^{\infty} dp_2 \int_{-\infty}^{\infty} dp'_1 \int_{-\infty}^{\infty} dp'_2 \int D\delta R(\tau) \exp(ip_1(r_1-\delta R(\tau)))$$

$$\exp(ip_2(r_2-\delta R(\tau')))\exp(ip'_1(r'_1-\delta R'(\tau)))\exp(ip'_2(r'_2-\delta R'(\tau')))$$ (C.11)

$$\exp\left(-\int_0^{L_b} d\tau \left(\frac{l_p}{4}\left(\frac{d^2\delta R(\tau)}{d\tau^2}\right)^2 + \frac{\beta_R}{2}\left(\frac{d\delta R(\tau)}{d\tau}\right)^2 + \frac{\alpha_R}{2}\delta R(\tau)^2\right)\right),$$

and $\Xi_{\eta,2}(\eta_1,\eta_2;\tau-\tau')$ is already given by Eq. (A.61). Evaluation of the integrals in Eq. (C.11) yields

$$\tilde{\Xi}_{R,2}(r_1,r_2,r'_1,r'_2;\tau-\tau') = \frac{1}{(2\pi)^2}\sqrt{\frac{1}{\det(\mathbf{M}(\tau-\tau'))}}\exp\left(-\frac{\mathbf{R}^T \mathbf{M}^{-1}(\tau-\tau')\mathbf{R}}{2}\right),$$ (C.12)

where

$$\mathbf{R} = \begin{pmatrix} r_1 \\ r_2 \\ r'_1 \\ r'_2 \end{pmatrix}, \qquad \mathbf{M}(\tau-\tau') = \begin{pmatrix} d_R^2 & G_R(\tau-\tau') & 0 & C_R(\tau-\tau') \\ G_R(\tau-\tau') & d_R^2 & -C_R(\tau-\tau') & 0 \\ 0 & -C_R(\tau-\tau') & \theta_R^2 & D_R(\tau-\tau') \\ C_R(\tau-\tau') & 0 & D_R(\tau-\tau') & \theta_R^2 \end{pmatrix},$$ (C.13)

and

$$C_R(\tau-\tau') = \frac{1}{\pi}\int_{-\infty}^{\infty}\frac{ik\exp(-ik(\tau-\tau'))}{l_p k^4 + 2\beta_R k^2 + 2\alpha_R}dk, \quad D_R(\tau-\tau') = \frac{1}{\pi}\int_{-\infty}^{\infty}\frac{k^2\exp(-ik(\tau-\tau'))}{l_p k^4 + 2\beta_R k^2 + 2\alpha_R}dk. \quad \text{(C.14)}$$

Here we will make a simplifying approximation and neglect $G_R(\tau-\tau')$, $C_R(\tau-\tau')$ and $D_R(\tau-\tau')$ in Eq. (C.13) as well as $G_\eta(\tau-\tau')$ in Eq. (A.76). This approximation should be valid again if the correlation ranges of both the fluctuations in $R$ and $\eta$ are sufficiently small. In Ref. [1], we derived expressions for the leading order corrections from these correlation functions, but we will not consider these here. Using this approximation, Eq. (C.9) simplifies to

$$\langle Wr_j \rangle_0 \approx \frac{L_b}{\langle Q \rangle_0 (2\pi)^3 R_0 d_R^2 d_\eta^2 \theta_R^2}\int_{-\infty}^{\infty}d\eta_1\int_{-\infty}^{\infty}d\eta_2\int_{-\infty}^{\infty}dr_1'\int_{-\infty}^{\infty}dr_2'\int_{-\infty}^{\infty}dr_1\int_{-\infty}^{\infty}dr_2\int_{-\infty}^{\infty}dx\exp\left(-\frac{\eta_1^2+\eta_2^2}{2d_\eta^2}\right)$$

$$\exp\left(-\frac{R_0^2(r_1^2+r_2^2)}{2d_R^2}\right)\exp\left(-\frac{r_1'^2+r_2'^2}{2\theta_R^2}\right)\tilde{\omega}_j(x\langle Q\rangle_0^{-1}; R_0(r_1+1), R_0(r_2+1), r_1', r_2', \eta_0+\eta_1, \eta_0+\eta_2)$$

$$\left(\frac{1}{2}(1+(-1)^j\cos x)+\frac{r_1^2}{4}+\frac{r_2^2}{4}+\frac{(r_1+r_2)}{2}(1+(-1)^j\cos x)+(-1)^j\frac{r_1 r_2}{2}\cos x+\tilde{P}^2 x^2\right)^{-3/2},$$

(C.15)

where
$$\tilde{P} = \frac{\left\langle\cos\left(\frac{\eta(\tau)}{2}\right)\right\rangle_0}{\langle Q\rangle_0 R_0}. \quad \text{(C.16)}$$

Also, in writing Eq. (C.15), we have assumed also the length of the braided section to be long. We will also assume that the fluctuations in $\eta_1$, $\eta_2$, $r_1'$ and $r_2'$ to be small, i.e. both $d_\eta^2$ and $\theta_R^2$ small. This allows us to expand out $\tilde{\omega}_j$ in powers of these variables. We retain both leading order terms in $\eta_1$, $\eta_2$, $r_1'$ and $r_2'$. Then, it is easy to perform the Gaussian integrations over these variables. We also make a change of variables $r_1 = R_0 r\cos\phi_r$ and $r_2 = R_0 r\sin\phi_r$. These steps allow us to approximate Eq. (C.15) (more details are given in Ref. [1]) as

$$\langle Wr_j \rangle_0 \approx -\frac{L_b}{2\pi d_R^2 \langle Q\rangle_0}\int_0^{2\pi}d\phi_r\int_0^{\infty}rdr\int_{-\infty}^{\infty}dx\exp\left(-\frac{R_0^2 r^2}{2d_R^2}\right)$$

$$\left[\left(\sin\left(\frac{\eta_0}{2}\right)\cos\left(\frac{\eta_0}{2}\right)(1+(-1)^j\cos x)(1+r\sin\phi_r)+(-1)^j\tilde{P}\sin^2\left(\frac{\eta_0}{2}\right)x\sin x\right)\left(1-\frac{d_\eta^2}{4}\right)\right.$$

$$\left.+\theta_R^2\left(-\frac{1}{8}\cot\left(\frac{\eta_0}{2}\right)(1+(-1)^j\cos x)(1+r\sin\phi_r)-\frac{(-1)^j}{4}\tilde{P}x\sin x\right)\right]$$

$$\left(\frac{1}{2}(1+(-1)^j\cos(x))+\frac{r^2}{4}+\frac{r(\cos\phi_r+\sin\phi_r)}{2}(1+(-1)^j\cos(x))+(-1)^j\frac{r^2\sin\phi_r\cos\phi_r}{2}\cos(x)+\tilde{P}^2 x^2\right)^{-3/2}.$$

(C.17)

This can be further rewritten as

$$\frac{d_R \langle Wr_j \rangle_0}{L_b} \equiv \overline{W}_j(\tilde{R}_0, \tilde{P}, \alpha_\eta, \theta_R, \eta_0) \approx -\frac{\tilde{P}\tilde{R}_0}{2\pi \cos\left(\frac{\eta_0}{2}\right)} \int_0^\infty r\,dr \exp\left(-\frac{\tilde{R}_0^2 r^2}{2}\right)$$

$$\left\{ \left[ \left( \sin\left(\frac{\eta_0}{2}\right)\cos\left(\frac{\eta_0}{2}\right) K_{j,1,0}(r,\tilde{P}) + (-1)^j K_{j,2,0}(r,\tilde{P}) \sin^2\left(\frac{\eta_0}{2}\right) \right)\left(1 - \frac{1}{8(2l_p\alpha_\eta)^{1/2}}\right) \right. \right.$$

$$\left. + \theta_R^2 \left( -\frac{1}{8}\cot\left(\frac{\eta_0}{2}\right) K_{j,1,0}(r,\tilde{P}) - \frac{(-1)^j}{4} K_{j,2,0}(r,\tilde{P}) \right) \right] + \left[ \sin\left(\frac{\eta_0}{2}\right)\cos\left(\frac{\eta_0}{2}\right)\left(1 - \frac{1}{8(2l_p\alpha_\eta)^{1/2}}\right) \right.$$

$$\left. \left. -\frac{\theta_R^2}{8}\cot\left(\frac{\eta_0}{2}\right) \right] K_{j,1,1}(r,\tilde{P}) \right\}, \tag{C.18}$$

where the functions $K_{j,i,k}(r,\tilde{P})$ are given by the double integrals

$$K_{j,1,k}(r,\tilde{P}) = r^k \int_0^{2\pi} d\phi_r \int_{-\infty}^\infty dx \left(1+(-1)^j \cos x\right) \sin^k \phi_r \left( \frac{1}{2}\left(1+(-1)^j \cos x\right) + \frac{r^2}{4} \right.$$

$$\left. + \frac{r(\cos\phi_r + \sin\phi_r)}{2}\left(1+(-1)^j \cos x\right) + (-1)^j \frac{r^2 \sin 2\phi_r}{4} \cos x + \tilde{P}^2 x^2 \right)^{-3/2}, \tag{C.19}$$

$$K_{j,2,k}(r,\tilde{P}) = r^k \int_0^{2\pi} d\phi_r \int_{-\infty}^\infty dx\, \tilde{P} x \sin x \sin^k \phi_r \left( \frac{1}{2}\left(1+(-1)^j \cos(x)\right) + \frac{r^2}{4} \right.$$

$$\left. + \frac{r(\cos\phi_r + \sin\phi_r)}{2}\left(1+(-1)^j \cos x\right) + (-1)^j \frac{r^2 \sin 2\phi_r}{4} \cos x + \tilde{P}^2 x^2 \right)^{-3/2}, \tag{C.20}$$

and we have that $\tilde{R}_0 = R_0 / d_R$.

Treatment of $K_{2,1,k}(r,\tilde{P})$ and $K_{2,2,k}(r,\tilde{P})$ is straightforward; they can be expanded out in power series in $r$. Therefore, they can simply be expressed as

$$K_{2,1,k}(r,\tilde{P}) \approx r^k \left( K_{2,1,k}^{(0)}(\tilde{P}) + K_{2,1,k}^{(1)}(\tilde{P})r + K_{2,1,k}^{(2)}(\tilde{P})r^2 + K_{2,1,k}^{(3)}(\tilde{P})r^3 + K_{2,1,k}^{(4)}(\tilde{P})r^4 + \ldots \right), \tag{C.21}$$

$$K_{2,2,k}(r,\tilde{P}) \approx r^k \left( K_{2,2,k}^{(0)}(\tilde{P}) + K_{2,2,k}^{(1)}(\tilde{P})r + K_{2,2,k}^{(2)}(\tilde{P})r^2 + K_{2,2,k}^{(3)}(\tilde{P})r^3 + K_{2,2,k}^{(4)}(\tilde{P})r^4 + \ldots \right). \tag{C.22}$$

Explicit expressions for the non-vanishing terms in the expansion are given in Appendix D. The integrals in Eq. (C.18) over $r$ can easily be done leading to a power series expansion in $1/\tilde{R}_0 = d_R / R_0$. This final integration yields the expression

$$\bar{W}_2(\tilde{R}_0,\tilde{P},\alpha_\eta,\theta_R,\eta_0) \approx -\frac{\tilde{P}}{2\pi\cos\left(\frac{\eta_0}{2}\right)}\left\{\left[\sin\left(\frac{\eta_0}{2}\right)\cos\left(\frac{\eta_0}{2}\right)\left(1-\frac{1}{8(2l_p\alpha_\eta)^{1/2}}\right)-\frac{\theta_R^2}{8}\cot\left(\frac{\eta_0}{2}\right)\right]\right.$$

$$\left(\frac{K_{2,1,0}^{(0)}(\tilde{P})}{\tilde{R}_0}+\frac{2\left(K_{2,1,0}^{(2)}(\tilde{P})+K_{2,1,1}^{(1)}(\tilde{P})\right)}{\tilde{R}_0^3}+\frac{8\left(K_{2,1,0}^{(4)}(\tilde{P})+K_{2,1,1}^{(3)}(\tilde{P})\right)}{\tilde{R}_0^5}\right)$$

$$\left.+\left[\sin^2\left(\frac{\eta_0}{2}\right)\left(1-\frac{1}{8(2l_p\alpha_\eta)^{1/2}}\right)-\frac{\theta_R^2}{4}\right]\left(\frac{K_{2,2,0}^{(0)}(\tilde{P})}{\tilde{R}_0}+\frac{2K_{2,2,0}^{(2)}(\tilde{P})}{\tilde{R}_0^3}+\frac{8K_{2,2,0}^{(4)}(\tilde{P})}{\tilde{R}_0^5}\right)\right\}.$$

(C.23)

Here we find that $K_{2,1,0}^1(\tilde{P})=K_{2,1,0}^3(\tilde{P})=K_{2,1,1}^0(\tilde{P})=K_{2,1,1}^2(\tilde{P})=K_{2,2,0}^1(\tilde{P})=K_{2,2,0}^3(\tilde{P})=0$. However, $K_{1,1,k}(r,\tilde{P})$ and $K_{1,2,k}(r,\tilde{P})$ cannot be handled this way due to divergent integrals in the expansion; they do not have a regular power series expansion. However, these integrals can be estimated for small $r$ by treating $r(\cos\phi_r+\sin\phi_r)(1-\cos x)/2+r^2\sin 2\phi_r(1-\cos x)/4$ as a correction, and expanding out Eqs. (C.19) and (C.20) in terms of it. The rational for such an expansion is that the leading order behaviour, for small $r$, should be when $x$ is small. This is because of the singular nature of these integrals in the limit $r\to 0$, when $x\to 0$, and the next to leading order terms are corrections to this dominant behaviour. This yields the expansions

$$K_{1,1,k}(r,\tilde{P})\approx\left(K_{1,1,k}^{(0)}(r,\tilde{P})+K_{1,1,k}^{(1)}(r,\tilde{P})\ldots\right),\tag{C.24}$$

$$K_{1,2,k}(r,\tilde{P})\approx\left(K_{1,2,k}^{(0)}(r,\tilde{P})+K_{1,2,k}^{(1)}(r,\tilde{P})\ldots\right),\tag{C.25}$$

where the terms in the expansion are given by

$$K_{1,1,k}^{(0)}(r,\tilde{P})=r^k\int_0^{2\pi}d\phi_r\int_{-\infty}^{\infty}dx(1-\cos x)\sin^k\phi_r\left(\frac{1}{2}(1-\cos x)+\frac{r^2}{4}-\frac{r^2\sin 2\phi_r}{4}+\tilde{P}^2x^2\right)^{-3/2},\tag{C.26}$$

$$K_{1,1,k}^{(1)}(r,\tilde{P})=-\frac{3}{2}r^k\int_0^{2\pi}d\phi_r\int_{-\infty}^{\infty}dx(1-\cos x)^2\sin^k\phi_r\left(\frac{r(\cos\phi_r+\sin\phi_r)}{2}+\frac{r^2\sin 2\phi_r}{4}\right)$$
$$\left(\frac{1}{2}(1-\cos x)+\frac{r^2}{4}-\frac{r^2\sin 2\phi_r}{4}+\tilde{P}^2x^2\right)^{-5/2},\tag{C.27}$$

$$K_{1,2,k}^{(0)}(r,\tilde{P})=r^k\int_0^{2\pi}d\phi_r\int_{-\infty}^{\infty}dx\tilde{P}x\sin x\sin^k\phi_r\left(\frac{1}{2}(1-\cos x)+\frac{r^2}{4}-\frac{r^2\sin 2\phi_r}{4}+\tilde{P}^2x^2\right)^{-3/2},\tag{C.28}$$

$$K^{(1)}_{1,2,k}(r,\tilde{P}) = -\frac{3}{2}r^k \int_0^{2\pi} d\phi_r \int_{-\infty}^{\infty} dx \tilde{P} x \sin x (1-\cos x) \sin^k \phi_r \left( \frac{r(\cos\phi_r + \sin\phi_r)}{2} + \frac{r^2 \sin 2\phi_r}{4} \right)$$
$$\left( \frac{1}{2}(1-\cos x) + \frac{r^2}{4} - \frac{r^2 \sin 2\phi_r}{4} + \tilde{P}^2 x^2 \right)^{-5/2}.$$

(C.29)

In these expressions, the angular integrals may now be evaluated analytically, leaving only an integral over $x$ in these functions, as well as an integral over $r$ in the expression for the writhe contribution. The resulting terms are given in Appendix D. Further analysis may be possible for small $r$ by considering the small $x$ behaviour of these terms, and perhaps a change of variables $x = y\cos\theta$ and $r = y\sin\theta$. Though, presently, we performed the other integrations numerically and used interpolation function representations of the resulting functions (of both $r$ and $\tilde{P}$) to calculate the contributions to the writhe. Such an analytical improvement may still leave an integral over $\theta$ that would still need to be attempted numerically.

We are left with

$$\overline{W}_1(\tilde{R}_0, \tilde{P}, \alpha_\eta, \theta_R, \eta_0) \approx -\frac{\tilde{P}\tilde{R}_0}{2\pi \cos\left(\frac{\eta_0}{2}\right)} \int_0^\infty r dr \exp\left( -\frac{\tilde{R}_0^2 r^2}{2} \right)$$

$$\left\{ \left[ \left( \sin\left(\frac{\eta_0}{2}\right) \cos\left(\frac{\eta_0}{2}\right) \left( K^{(0)}_{1,1,0}(r,\tilde{P}) + K^{(1)}_{1,1,0}(r,\tilde{P}) \right) - \left( K^{(0)}_{1,2,0}(r,\tilde{P}) + K^{(1)}_{1,2,0}(r,\tilde{P}) \right) \sin^2\left(\frac{\eta_0}{2}\right) \right) \left( 1 - \frac{1}{8(2l_p\alpha_\eta)^{1/2}} \right) \right. \right.$$

$$\left. + \theta_R^2 \left( -\frac{1}{8}\cot\left(\frac{\eta_0}{2}\right) \left( K^{(0)}_{1,1,0}(r,\tilde{P}) + K^{(1)}_{1,1,0}(r,\tilde{P}) \right) + \frac{1}{4}\left( K^{(0)}_{1,2,0}(r,\tilde{P}) + K^{(1)}_{1,2,0}(r,\tilde{P}) \right) \right) \right] + \left[ \sin\left(\frac{\eta_0}{2}\right) \cos\left(\frac{\eta_0}{2}\right) \left( 1 - \frac{1}{8(2l_p\alpha_\eta)^{1/2}} \right) \right.$$

$$\left. -\frac{\theta_R^2}{8}\cot\left(\frac{\eta_0}{2}\right) \right] \left( K^{(0)}_{1,1,1}(r,\tilde{P}) + K^{(1)}_{1,1,1}(r,\tilde{P}) \right) \right\}.$$

(C.30)

Last of all, we approximate (Ref [1])

$$\tilde{P} \approx \frac{1}{2\tan\left(\frac{\eta_0}{2}\right)} f(\tilde{R}_0) \left( 1 + \frac{\theta_R^2}{8\sin^2\left(\frac{\eta_0}{2}\right)} \right),$$

(C.31)

where

$$f(\tilde{R}_0) \approx 1 - \frac{1}{\tilde{R}_0^2} - \frac{2}{\tilde{R}_0^4}.$$

(C.32)

Partial differentiation with respect to the various variational parameters is straight forward, and is most accurately done before performing numerical integrations in obtaining equations that minimize the free energy (Eq. (2.67) of the main text) with respect to the various variational parameters. We refrain from giving particular details of this procedure, for those see Ref. [1].

## D. Functions in the expansion of the average writhe

Here, we present functions that occur in the expansion of the average writhe considered in the Appendix C. We have for the non-vanishing terms in the expansion of $K_{2,1,k}(r,\tilde{P})$ and $K_{2,2,k}(r,\tilde{P})$ the following:

$$K_{2,1,0}^{(0)}(\tilde{P}) = 2\pi \int_{-\infty}^{\infty} dx \frac{(1+\cos x)}{\left(\frac{1}{2}(1+\cos x)+\tilde{P}^2 x^2\right)^{3/2}}, \qquad (D.1)$$

$$K_{2,1,0}^{(2)}(\tilde{P}) = 2\pi \int_{-\infty}^{\infty} dx \left( \frac{15(1+\cos x)^3}{32\left(\frac{1}{2}(1+\cos x)+\tilde{P}^2 x^2\right)^{7/2}} - \frac{3(1+\cos x)}{8\left(\frac{1}{2}(1+\cos x)+\tilde{P}^2 x^2\right)^{5/2}} \right), \qquad (D.2)$$

$$K_{2,1,0}^{(4)}(\tilde{P}) = \frac{15\pi}{128} \int_{-\infty}^{\infty} dx \left( \frac{(1+\cos x)(2+\cos^2 x)}{\left(\frac{1}{2}(1+\cos x)+\tilde{P}^2 x^2\right)^{7/2}} - \frac{7(2+\cos x)(1+\cos x)^3}{2\left(\frac{1}{2}(1+\cos x)+\tilde{P}^2 x^2\right)^{9/2}} + \frac{63(1+\cos x)^5}{16\left(\frac{1}{2}(1+\cos x)+\tilde{P}^2 x^2\right)^{11/2}} \right), \qquad (D.3)$$

$$K_{2,1,1}^{(1)}(\tilde{P}) = -\frac{3\pi}{4} \int_{-\infty}^{\infty} dx \frac{(1+\cos x)^2}{\left(\frac{1}{2}(1+\cos x)+\tilde{P}^2 x^2\right)^{5/2}}, \qquad (D.4)$$

$$K_{2,1,1}^{(3)}(\tilde{P}) = \frac{\pi}{2} \int_{-\infty}^{\infty} dx \left( \frac{15(2+\cos x)(1+\cos x)^2}{32\left(\frac{1}{2}(1+\cos x)+\tilde{P}^2 x^2\right)^{7/2}} - \frac{105(1+\cos x)^4}{128\left(\frac{1}{2}(1+\cos x)+\tilde{P}^2 x^2\right)^{9/2}} \right), \qquad (D.5)$$

$$K_{2,2,0}^{(0)}(\tilde{P}) = 2\pi \int_{-\infty}^{\infty} dx \frac{\tilde{P} x \sin x}{\left(\frac{1}{2}(1+\cos x)+\tilde{P}^2 x^2\right)^{3/2}}, \qquad (D.6)$$

$$K_{2,2,0}^{(2)}(\tilde{P}) = 2\pi\tilde{P} \int_{-\infty}^{\infty} dx\, x \sin x \left( \frac{15(1+\cos x)^2}{32\left(\frac{1}{2}(1+\cos x)+\tilde{P}^2 x^2\right)^{7/2}} - \frac{3}{8\left(\frac{1}{2}(1+\cos x)+\tilde{P}^2 x^2\right)^{5/2}} \right), \quad \text{(D.7)}$$

$$K_{2,2,0}^{(4)}(\tilde{P}) = \frac{15\pi\tilde{P}}{128} \int_{-\infty}^{\infty} dx\, x \sin x$$

$$\left( \frac{2+\cos^2 x}{\left(\frac{1}{2}(1+\cos x)+\tilde{P}^2 x^2\right)^{7/2}} - \frac{7(1+\cos x)^2(2+\cos x)}{2\left(\frac{1}{2}(1+\cos x)+\tilde{P}^2 x^2\right)^{9/2}} + \frac{63(1+\cos x)^4}{16\left(\frac{1}{2}(1+\cos x)+\tilde{P}^2 x^2\right)^{11/2}} \right). \quad \text{(D.8)}$$

We have for the terms in the expansion of $K_{1,1,k}(r,\tilde{P})$ and $K_{1,2,k}(r,\tilde{P})$ the following:

$$K_{1,1,0}^{(0)}(r,\tilde{P}) = 2\int_0^{\infty} dx (1-\cos x) U(x,r,\tilde{P}) \quad \text{(D.9)}$$

$$K_{1,2,0}^{(0)}(r,\tilde{P}) = 2\tilde{P} \int_0^{\infty} dx\, x \sin x\, U(x,r,\tilde{P}), \quad \text{(D.10)}$$

where

$$U(x,r,\tilde{P}) = \left\{ \frac{4}{\left[\left(\tilde{P}^2 x^2 + \frac{1}{2}(1-\cos x)\right)\left(\frac{r^2}{2}+\tilde{P}^2 x^2 + \frac{1}{2}(1-\cos x)\right)\right]^{1/2}} \right.$$

$$\left[ E\left(\arcsin\left(\sqrt{\frac{r^2 + 2\tilde{P}^2 x^2 + (1-\cos x)}{r^2 + 4\tilde{P}^2 x^2 + 2(1-\cos x)}}\right), \frac{r^2}{r^2 + 2\tilde{P}^2 x^2 + (1-\cos x)}\right) + E\left(\frac{\pi}{4}, \frac{r^2}{r^2 + 2\tilde{P}^2 x^2 + (1-\cos x)}\right) \right]$$

$$\left. - \frac{r^2}{\left(\tilde{P}^2 x^2 + \frac{1}{2}(1-\cos x)\right)\left(\tilde{P}^2 x^2 + \frac{1}{2}(1-\cos x)+\frac{r^2}{2}\right)\left(\tilde{P}^2 x^2 + \frac{1}{2}(1-\cos x)+\frac{r^2}{4}\right)^{1/2}} \right\}$$

(D.11)

In Eq. (D.11), the functions $E(\phi,k)$ are incomplete elliptic integrals of the second kind defined through

$$E(\phi,k) = \int_0^{\phi} (1-k\sin^2\theta)^{1/2} d\theta \quad \text{(D.12)}$$

Next,

$$K^{(0)}_{1,1,1}(r, \tilde{P}) = 0 \tag{D.13}$$

$$K^{(1)}_{1,1,0} = 2\int_0^\infty dx (1 - \cos x)^2 \left.\frac{dV_1(s,r)}{ds}\right|_{s=\frac{1}{2}(1-\cos x) + \tilde{P}^2 x^2}, \tag{D.14}$$

$$K^{(1)}_{1,2,0} = 2\tilde{P}\int_0^\infty dx\, x \sin x (1 - \cos x) \left.\frac{dV_1(s,r)}{ds}\right|_{s=\frac{1}{2}(1-\cos x) + \tilde{P}^2 x^2}, \tag{D.15}$$

where

$$V_1(s,r) = \left[\left(\frac{r^2}{4} + s\right)\frac{4}{s}\frac{1}{\sqrt{\frac{r^2}{2} + s}}\left[E\left(\arcsin\left(\sqrt{\frac{r^2 + 2s}{r^2 + 4s}}\right), \frac{r^2}{r^2 + 2s}\right) + E\left(\frac{\pi}{4}, \frac{r^2}{r^2 + 2s}\right)\right]\right.$$

$$\left. - \frac{r^2}{s}\frac{1}{\left(s + \frac{r^2}{2}\right)}\frac{1}{\left(s + \frac{r^2}{4}\right)^{1/2}}\right] - \frac{4}{\left(\frac{r^2}{2} + s\right)^{1/2}}\left[F\left(\arcsin\left(\sqrt{\frac{r^2 + 2s}{r^2 + 4s}}\right), \frac{r^2}{r^2 + 2s}\right) + F\left(\frac{\pi}{4}, \frac{r^2}{r^2 + 2s}\right)\right]$$

$$\tag{D.16}$$

Here, the functions $F(\phi, k)$ are incomplete elliptic integrals of the first kind defined through

$$F(\phi, k) = \int_0^\phi (1 - k \sin^2 \theta)^{-1/2} d\theta \tag{D.17}$$

Next,

$$K^{(1)}_{1,1,1} = 2\int_0^\infty dx (1 - \cos x)^2 \left.\frac{dV_2(s,r)}{ds}\right|_{s=\frac{1}{2}(1-\cos x) + \tilde{P}^2 x^2} \tag{D.18}$$

where

$$V_2(s,r) = \left[\left(\frac{r^2}{2}+s\right)\frac{4}{s}\frac{1}{\sqrt{\frac{r^2}{2}+s}}\left[E\left(\arcsin\left(\sqrt{\frac{r^2+2s}{r^2+4s}}\right),\frac{r^2}{r^2+2s}\right)+E\left(\frac{\pi}{4},\frac{r^2}{r^2+2s}\right)\right]\right.$$

$$\left.-\frac{r^2}{s}\frac{1}{\left(s+\frac{r^2}{2}\right)}\frac{1}{\left(s+\frac{r^2}{4}\right)^{1/2}}\right]-\frac{4}{\left(\frac{r^2}{2}+s\right)^{1/2}}\left[F\left(\arcsin\left(\sqrt{\frac{r^2+2s}{r^2+2=4s}}\right),\frac{r^2}{r^2+2s}\right)+F\left(\frac{\pi}{4},\frac{r^2}{r^2+2s}\right)\right]\right].$$

(D.19)

## E. Mean field electrostatics

Here, we give expressions for the effective interaction terms in the free energy using the mean field electrostatics of the KL-theory [2,3, 4,5]. These results we use to numerical generate the supercoil geometry and degree of configurational fluctuations in the variational theory. First, before presenting explicit terms we first use the fact that the infinite series (Eq. (2.48) of the main text) can be truncated to good accuracy to $-2 \leq n \leq 2$. It can be rewritten to be (when considering Refs. [2,3,4,5]).

$$\tilde{E}_{int} = \sum_{n=-2}^{2}\int_0^{L_b} d\tau [\bar{\varepsilon}_{int,KL}^{(0)}(R_0,\delta R(\tau),g_{av},n)+\sin\eta(s)\bar{\varepsilon}_{int,KL}^{(1)}(R_0,\delta R(\tau),g_{av},n)(1-\delta_{n,0})]\exp(-in\Delta\xi(\tau)),$$

(E.1)

as well as the fact that the interaction terms obey the relation

$$\bar{\varepsilon}_{int,KL}^{(0)}(R_0,\delta R(\tau)g_{av},-n) = \bar{\varepsilon}_{int,KL}^{(0)}(R_0,\delta R(\tau)g_{av},n).$$

(E.2)

Note we have not considered, here, an additional $\sin^2\eta(s)$ term, which was argued of geometrical grounds in Ref. [6] for strong helix specific forces. This is because the helix specific forces are considered to be weak and such a limiting term does not seem appropriate to use within the expansion (Eq. (2.57) of main text). We treat $\eta(s)-\eta_0$ as small, allowing us first to approximate

$$\tilde{E}_{int} \approx \sum_{n=-2}^{2}\int_0^{L_b} d\tau \left\{[\bar{\varepsilon}_{int,KL}^{(0)}(R_0,\delta R(\tau)g_{av},n)+\sin\eta_0\bar{\varepsilon}_{int,KL}^{(1)}(R_0,\delta R(\tau)g_{av},n)(1-\delta_{n,0})]\exp(-in\Delta\xi(\tau))\right.$$

$$\left.+\delta\eta(\tau)\cos\eta_0\bar{\varepsilon}_{int,KL}^{(1)}(R_0,\delta R(\tau)g_{av},n)(1-\delta_{n,0})-\frac{\delta\eta(\tau)^2}{2}\sin\eta_0\bar{\varepsilon}_{int,KL}^{(1)}(R_0,\delta R(\tau)g_{av},n)(1-\delta_{n,0})\right\}$$

(E.3)

Using Eqs. (E.2) and (E.3), our expressions for the effective interaction (Eqs. (A.103) -(A.107) and (B.9)) further simplify to

$$F_{un} = \frac{1}{d_R\sqrt{2\pi}} \int_{-\infty}^{\infty} dr \bar{\varepsilon}_{\mathrm{int},KL}^{(0)}\left(R_0, r, 0, \langle g_{av}\rangle_0, 0\right) \exp\left(-\frac{r^2}{2d_R^2}\right), \tag{E.4}$$

$$\frac{F_{cor}^{(1)}}{k_B T} = F_{cor,1}^{(1)} + F_{cor,2}^{(1)} + F_{cor,3}^{(1)}, \tag{E.5}$$

where

$$F_{cor,1}^{(1)} = \frac{2L_b \lambda_c}{(k_B T)^2}\left[\Delta_{0,0}\left(R_0, \eta_0, \langle g_{av}\rangle_T, d_R, d_\eta, 1\right)^2 + \frac{1}{4}\Delta_{0,0}\left(R_0, \eta_0, \langle g_{av}\rangle_T, d_R, d_\eta, 2\right)^2\right], \tag{E.6}$$

$$\begin{aligned}F_{cor,2}^{(1)} = \frac{l_p L_b}{(k_B T)^2}\bigg[&\Delta_{1,0}\left(R_0, \eta_0, \langle g_{av}\rangle_0, d_R, d_\eta, 1\right)^2 \Omega_{1,\eta}\left(\frac{l_p}{\lambda_c}, l_p \alpha_\eta\right) \\ &+\Delta_{1,0}\left(R_0, \eta_0, \langle g_{av}\rangle_0, d_R, d_\eta, 2\right)^2 \Omega_{1,\eta}\left(\frac{4l_p}{\lambda_c}, l_p \alpha_\eta\right)\bigg],\end{aligned} \tag{E.7}$$

$$\begin{aligned}F_{cor,3}^{(1)} = \frac{l_p^3 L_b}{(k_B T)^2}\bigg[&\Delta_{0,1}\left(R_0, \eta_0, \langle g_{av}\rangle_T, d_R, d_\eta, 1\right)^2 \Omega_{1,R}\left(\frac{l_p}{\lambda_c}, \alpha_R l_p^3, \beta_R l_p\right) \\ &+\Delta_{0,1}\left(R_0, \eta_0, \langle g_{av}\rangle_T, d_R, d_\eta, 2\right)^2 \Omega_{1,R}\left(\frac{4l_p}{\lambda_c}, \alpha_R l_p^3, \beta_R l_p\right)\bigg],\end{aligned} \tag{E.8}$$

and

$$\frac{F_{cor}^{(2)}}{k_B T} = \frac{3\lambda_c^2 L_b}{(k_B T)^3}\Delta_{0,0}\left(R_0, \eta_0, \langle g_{av}\rangle_T, d_R, d_\eta, 1\right)^2 \Delta_{0,0}\left(R_0, \eta_0, \langle g_{av}\rangle_T, d_R, d_\eta, 2\right), \tag{E.9}$$

as well as

$$\frac{F_{cor}^{(3)}}{k_B T} = F_{cor,1}^{(3)} + F_{cor,2}^{(3)} + F_{cor,3}^{(3)}, \tag{E.10}$$

where

$$F_{cor,1}^{(3)} = \frac{L\lambda_c^3}{(k_B T)^4}\left(\frac{7}{2} + \frac{4\lambda_c^2}{(\lambda_c^{(0)})^2 - \lambda_c^2}\right)\Delta_{0,0}\left(R_0, \eta_0, \langle g_{av}\rangle_T, d_R, d_\eta, 1\right)^4, \tag{E.11}$$

$$F_{cor,2}^{(3)} = \frac{L\lambda_c^3}{(k_B T)^4}\left[\frac{20\lambda_c^2}{25(\lambda_c^{(0)})^2 - 16\lambda_c^2} - \frac{20}{9}\right]\Delta_{0,0}\left(R_0, \eta_0, \langle g_{av}\rangle_T, d_R, d_\eta, 1\right)^2 \Delta_{0,0}\left(R_0, \eta_0, \langle g_{av}\rangle_T, d_R, d_\eta, 2\right)^2, \tag{E.12}$$

$$F_{cor,3}^{(3)} = \frac{L\lambda_c^3}{(k_B T)^4}\left[\frac{7}{128} + \frac{\lambda_c^2}{16\left((\lambda_c^{(0)})^2 - \lambda_c^2\right)}\right]\Delta_{0,0}\left(R_0,\eta_0,\langle g_{av}\rangle_T,d_R,d_\eta,2\right)^4. \tag{E.13}$$

The forms of $\Delta_{0,0}\left(R_0,\eta_0,\langle g_{av}\rangle_T,d_R,d_\eta,n\right)$, $\Delta_{1,0}\left(R_0,\eta_0,\langle g_{av}\rangle_T,d_R,d_\eta,n\right)$ and $\Delta_{0,1}\left(R_0,\eta_0,\langle g_{av}\rangle_T,d_R,d_\eta,n\right)$ reduce to

$$\Delta_{0,0}\left(R_0,\eta_0,\langle g_{av}\rangle_T,d_R,d_\eta,n\right) \approx \frac{1}{\sqrt{2\pi}d_R}\int_{-\infty}^{\infty} dr\exp\left(-\frac{r^2}{2d_R^2}\right)$$
$$\left\{[\bar{\varepsilon}_{int,KL}^{(0)}(R_0,r,g_{av},n) + \sin\eta_0\left(1 - d_\eta^2/2\right)\bar{\varepsilon}_{int,KL}^{(1)}(R_0,r,g_{av},n)]\right. \tag{E.14}$$
$$\equiv \Delta_{0,0,0}\left(R_0,\eta_0,\langle g_{av}\rangle_T,d_R,d_\eta,n\right) - \frac{d_\eta^2}{2}\Delta_{0,0,1}\left(R_0,\eta_0,\langle g_{av}\rangle_T,d_R,d_\eta,n\right),$$

$$\Delta_{1,0}\left(R_0,\eta_0,\langle g_{av}\rangle_T,d_R,d_\eta,n\right) \approx \frac{1}{\sqrt{2\pi}d_R}\int_{-\infty}^{\infty} dr\exp\left(-\frac{r^2}{2d_R^2}\right)\cos\eta_0\bar{\varepsilon}_{int,KL}^{(1)}(R_0,r,g_{av},n), \tag{E.15}$$

$$\Delta_{0,1}\left(R_0,\eta_0,\langle g_{av}\rangle_0,d_R,d_\eta,n\right) \approx \frac{1}{\sqrt{2\pi}d_R^3}\int_{-\infty}^{\infty} dr\, r\exp\left(-\frac{r^2}{2d_R^2}\right)[\bar{\varepsilon}_{int,KL}^{(0)}(R_0,r,g_{av},n)$$
$$+ \sin\eta_0\bar{\varepsilon}_{int,KL}^{(1)}(R_0,r,g_{av},n)]. \tag{E.16}$$

We then expand out for small $d_\eta$ Eqs. (E.6) and (E.11)-(E.13) yielding the final results

$$F_{cor,1}^{(1)} \approx \frac{2L_b\lambda_c}{(k_B T)^4}\left[\Delta_{0,0,0}\left(R_0,\eta_0,\langle g_{av}\rangle_T,d_R,d_\eta,1\right)^2 - \frac{1}{4}\Delta_{0,0,0}\left(R_0,\eta_0,\langle g_{av}\rangle_T,d_R,d_\eta,2\right)^2\right]$$
$$+ 2d_\eta^2 L_b\lambda_c\left[\Delta_{0,0,0}\left(R_0,\eta_0,\langle g_{av}\rangle_T,d_R,d_\eta,1\right)\Delta_{0,0,1}\left(R_0,\eta_0,\langle g_{av}\rangle_T,d_R,d_\eta,1\right)\right. \tag{E.17}$$
$$\left.- \frac{1}{4}\Delta_{0,0,0}\left(R_0,\eta_0,\langle g_{av}\rangle_T,d_R,d_\eta,2\right)\Delta_{0,0,1}\left(R_0,\eta_0,\langle g_{av}\rangle_T,d_R,d_\eta,2\right)\right],$$

$$\frac{F_{cor}^{(2)}}{k_B T} = \frac{3\lambda_c^2 L_b}{(k_B T)^3}\Delta_{0,0,0}\left(R_0,\eta_0,\langle g_{av}\rangle_T,d_R,d_\eta,1\right)\left[\Delta_{0,0,0}\left(R_0,\eta_0,\langle g_{av}\rangle_T,d_R,d_\eta,1\right)\Delta_{0,0,0}\left(R_0,\eta_0,\langle g_{av}\rangle_T,d_R,d_\eta,2\right)\right.$$
$$- d_\eta^2\Delta_{0,0,1}\left(R_0,\eta_0,\langle g_{av}\rangle_T,d_R,d_\eta,1\right)\Delta_{0,0,0}\left(R_0,\eta_0,\langle g_{av}\rangle_T,d_R,d_\eta,2\right)$$
$$\left.- \frac{d_\eta^2}{2}\Delta_{0,0,0}\left(R_0,\eta_0,\langle g_{av}\rangle_T,d_R,d_\eta,1\right)\Delta_{0,0,1}\left(R_0,\eta_0,\langle g_{av}\rangle_T,d_R,d_\eta,2\right)\right],$$
$$\tag{E.18}$$

$$F_{cor,1}^{(3)} = \frac{L\lambda_c^3}{(k_B T)^4}\left(\frac{7}{2} + \frac{4\lambda_c^2}{\left((\lambda_c^{(0)})^2 - \lambda_c^2\right)}\right)\Delta_{0,0,0}\left(R_0,\eta_0,\langle g_{av}\rangle_T,d_R,d_\eta,1\right)^3\left[\Delta_{0,0,0}\left(R_0,\eta_0,\langle g_{av}\rangle_T,d_R,d_\eta,1\right)\right.$$
$$\left.- 2d_\eta^2\Delta_{0,0,1}\left(R_0,\eta_0,\langle g_{av}\rangle_T,d_R,d_\eta,1\right)\right],$$

(E.19)

$$F^{(3)}_{cor,2} = \frac{L\lambda_c^3}{(k_BT)^4}\left[\frac{20\lambda_c^2}{\left(25(\lambda_c^{(0)})^2 - 16\lambda_c^2\right)} - \frac{20}{9}\right]\Delta_{0,0,0}\left(R_0,\eta_0,\langle g_{av}\rangle_T,d_R,d_\eta,1\right)\Delta_{0,0,0}\left(R_0,\eta_0,\langle g_{av}\rangle_T,d_R,d_\eta,2\right)$$

$$\left[\Delta_{0,0,0}\left(R_0,\eta_0,\langle g_{av}\rangle_T,d_R,d_\eta,1\right)\Delta_{0,0,0}\left(R_0,\eta_0,\langle g_{av}\rangle_T,d_R,d_\eta,2\right) - d_\eta^2\left(\Delta_{0,0,0}\left(R_0,\eta_0,\langle g_{av}\rangle_T,d_R,d_\eta,1\right)\right.\right.$$

$$\left.\left.\Delta_{0,0,1}\left(R_0,\eta_0,\langle g_{av}\rangle_T,d_R,d_\eta,2\right) + \Delta_{0,0,1}\left(R_0,\eta_0,\langle g_{av}\rangle_T,d_R,d_\eta,1\right)\Delta_{0,0,0}\left(R_0,\eta_0,\langle g_{av}\rangle_T,d_R,d_\eta,2\right)\right)\right],$$

(E.20)

$$F^{(3)}_{cor,3} = \frac{L\lambda_c^3}{(k_BT)^4}\left[\frac{7}{128} + \frac{\lambda_c^2}{16\left((\lambda_c^{(0)})^2 - \lambda_c^2\right)}\right]\Delta_{0,0,0}\left(R_0,\eta_0,\langle g_{av}\rangle_T,d_R,d_\eta,2\right)^3\left[\Delta_{0,0,0}\left(R_0,\eta_0,\langle g_{av}\rangle_T,d_R,d_\eta,2\right)\right.$$

$$\left. -2d_\eta^2\Delta_{0,0,1}\left(R_0,\eta_0,\langle g_{av}\rangle_T,d_R,d_\eta,2\right)\right],$$

(E.21)

along with Eqs. (E.7) and (E.8).

Last of all, $\bar{\varepsilon}^{(0)}_{int,KL}(R_0,r,g_{av},n)$ and $\bar{\varepsilon}^{(1)}_{int,KL}(R_0,r,g_{av},n)$ are specified by the results of Refs. [2,3,4], for the KL theory of mean-field electrostatics. They take the form

$$\bar{\varepsilon}^{(0)}_{int,KL}(R_0,\delta R(\tau),\langle g_{av}\rangle_T,n) = \frac{2l_B\zeta(n)^2}{l_c^2}\frac{(-1)^n K_0(\kappa_n R)}{(a\kappa_n)^2 K_n'(\kappa_n a)^2} + \delta_{n,0}\sum_{m=-4}^{4}\frac{2l_B\zeta(m)^2\Omega_n(\kappa_m R,\kappa_m a)}{l_c^2(a\kappa_m)^2 K_m'(\kappa_m a)^2},$$

(E.22)

$$\Omega_n(\kappa_n R,\kappa_n a) = -\sum_{j=-\infty}^{\infty}K_{j-n}(\kappa_n R)K_{j-n}(\kappa_n R)\frac{I_j'(\kappa_n a)}{K_j'(\kappa_n a)},$$

(E.23)

$$\bar{\varepsilon}^{(1)}_{int,KL}(R_0,\delta R(\tau),\langle g_{av}\rangle_T,n) = \frac{2l_B n^2\zeta(n)^2}{l_c^2}\frac{\langle g_{av}\rangle}{\kappa_n}\frac{(-1)^n K_1(\kappa_n R)}{(a\kappa_n)^2 K_n'(\kappa_n a)^2},$$

(E.24)

where

$$\kappa_n = \sqrt{\kappa_D^2 + n^2\langle g_{av}\rangle_T^2}, \qquad \text{and} \qquad \zeta(n) = \theta\left(f_1 + (-1)^n f_2\right) - \cos\tilde{\phi}_s.$$

(E.25)

Here, $I_n(x)$ and $K_n(x)$ are modified Bessel functions of the first and second kind, and $I_n'(x)$ and $K_n'(x)$ are their derivatives with respect to argument. In Eqs. (E.22)-(E.25), there are the following fixed parameters: the Bjerrum length $l_B \approx 6.9\text{Å}$; $l_c \approx 1.7\text{Å}$, where $e/l_c$ is the linear phosphate charge density (in neutral solution); the effective electrostatic radius of DNA $a \approx 11.2\text{Å}$; and last of all, the half width of the minor groove $\tilde{\phi}_s \approx 0.4\pi$. Variable parameters are: $\kappa_D$, the Debye Screening length; $\theta$, the faction of DNA phosphate charge neutralized by counter-ions condensed near (or bound at) the DNA surface; $f_1$, the faction of condensed (or bound) ions localized near the centre of the minor groove; and finally $f_2$, the faction of condensed (or bound) ions localized near the centre of the major groove.

## F. Additional Numerical Results

### F.1. Changing the degree of charge localization

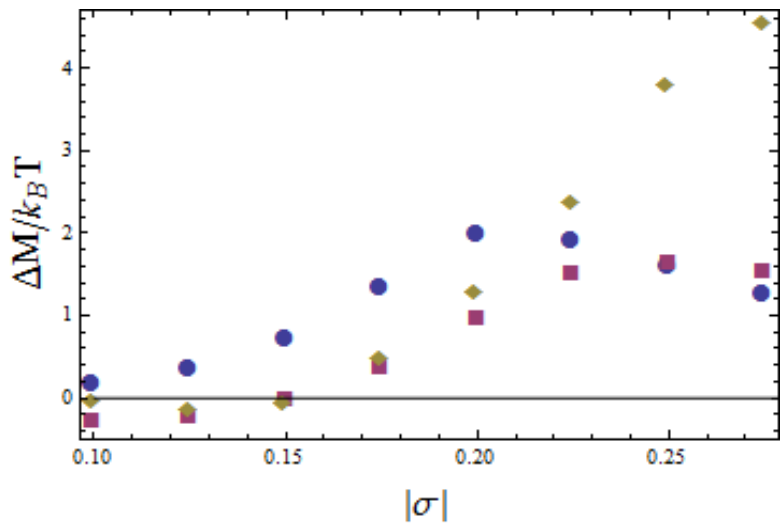

Fig F.1. Difference between moments required to produce left and right handed supercoils, $\Delta M = M(\sigma) + M(-\sigma)$. In all plots we set $\theta = 0.6$ in the interaction energy. We keep the ratio $f_1/f_2$ fixed to $0.7/0.3$ but change the value of $f_1 + f_2$, a measure of the degree of charge localization near the grooves of the DNA. The solid dark yellow points, purple points and blue points correspond to the values $f_1 + f_2 = 0, 0.5$ and $1$, respectively.

Here, we now investigate changing the amount of ions localized in the vicinity of the DNA grooves, namely the parameter $f_1 + f_2$. The value $f_1 + f_2 = 0$ corresponds to all the condensed (bound) counter-ions being uniformly smeared over the DNA surface, whereas $f_1 + f_2 = 1$ corresponds to them being completely localized in the grooves. As the results presented in the main text, we first look at $\Delta M$, the difference in the magnitude of the moment, $|M|$ between positive and negative supercoils with the same value of $|\sigma|$. As when varying $\theta$, (c.f. Fig. F1.) over a wide range of values it is easier to form additional negative supercoils than positive ones; only when $|\sigma|$ is small (for the values $f_1 + f_2 = 0$ and $0.5$) is it easier to increase the number of positive supercoils. We notice, for relatively small values of $|\sigma|$, that the difference in $\Delta M$ between left and right supercoils is less for supercoils where the counter-ions are more delocalized from the grooves (those values of $f_1 + f_2 = 0$ and $0.5$) than those with completely groove localized ones. It seems that there is not much difference in the $\Delta M$ values between $f_1 + f_2 = 0$ and $0.5$ below $|\sigma| = 0.2$. However, for values of $|\sigma| \geq 0.2$, it starts to become far easier to build negative supercoils, when the ions are completely delocalized from the grooves ($f_1 + f_2 = 0$) than other parameter values. The moment curves as functions of $\sigma$, for both left and right supercoils, are shown in Fig F.9.

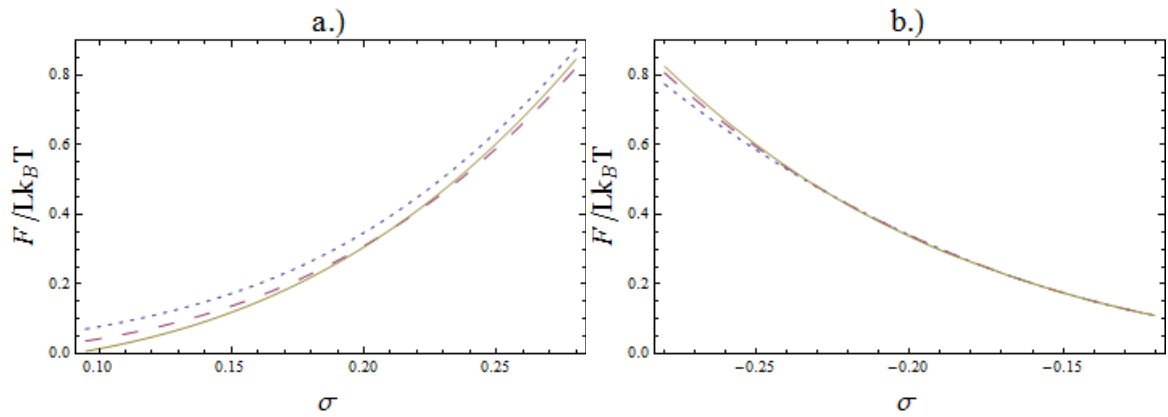

Fig F.2. Plots of $F / Lk_B T$, the supercoiling free energy per unit length and $k_B T$. On the left hand side we present curves for positive values of $\sigma$, whereas on the right we present curves for negative values. In panels a.) and b.) we fix $\theta = 0.6$ and $f_1 / f_2 = 7/3$, so varying $f_1 + f_2$. In these plots, the blue short dashed line, red medium dashed line and dark yellow solid lines refer to the values $f_1 + f_2 = 0, 0.5$ and $1$.

In Fig F.2, we show plots of $F / kT \approx F_T / kT$ with different values of $f_1 + f_2$ (with $f_1 / f_2 = 0.7 / 0.3$ and $\theta = 0.6$) for both positive and negative supercoils. Again the asymmetry is seen, but there are only slight differences with changing $f_1 + f_2$. The most difference is seen for positive values of $\sigma$, for negative values there is effectively no difference at all, except when $\sigma < -0.25$.

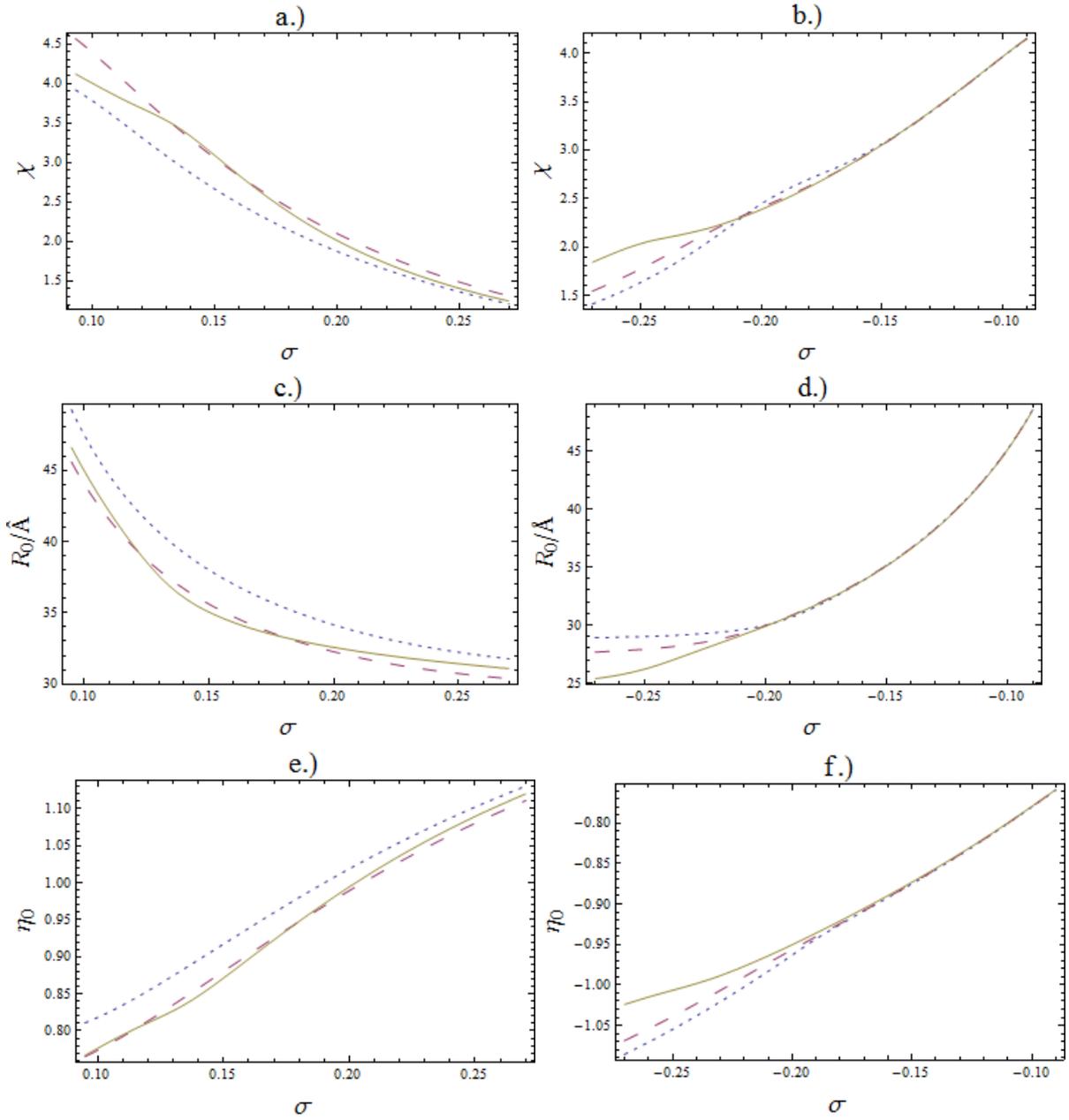

Fig. F.3. Mean supercoil structural parameters as a function of supercoiling density. In all plots we set $\theta = 0.6$ in the interaction energy. We keep the ratio $f_1/f_2$ fixed to the value $0.7/0.3$, but change the value of $f_1 + f_2$, which measures the degree of charge localization near the DNA grooves. In panels a.) (positive $\sigma$ values) and b.) (negative $\sigma$ values) we plot $\chi$, the ratio of average writhe $\langle Wr \rangle$ to average twist difference $\langle \Delta Tw \rangle$, away from torsionally relaxed DNA, . In panels c.) (positive $\sigma$) and d.) (negative $\sigma$) we plot the average inter-axial separation, $R_0$, between the two segments in the plectoneme braid. Finally in e.) (positive $\sigma$) and f.) (negative $\sigma$) we plot the average tilt angle, $\eta_0$ the angle between the tangents of the molecular centre lines of the two segments in the braid. In all plots the solid dark yellow, long dashed purple and short dashed blue lines correspond to $f_1 + f_2 = 0, 0.5$ and $1$, respectively.

In Fig. F.3, we examine how the average supercoil geometric parameters change, when $f_1 + f_2$ is changed. Interestingly enough, positive supercoils are affected more by changing $f_1 + f_2$, unlike when $\theta$ was changed. For the most part, for positive $\sigma$ values, the curves for both $f_1 + f_2 = 0$ and $0.5$ lie close to each other, while the curves for $f_1 + f_2 = 1$ lie further apart from the other ones. Supercoils with the lower $f_1 + f_2$ values seem to have larger $\chi$ values, as well as both smaller values of $R_0$ and $\eta_0$, at fixed values of $\sigma$. Again, decreases in both $R_0$ and $\eta_0$ are probably linked together through elastic forces, which favour smaller values of $\eta_0$ for small $R_0$. For negative values of $\sigma$, the supercoil geometric parameters are only affected significantly in the same way when $\sigma < -0.2$.

The $R_0$ results suggest that reducing $f_1 + f_2$ favours reduced repulsion. Part of the explanation of this might be due to the fact that, in the case of weak helix specific forces, there is no preferred average optimal azimuthal orientation $\langle \Delta \xi(\tau) \rangle$. Having, in this case, a majority of ions in minor groove would favour states with a particular value $\langle \Delta \xi(\tau) \rangle \neq 0$ for increased attraction between the segments, if thermal fluctuations were sufficiently low to allow for such a state. On the other hand, if the ions were smeared there might be less of an energetic preference for a particular value of $\langle \Delta \xi(\tau) \rangle$, and an increased entropy due to fluctuations in $\Delta \xi(\tau)$. Thus, in the case of strong thermal fluctuations, where all azimuthal orientations are more or less equally distributed, smearing out of the ions could indeed lead to increased attraction on average, when $f_1 + f_2 = 0$ at fixed values $f_1 / f_2 = 0.7 / 0.3$, $\theta = 0.6$. Also smearing out of the ions would have the tendency to reduce short ranged repulsive image charge interactions [2,3], which would increase as counter-ion charge is localized.

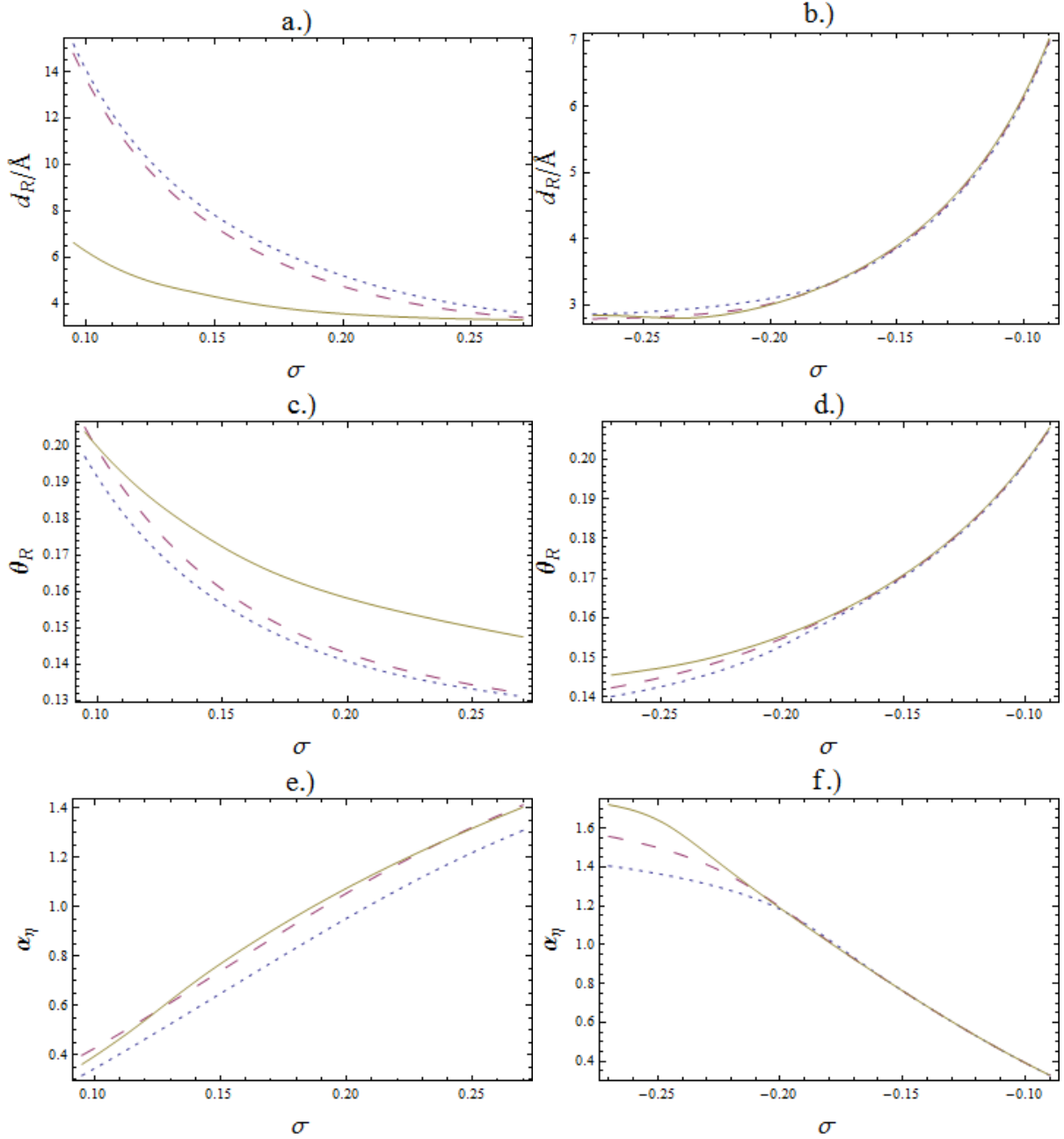

Fig.F.4. Supercoiling fluctuation parameters as functions of $\sigma$, keeping fixed $\theta = 0.6$ and $f_1/f_2 = 7/3$, but varying $f_1 + f_2$. On the left hand side we present curves for positive values of $\sigma$, whereas on the right we present curves for negative $\sigma$ values. In panels a.) and b.) we present plots of the mean squared amplitude of undulations in the supercoil $d_R = \langle \delta R(\tau)^2 \rangle^{1/2}$. In panels c.) and d.) we present the angular fluctuation amplitude $\theta_R = \langle (dR(\tau)/d\tau)^2 \rangle^{1/2}$. Finally, in panels e.) and f.) we present plots of the variational 'spring constant' $\alpha_\eta$ that determines the size of the fluctuations in the tilt angle $\eta(\tau)$. In all plots the blue short dashed, red medium dashed and dark yellow lines correspond to the values $f_1 + f_2 = 1, 0.5$ and $0$.

In Fig. F.4, we also plot the fluctuation parameters $d_R$, $\theta_R$ and $\alpha_\eta$. Noticeably, the degree of asymmetry in them is much less for the delocalized case $f_1 + f_2 = 0$ than in other cases. For negative supercoils $d_R$ is markedly reduced for $f_1 + f_2 = 0$, and is far more symmetric between left and right supercoils. This could be due to the fact that, for this distribution of ions, the interaction energy is least sensitive to the value of $\eta_0$. Instead, what could account for the asymmetry in $R_0$ is that for positive supercoils that the entropy associated with fluctuations in $\Delta \xi(\tau)$ decreases more rapidly, as the mean separation between two segments is reduced, than for negative ones. Thus, a larger value of $R_0$ might be favoured due to entropic repulsion due to fluctuations in $\Delta \xi(\tau)$. Also, again, changing $g_{av}$ may have a role to play.

**F.2. Changing the proportions of localized ions near the grooves**

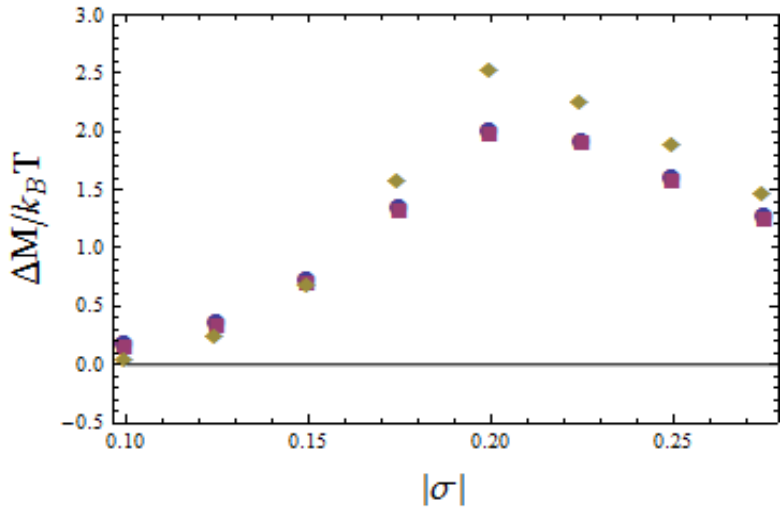

Fig. F.5. Difference between moments required to produce left and right handed supercoils, $\Delta M = M(\sigma) + M(-\sigma)$. In all plots we set $\theta = 0.6$ in the interaction energy, and also keep the sum $f_1 + f_2$ fixed to 1, but change the value of $f_1 / f_2$, the proportion of ions localized between the minor and major grooves. The solid dark yellow points, purple points and blue points correspond to the values $f_1 / f_2 = 1, 0.6 / 0.4$ and $0.7 / 0.3$, respectively.

We now examine what happens when we keep both fixed $\theta = 0.6$ and $f_1 + f_2 = 1$, but change the proportion of counter-ions within the grooves. Unfortunately, we have not been able to examine what happens when most the ions are localized in the major groove. This, due to an instability in free energy, for negative $\sigma$, in the equations that minimize the free energy, that arises from not including higher order terms in $d_\eta^2 = 1/(2l_p \alpha_\eta)^{1/2}$. In addition, we know from previous work on mechanical braiding [6] that, for such parameter values ($f_1 / f_2 < 1$, $f_1 + f_2 = 1$, $\theta = 0.6$), states with preferred average azimuthal orientation, $\langle \Delta \xi(\tau) \rangle$ between helices could be favoured. This situation has yet to be considered. Thus, we investigate the parameter values $f_1 = 0.5$ ($f_2 = 0.5$), $f_1 = 0.6$ ($f_2 = 0.4$) and $f_1 = 0.7$ ($f_2 = 0.3$). We notice that in all plots, across this range, that the plots of $\Delta M$ and the geometric parameters for the supercoils are most insensitive when compared to other variations in the parameters.

For the plots of $\Delta M$, presented in Fig. F.5, the most asymmetry between left and right supercoils is seen for $f_1 = 0.5$ ($f_2 = 0.5$) when $|\sigma|$ is large, and it is only slightly different from the other values. In fact, there is very little difference between the values for $f_1 = 0.6$ ($f_2 = 0.4$) and $f_1 = 0.7$ ($f_2 = 0.3$). The points for these two sets of values practically coincide with each other.

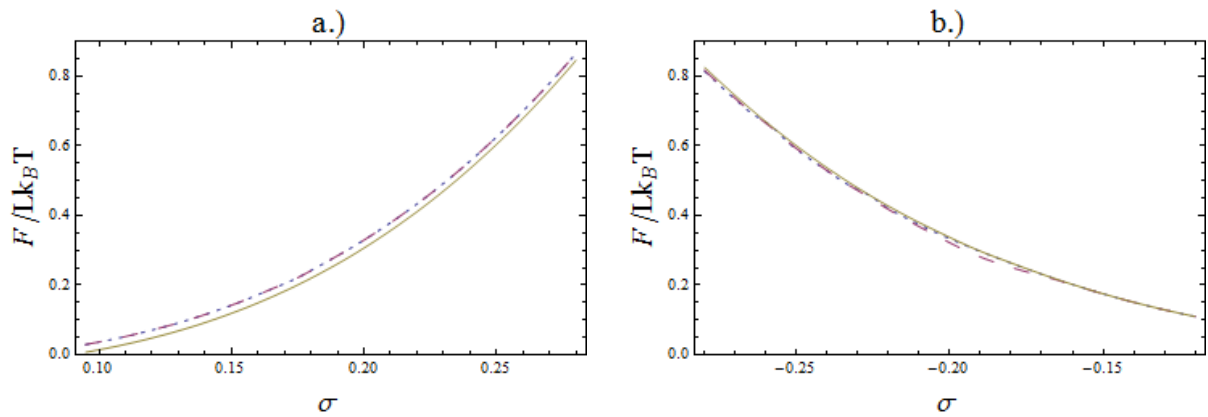

Fig. F.6. Plots of $F / Lk_BT$ the supercoiling free energy per unit length and $k_BT$. On the left hand side we present curves for positive values of $\sigma$, whereas on the right we present curves for negative values. In panels a.) and b.) we fix $\theta = 0.6$ and $f_1 + f_2 = 1$, but vary $f_1 / f_2$. In these plots, the blue short dashed line, red medium dashed line and dark yellow solid lines refer to the values $f_1 / f_2 = 1, 1.5$ and $7/3$.

For the plots of free energy, presented in Fig. F.6, there is practically no change at all as the proportions of ions localized in the grooves is changed.

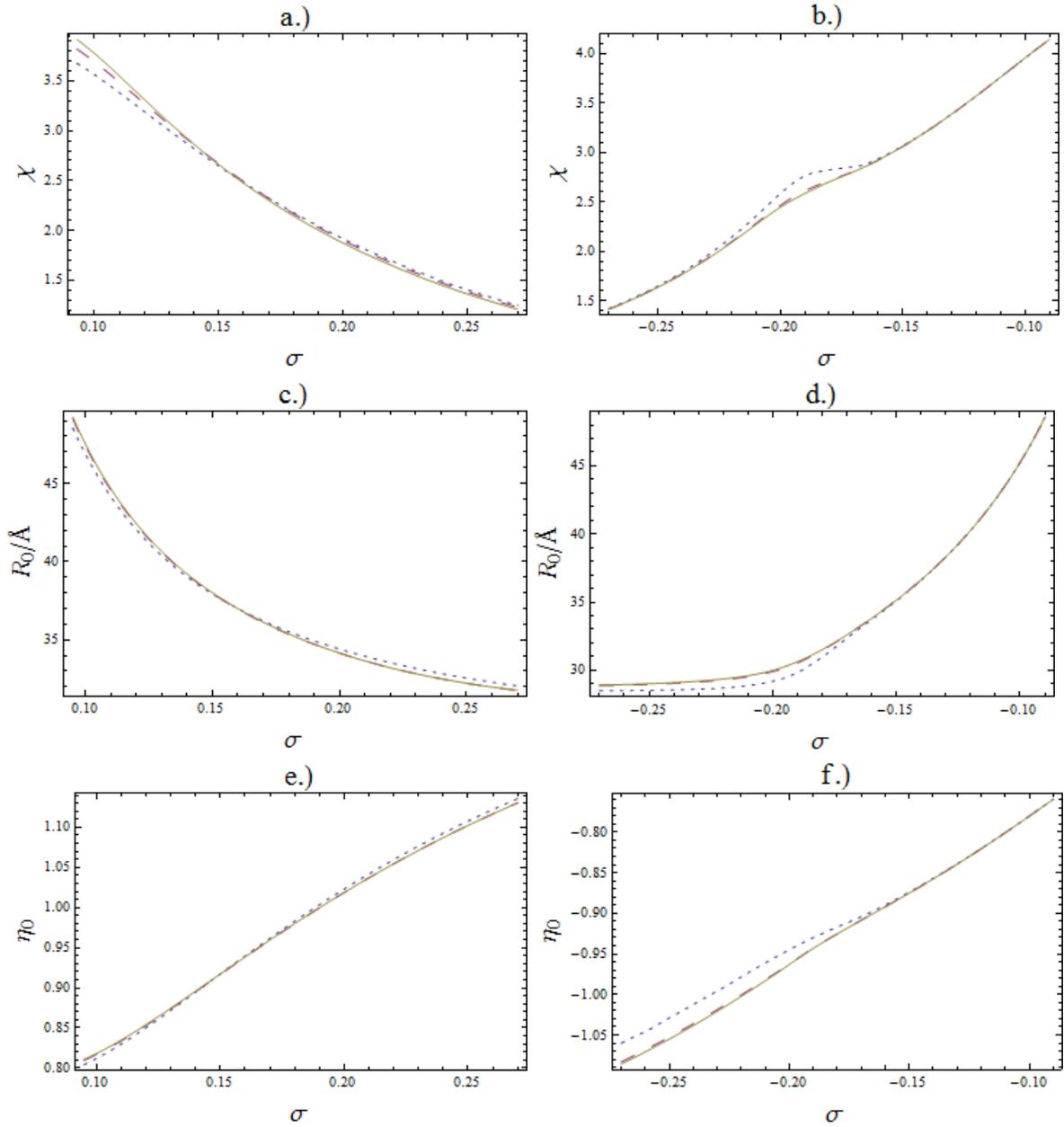

Fig. F.7. Mean supercoil structural parameters as a function of supercoiling density. In all plots we set $\theta = 0.6$ in the interaction energy, as well as the sum $f_1 + f_2$ being fixed to $1$. However, we change the value of $f_1/f_2$, the relative proportion of ions localized between the minor and major grooves. In panels a.) (positive $\sigma$ values) and b.) (negative $\sigma$ values) we plot $\chi$, the ratio of average writhe $\langle Wr \rangle$ to average twist difference $\langle \Delta Tw \rangle$, away from the twist of torsionally relaxed DNA. In panels c.) (positive $\sigma$) and d.) (negative $\sigma$) we plot the average inter-axial separation $R_0$ between the two segments in the plectoneme braid. Finally in e.) (positive $\sigma$) and f.) (negative $\sigma$) we plot the tilt angle $\eta_0$, the angle between the tangents of the molecular centre lines of the two segments in the braid. In all plots the solid dark yellow, long dashed purple and short dashed blue lines correspond to the values of $f_1/f_2 = 0.7/0.3, 0.6/0.4$ and $1$ respectively.

In Fig.7 we examine how the supercoiling geometric parameters change with the changing ratio $f_2/f_1$. For positive $\sigma$ values, there is very little change in the curves for $\chi$, $R_0$ and $\eta_0$ from

$f_1 = 0.5$ ($f_2 = 0.5$) to $f_1 = 0.7$ ($f_2 = 0.3$). For negative $\sigma$ values, for $\sigma < -1.6$, the case where $f_1 = 0.5$ ($f_2 = 0.5$) is slightly different. Here $R_0$ is slightly less for the value $f_1 = 0.5$, as well as the magnitude of $\eta_0$, suggesting less repulsion between the braided segments at this value. This could be due to a reduction in short range image charge repulsion, as well as a greater mean separation between positive and negative charges resulting also in increased attraction. For the values of $\chi$, for negative $\sigma$, there a bump where the gradient flattens, with respect to $\sigma$, before steepening again with increasing $|\sigma|$. This also becomes more pronounced when $f_1 = 0.5$ ($f_2 = 0.5$). The increase in this bump can be attributed to a steeper drop in $R_0$ at that point, thus increasing the writhe. Also, seen here is a kink in the moment curves presented in Fig F.9, below. This could be the precursor of two distinct braid states with different values of $R_0$, coexisting at a critical value of $M = M_c$, when $f_1 / f_2$ is reduced below 1– yet to be investigated. Such braid collapse and coexistence was discussed in Ref. [6] in the context of the mechanical braiding of two DNA molecules.

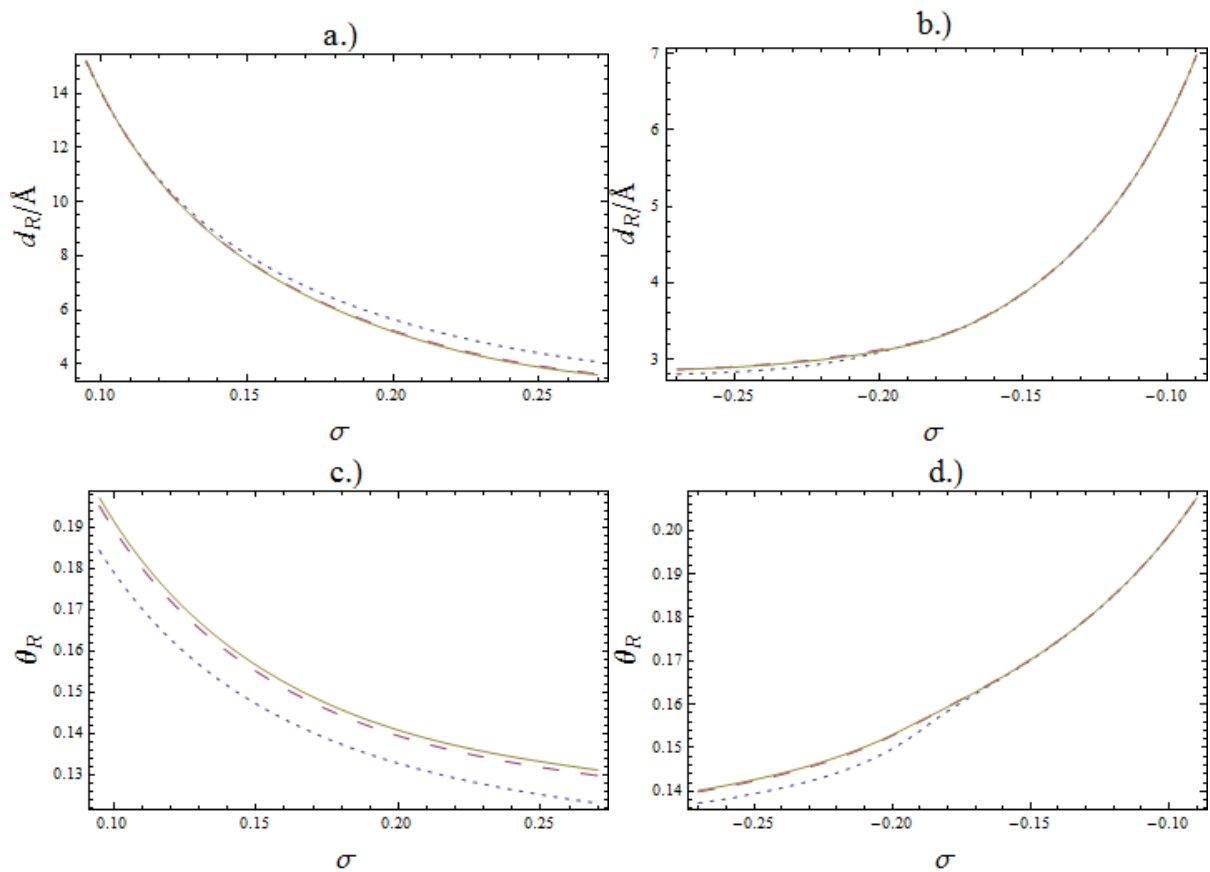

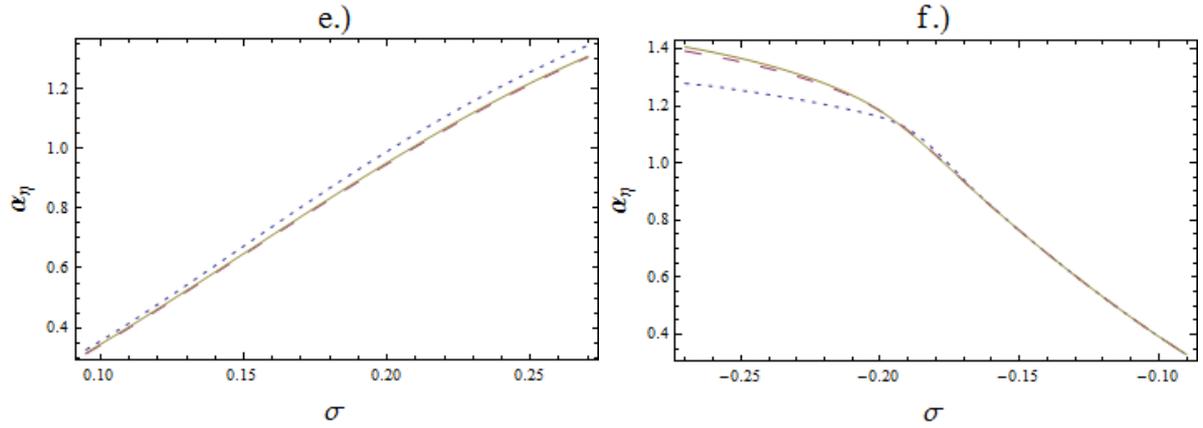

Fig. F.8. Supercoiling fluctuation parameters as functions of $\sigma$, keeping fixed $\theta = 0.6$ and $f_1 + f_2 = 1$, but varying $f_1/f_2$. On the left hand side we present curves for positive values of $\sigma$, whereas on the right we present curves for negative values. In panels a.) and b.) we present plots of the mean squared amplitude of undulations in the supercoil $d_R = \langle \delta R(\tau)^2 \rangle^{1/2}$. In panels c.) and d.) we present the angular fluctuation amplitude $\theta_R = \langle (dR(\tau)/d\tau)^2 \rangle^{1/2}$. Finally, in panels e.) and f.) we present plots of the variational 'spring constant' $\alpha_\eta$ that determines the size of the fluctuations in the tilt angle $\eta(\tau)$. In all plots the blue short dashed, red medium dashed and dark yellow lines correspond to the values $f_1/f_2 = 1, 1.5$ and $7/3$.

In Fig. F.8., we also present plots of $d_R$, $\theta_R$, and $\alpha_\eta$ as functions of both positive and negative $\sigma$. A distinct difference is found in $\theta_R$: for positive supercoils the curve is slightly lower for $f_1 = 0.5$ ($f_2 = 0.5$) than for the other parameter values. Also, a significant difference is seen for the values of $\alpha_\eta$ for negative supercoils for $\sigma \leq -0.2$, where $\alpha_\eta$ is smaller (for the case $f_1/f_2 = 1$).

### F.3 Moment Curves

Last of all, for completeness, in Fig. F.9 we show curves for $M$ as a function of $\sigma$.

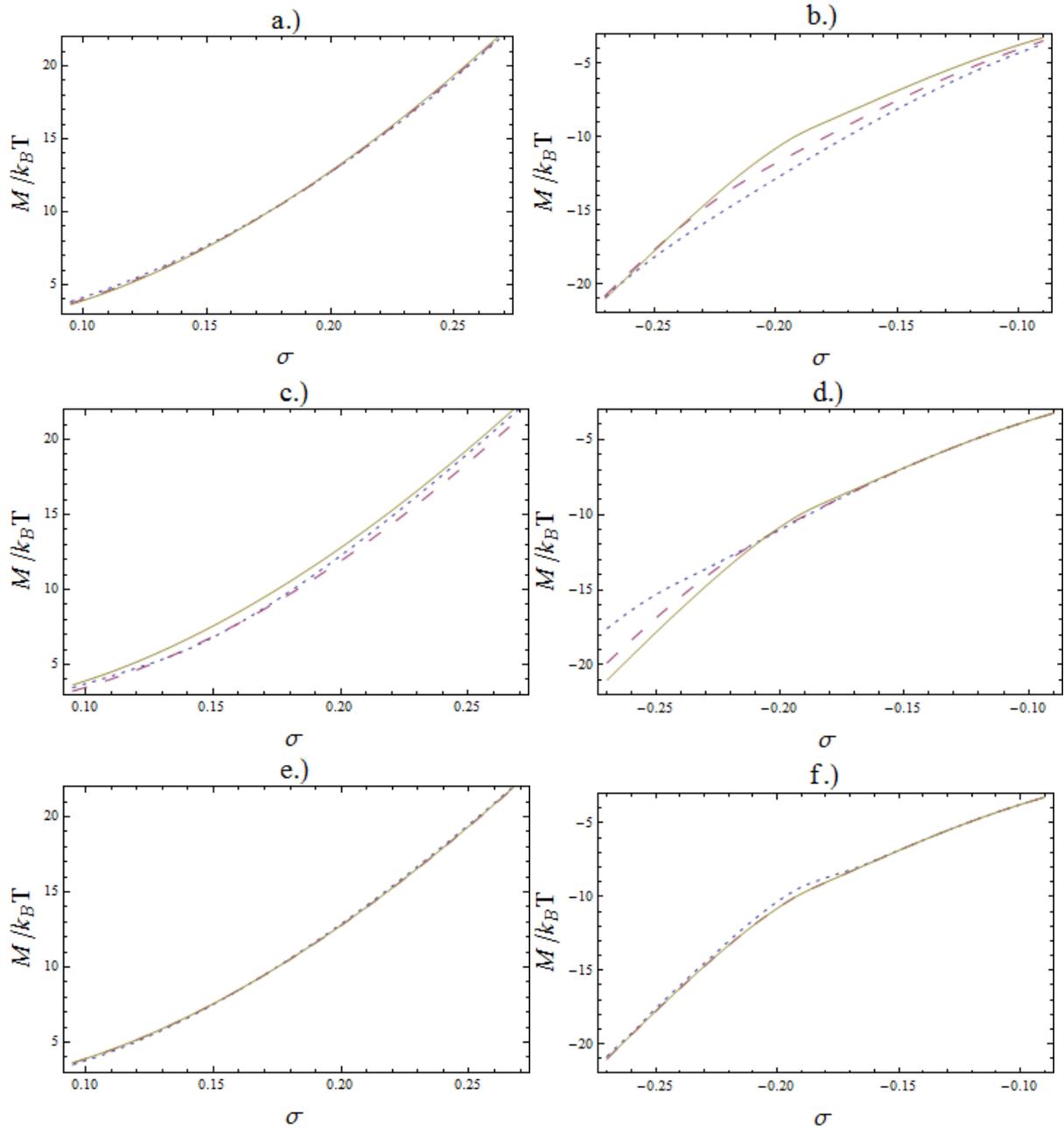

Fig F.9. Plots of the moment $M$ required to close the ends of the supercoil as a function of supercoiling density $\sigma$. On the left hand side we present curves for positive values of $\sigma$, whereas on the right we present curves for negative values. In panels a.) and b.) we investigate changing the value of $\theta$, while keeping fixed $f_1 = 0.7$ and $f_2 = 0.3$. Here, the blue short dashed line, red medium dashed line and dark yellow solid lines refer to the values $\theta = 0.4, 0.5$ and $0.6$, respectively. In panels c.) and d.) we fix $\theta = 0.6$ and $f_1/f_2 = 7/3$ and vary $f_1 + f_2$. In these plots, the blue short dashed line, red medium dashed line and dark yellow solid lines refer to the values $f_1 + f_2 = 0, 0.5$ and $1$. Lastly, in panels e.) and f.) we fix $\theta = 0.6$ and $f_1 + f_2 = 1$, but vary $f_1/f_2$. In these plots, the blue short dashed line, red medium dashed line and dark yellow solid lines refer to the values $f_1/f_2 = 1, 1.5$ and $7/3$.